\documentclass[11pt,a4paper]{article}
\usepackage[sc,noBBpl]{mathpazo}
\usepackage[mathscr]{eucal}
\usepackage{amsmath,amsfonts,amssymb,amsthm}
\usepackage{graphicx}
\newcommand\scA{{\mathscr A}}

\newcommand\scC{{\mathscr C}}
\newcommand\scD{{\mathscr D}}

\newcommand\scG{{\mathscr G}}

\newcommand\scI{{\mathscr I}}
\newcommand\scJ{{\mathscr J}}

\newcommand\scL{{\mathscr L}}
\newcommand\scM{{\mathscr M}}

\newcommand\scS{{\mathscr S}}

\newcommand\scV{{\mathscr V}}

\newcommand\scX{{\mathscr X}}

\newcommand\scZ{{\mathscr Z}}
\newcommand\mvector{\boldsymbol}

\newcommand\va{\mvector{a}}
\newcommand\vb{\mvector{b}}
\newcommand\vc{\mvector{c}}
\newcommand\vd{\mvector{d}}
\newcommand\ve{\mvector{e}}
\newcommand\vf{\mvector{f}}
\newcommand\vg{\mvector{g}}
\newcommand\vh{\mvector{h}}

\newcommand\vp{\mvector{p}}
\newcommand\vq{\mvector{q}}

\newcommand\vs{\mvector{s}}

\newcommand\vx{\mvector{x}}
\newcommand\vy{\mvector{y}}
\newcommand\vz{\mvector{z}}
\newcommand\vA{\mvector{A}}

\newcommand\vE{\mvector{E}}
\newcommand\vF{\mvector{F}}

\newcommand\veta{\mvector{\eta}}
\newcommand\vxi{\mvector{\xi}}

\newcommand\vlambda{\mvector{\lambda}}
\newcommand\vLambda{\mvector{\Lambda}}

\newcommand\vvarphi{\mvector{\varphi}}

\newcommand\vzero{\mvector{0}}

\newcommand\field{\mathbb}

\newcommand\C{\field{C}}
\newcommand\Z{\field{Z}}

\newcommand\N{\field{N}}
\newcommand\Q{\field{Q}}

\newcommand\bbP{\mathbb{P}}

\newcommand\rank{\operatorname{rank}}

\newcommand\tr{\operatorname{Tr}}
\newcommand\res{\operatorname{res}}
 
 \newcommand\spectr{\operatorname{spectr}}

\newcommand\card{\operatorname{card}}

\newcommand\rmd{\mathrm{d}\mspace{1mu}}

\newcommand\CP{\ensuremath{\C\bbP}}
\newcommand\rmi{\mathrm{i}\mspace{1mu}}

\newcommand\Dt{\frac{\mathrm{d}\phantom{t} }{\mathrm{d}\mspace{1mu} t}}

\newcommand\pder[2]{\dfrac{\partial #1 }{\partial #2}}
\newcommand\abs[1]{\lvert #1 \rvert}

\newcommand\bfi[1]{{\bfseries\itshape{#1}}}

\newcommand\mtext[1]{\quad\text{#1}\quad}

\newcommand\defset[2]{\left\{{#1}\;\vert \;\; {#2} \,\right\}}
\newcommand\deftuple[2]{\left({#1}\;\vert \;\; {#2} \,\right)}

\usepackage[square]{natbib}
\theoremstyle{plain}

\newtheorem{theorem}{Theorem}
\newtheorem{lemma}{Lemma}
\newtheorem{proposition}{Proposition}

\theoremstyle{definition}

\newtheoremstyle{note}{\topsep}{\topsep}{\slshape}{}{\scshape}{}{ }{}
\theoremstyle{note}

\newtheorem{assumption}{Assumption}
\newtheorem{remark}{Remark}
\newtheorem{example}{Example}

\numberwithin{equation}{section}
\numberwithin{theorem}{section}
\numberwithin{lemma}{section}
\numberwithin{proposition}{section}
\numberwithin{remark}{section}
\numberwithin{example}{section}
\numberwithin{assumption}{section}

\numberwithin{figure}{section}
\numberwithin{table}{section}

\begin{document}
\thispagestyle{empty}
\vspace*{1em}
\begin{center}
\LARGE\textbf{Darboux points and integrability of homogeneous Hamiltonian
systems with
three and more degrees of freedom}
\end{center}
\vspace*{0.5em}
\begin{center}
\large  Maria Przybylska
\end{center}
\vspace{2em}
\hspace*{2em}\begin{minipage}{0.8\textwidth}
\small
Toru\'n Centre for Astronomy,
  N.~Copernicus University, \\
  Gagarina 11, PL-87--100 Toru\'n, Poland, \\
  (e-mail: Maria.Przybylska@astri.uni.torun.pl)
\end{minipage}\\[2.5em]
%\maketitle
%\vspace*{1.5em}
{\small \textbf{Abstract.} We consider natural complex Hamiltonian systems with $n$ degrees of freedom given by
 a Hamiltonian function which is a sum of the standard kinetic energy and a homogeneous polynomial potential
 $V$  of degree $k>2$.  The well known Morales-Ramis theorem gives the strongest known necessary conditions
for the Liouville integrability of such systems.  It states that  for each  $k$ there exists an explicitly known
infinite set $\scM_k\subset\Q$  such that if the system is integrable, then all eigenvalues of the Hessian
matrix $V''(\vd)$  calculated at a non-zero $\vd\in\C^n$ satisfying
$V'(\vd)=\vd$, belong to  $\scM_k$.

The aim of this paper is, among others, to sharpen this result.  Under certain genericity assumption concerning
 $V$ we prove the following fact. For each  $k$  and $n$ there exists a finite set $\scI_{n,k}\subset\scM_k$
such that if the system is integrable, then all eigenvalues of the Hessian matrix $V''(\vd)$  belong to
  $\scI_{n,k}$.  We give an algorithm which allows to find sets $\scI_{n,k}$.

We applied this results for the case $n=k=3$ and we found all integrable potentials satisfying the
 genericity assumption.  Among them several are new and they are integrable in a highly non-trivial way.
We found three potentials for which the additional first integrals are of degree 4 and 6 with respect to the
momenta.
}

\section{Introduction}

The problem of the integrability of low-dimensional Hamiltonian systems has been
a very active field of research
in the last half century. Let us recall that a Hamiltonian system with $n$
degrees of freedom defined by Hamilton's function $H=H(\vq, \vp)$, where $\vq
=(q_1,\ldots,q_n)$ and $\vp=(p_1,\ldots,p_n)$, is integrable in the Liouville
sense if
the canonical  equations
\begin{equation}
\label{eq:eqhamcan}
\Dt q_i= \dfrac{\partial H}{\partial p_i}, \qquad \Dt p_i =- \dfrac{\partial
H}{\partial q_i},\qquad i=1,\ldots,n,
\end{equation}
 admit $n$
functionally independent and commuting first integrals.
Investigations of physically important systems and more or less systematic
analysis  of systems depending on a certain number of parameters gave rise
to several new integrable systems. Among the tools used for such investigations two
are most popular: the direct method and the Painlev\'e test. In spite of
their great successes, these techniques are rather limited and  are not
appropriate for a more ambitious program which can be formulated in the following way:\\
\textit{Give a complete list of integrable Hamiltonian systems of a given class. In other words:
formulate  necessary and sufficient conditions for the integrability of a given
class of Hamiltonian systems.}

Such a program could be attractive if the chosen class contains, on the one hand,
physically interesting systems, and on the other hand, it is large enough, and moreover is,
in some sense, naturally distinguished. In this paper we consider natural
Hamiltonian systems given by the following Hamiltonian
\begin{equation}
H=\dfrac{1}{2}\sum_{i=1}^np_i^2+V(\vq),
\label{eq:ham}
\end{equation}
where $V(\vq)$ is a homogeneous polynomial of degree $k>2$. The canonical equations
corresponding to the above Hamiltonian  have the form
\begin{equation}
\label{eq:eqham}
\Dt \vq= \vp, \qquad \Dt \vp =-V'(\vq),
\end{equation}
where $V'(\vq)$ denotes the gradient of $V(\vq)$. We say that a potential $V$ is
integrable if the above canonical equations are integrable in the Liouville
sense.

The class of systems given by~\eqref{eq:ham} satisfies the imposed requirements.
Among homogeneous potentials there are many important systems which come
from physics, astronomy and other natural sciences, see examples in paper
\cite{Almeida:98::} and references therein. Furthermore, thanks to works
\cite{Hietarinta:87::,Yoshida:88::b} we know that if a non-homogeneous
polynomial potential $V=V_{\mathrm{min}}+\cdots +V_{\mathrm{max}}$, where
$V_{\mathrm{min}}$ and $V_{\mathrm{max}}$ are homogeneous terms of the lowest
and the highest degree, respectively, is integrable, then potentials
$V_{\mathrm{min}}$ and $V_{\mathrm{max}}$ are also integrable. Thus the integrable
homogeneous potentials can be considered as building blocks of more complicated
non-homogeneous integrable models.

Obviously, having in mind the described program, we must be equipped with
tools strong enough  to guarantee its realisation. Fortunately, we have at
our disposal two methods: the Ziglin \cite{Ziglin:82::b,Ziglin:83::b} and
Morales-Ramis \cite{Morales:01::a,Morales:99::c} theories which are very strong
and effective. In both theories obstructions to the integrability are derived
from an analysis of the variational equations along a non-equilibrium particular solution
of the considered system. In this paper we will use the Morales-Ramis theory that can be considered as algebraic extension of the Ziglin method and its detailed description with many examples one can find
in  books~\cite{Morales:99::c,Audin:01::c}, see also
\cite{Morales:01::b1,Morales:01::b2,Morales:06::,Morales:07::}.

The main theorem of the Morales-Ramis theory is following.
\begin{theorem}[Morales,1999]
\label{thm:mo}
  If a complex Hamiltonian system is integrable in the Liouville sense
  in a neighbourhood of a particular non-equilibrium solution, then the identity
  component of the differential Galois group of the variational
  equations along this solution is Abelian.
\end{theorem}
For a proof, examples and the discussion,  see~\cite{Morales:99::c}.
\begin{remark}
In the above theorem we have to distinguish two cases. If the variational
equations are Fuchsian, or its irregular singularities correspond to finite points in the phase space,  then we assume the integrability with first integrals
which are meromorphic in a connected neighbourhood of the phase curve. If the
variational equations have an irregular singularity corresponding to a point in the phase space lying at the infinity, then we assume the integrability with
 first integrals which are  meromorphic in a connected neighbourhood of the
phase curve and additionally have a meromorphic growth at the infinity. In the
case of Hamiltonian systems considered in this paper it means that first
integrals are rational.
\end{remark}
Here we underline that the above theorem is very well suited to a study of the
integrability of homogeneous potentials, because  system~\eqref{eq:eqham}
admits, at least in a generic case, a certain number of particular solutions.
For  general Hamiltonian systems~\eqref{eq:eqhamcan} it is not completely
obvious how to find particular solutions in order to have a chance to apply the
Ziglin or Morales-Ramis method.

In \cite{Morales:99::c,Morales:01::a} Morales and Ramis obtained  very nice
 general integrability obstructions for homogeneous potentials with an arbitrary
 number $n\geq 2$ of degrees of freedom and an arbitrary degree of homogeneity $k$.
 As it was observed by H.~Yoshida, system~\eqref{eq:eqham} has a particular
 solution of the form
\begin{equation}
\label{eq:part}
  \vq(t) = \varphi(t) \vd, \qquad \vp(t)= \dot\varphi(t) \vd,
\end{equation}
 provided that the scalar function
$\varphi(t)$ satisfies the following equation
\begin{equation*}
\ddot\varphi = -\varphi^{k-1},
\end{equation*}
and  $\vd\in\C^n$ is a solution of polynomial equations
\begin{equation}
\label{eq:darboux}
 V'(\vd)=\vd.
\end{equation}
A vector $\vd$ satisfying the above equations is called the proper Darboux point of the potential.

The variational equations along solution~\eqref{eq:part} have the form
\begin{equation}
  \label{eq:var}
  \dot \vx = \vy, \qquad \dot \vy = -\varphi(t)^{k-2}V''(\vd) \vx.
\end{equation}
Hessian $V''(\vd)$ of the potential $V$ calculated at a proper Darboux point
$\vd$ is a symmetric matrix. Let us assume that $V''(\vd)$  is diagonalisable.
Then there exists a
complex orthogonal $n\times n$ matrix $A$ such that the canonical
transformation
\begin{equation*}
\vx = A \veta, \qquad  \vy = A \vxi,
\end{equation*}
transforms system~\eqref{eq:var} to the form
\begin{equation*}
%  \label{eq:unve}
  \dot \eta_i = \xi_i, \qquad \dot \xi_i =
-\lambda_i \varphi(t)^{k-2}\eta_i,
\quad i=1,\ldots, n,
\end{equation*}
or simply
\begin{equation}
  \label{eq:unve}
  \ddot \eta_i = -\lambda_i \varphi(t)^{k-2}\eta_i,
\qquad i=1,\ldots, n,
\end{equation}
where $(\lambda_1,\ldots, \lambda_n)$ are eigenvalues
of $V''(\vd)$. It can be shown that if the identity component of the differential Galois group  of the above
system is Abelian, then the identity component of the differential Galois group  of each equation in the system
 is Abelian.
\begin{remark}
\label{rem:diag}
As it was explained in \cite{Duval:08::}  the assumption that $V''(\vd)$ is
diagonalisable is irrelevant. That is, the necessary conditions for the
integrability are the same: if the potential is integrable, then for each
$\lambda\in\spectr V''(\vd)$ the identity component of the differential Galois
group of  equation $\ddot \eta=-\lambda\varphi(t)^{k-2}\eta$, is Abelian.
\end{remark}
For a given energy $e\in\C^\star$ the phase curve
$\Gamma_e$ associated with the particular solution $\varphi(t)$ is, for
$k=3,4$ an elliptic, or for $k>4$ a hyperelliptic curve given by
\begin{equation}
\label{eq:hypel}
{\dot\varphi}^2 = \frac{2}{k}\left(\varepsilon - \varphi^k\right),
\qquad
   e = \frac{1}{k}\varepsilon .
\end{equation}
As it was observed by H.~Yoshida \cite{Yoshida:87::a}, each of
equations~\eqref{eq:unve} can be transformed to the hypergeometric
equation. It can be done by the following change of the independent
variable
\begin{equation}
\label{eq:yoshi}
t\longrightarrow z := \frac{1}{ \varepsilon} \varphi(t)^k.
\end{equation}
After this transformation equations ~\eqref{eq:unve} read
\begin{equation}
\label{eq:hypi}
z(1-z)\eta_i'' + \left(\frac{k-1}{k} - \frac{3k-2}{2k}z\right)\eta_i' +
\frac{\lambda_i}{2k} \eta_i =0 ,
\end{equation}
where $i=1, \ldots, n$.
Equation~\eqref{eq:hypi} is a special case of the hypergeometric differential
equation, for which, thanks to works of Schwarz \cite{Schwarz:1872::}, Kimura
\cite{Kimura:69::} and others, the differential Galois group  is well known.

This  fact combined with Theorem~\ref{thm:mo} has allowed  J.~J.~Morales-Ruiz
and J.~P.~Ramis to formulate in \cite{Morales:01::a} a general theorem
concerning the integrability of Hamiltonian systems with a homogeneous
potential.  Here, we formulate this theorem for a polynomial
homogeneous potential.
\begin{theorem}
\label{thm:MoRa}
If Hamiltonian system~\eqref{eq:eqham} with polynomial homogeneous
potential $V(\vq)$ of degree $k>2$ is meromorphically integrable in
the Liouville sense, then for a proper Darboux point the values of
$(k,\lambda_i)$ for $i=1,\ldots,n$ belong to the following list
\begin{equation}
\label{eq:tabMoRa}
\begin{tabular}{clcl}
1.& $\left( k, p + \dfrac{k}{2}p(p-1)\right)$,&2.& $\left(k,\dfrac 1
{2}\left[\dfrac {k-1} {k}+p(p+1)k\right]\right)$, \\[1em]
3.& $\left(3,-\dfrac 1 {24}+\dfrac 1 {6}\left( 1 +3p\right)^2\right)$, &
4.& $\left(3,-\dfrac 1 {24}+\dfrac 3 {32}\left(  1  +4p\right)^2\right)$,
\\[1em]
5.& $\left(3,-\dfrac 1 {24}+\dfrac 3 {50}\left(  1  +5p\right)^2\right)$,&
6.& $\left(3,-\dfrac 1 {24}+\dfrac{3}{50}\left(2 +5p\right)^2\right)$,\\[1em]
7.& $ \left(4,-\dfrac 1 8 +\dfrac{2}{9} \left( 1+ 3p\right)^2\right)$,&
8.& $\left(5,-\dfrac 9 {40}+\dfrac 5 {18}\left(1+ 3p\right)^2\right)$,\\[1em]
9.& $\left(5,-\dfrac 9 {40}+\dfrac 1 {10}\left(2+5p\right)^2\right)$,&
 &
\end{tabular}
\end{equation}
where $p$ is an integer.
\end{theorem}
\begin{remark}
\label{rem:j}
By Remark~\ref{rem:diag}, Theorem~\ref{thm:MoRa} is true independently of the
fact whether matrix $V''(\vd)$ is diagonalisable or not. However, if $V''(\vd)$ is
not diagonalisable, new obstacles for the integrability appear. Namely, if the
Jordan form of $V''(\vd)$ has a block
\begin{equation*}
J_3(\lambda):=
\begin{bmatrix}
\lambda &1&0\\
0&\lambda &1\\
0&0&\lambda
\end{bmatrix},
\end{equation*}
 then the system is not integrable. Moreover, if the Jordan form of $V''(\vd)$
 has a two dimensional block $J_2(\lambda)$, and $\lambda$ belongs to the first
 item of table~\eqref{eq:tabMoRa}, then the system is not integrable. This fact was proved in~\cite{Duval:08::}.
\end{remark}
We denote by $\mathscr{M}_k$ a subset of rational numbers $\lambda$ specified by
the table in the above theorem for a given $k$, e.g., for $k>5$ we have
\begin{equation}
 \mathscr{M}_k=\defset{ p + \dfrac{k}{2}p(p-1)}{ p\in\Z}\cup
\defset{\dfrac 1
{2}\left[\dfrac {k-1} {k}+p(p+1)k\right]}{p\in\Z}.
\end{equation}
Let us make some remarks about practical aspects of applications of
Theorem~\ref{thm:MoRa}. for a study of the integrability of a potential depending on
several parameters. At first, we have to find a proper Darboux point, i.e., to
solve nonlinear equations~\eqref{eq:darboux}. And just at this very initial
step we meet a very fundamental problem. Namely, even if
equations~\eqref{eq:darboux} have solutions, in most cases we cannot find their
explicit form. Unfortunately, it is not the only problem that we have. Even if we
find the explicit form of a Darboux point $\vd$, its coordinates depend on
parameters in a very complicated way and, moreover, we obtain much more intricate
expressions for eigenvalues $\lambda_i$ of $V''(\vd)$. So, it is at least
doubtful if we can extract useful information from the restriction
$\lambda_i\in\mathscr{M}_k$, and this is not only a technical problem. In fact, a
restriction of the form $\lambda_i\in\mathscr{M}_k$ gives rise to an infinite
number of potentials which form co-dimension one families in the parameters
space. For each of these families, the necessary conditions for the
integrability are satisfied, so we need to apply stronger theorems to check if
within these families there are integrable potentials. Hence, it seems that we
have to restrict applications of Theorem~\ref{thm:MoRa} to  very limited
classes of homogeneous potentials.

 Our aim in this paper is to show how to
overcome the above mentioned difficulties. The basic observation is following.
The existence of a proper Darboux point together with the necessary conditions
given by Theorem~\ref{thm:MoRa} put restrictions on the potential no matter if
 we know explicitly the proper Darboux point or not. The question is, if we
can deduce from this fact a kind of a global restriction on $V$. We give a
positive answer to this question. But the most amazing fact is that this answer
implies the following statement which is valid for all homogeneous polynomial potentials of degree $k>2$ which
satisfy certain genericity conditions: \\
\textit{For each $n$ and $k$  there exists a \emph{finite} set $\scI_{n,k}\subset\scM_k$ such that if $V$ is
integrable, then for each Darboux point $\vd$ of $V$ all eigenvalues of $V''(\vd)$ belong to  $\scI_{n,k}$.
Moreover there is a constructive algorithm which allows
to find sets  $\scI_{n,k}$. }

 The described general idea has found successful applications for systems with two
 degrees of freedom, see \cite{mp:04::d,mp:05::d,mp:07::h}. So nowadays all
 integrable homogeneous potentials of degree $3$ and $4$ are known\footnote{For
 $k=4$ there exists a family of potentials depending on one discrete parameter and
 it is unknown if this family contains an integrable case or not,
 see~\cite{mp:05::c} for details.}. In this paper we show that the above idea
 has a nice generalisation for systems with $n>2$ degrees of freedom. Moreover,
 we apply the obtained results to a study of the integrability of homogeneous
 potentials  with three degrees of freedom of the homogeneity degree equal to~3.

 The methods applied in
 this paper force us to work with complex Hamiltonian systems. So, we assume
 that our system given by Hamiltonian~\eqref{eq:ham} is defined on a complex
 linear symplectic space $(\C^{2n}, \omega)$ equipped with the canonical
 symplectic form
\begin{equation}
 \omega = \sum_{i=1}^n\rmd q_i\wedge\rmd p_i.
\end{equation}
 Moreover, we assume that potential $V$ is also complex, i.e.,
 $V\in\C_k[\vq]$, where $\C_k[\vq]$ denotes $\C$-linear space of homogeneous
 polynomials of degree~$k$.

In order to make a reasonable classification, we have to divide all considered
potentials into equivalent classes. To this end we proceed as
in~\cite{mp:05::c}. Namely, let $\mathrm{PO}(n,\C)$ be the complex projective
orthogonal subgroup of $\mathrm{GL}(n,\C)$, i.e.,
\begin{equation}
\mathrm{PO}(n,\C)=\{\vA\in\mathrm{GL}(n,\C),\ |\ \vA\vA^T=\alpha \vE_n,\;
\alpha\in\C^\star\},
\end{equation}
where $\vE_n$ is $n$-dimensional identity matrix.
We say that $V$ and $\widetilde V$ are equivalent if there exists $\vA\in
\mathrm{PO}(n,\C)$
such that $\widetilde V(\vq)=V_{\vA}(\vq):=V(\vA\vq)$.  Later a
potential means a class of equivalent potentials in the above sense.

\section{Darboux points}
\label{sec:dar}
In the Morales-Ramis Theorem~\ref{thm:MoRa} the considered particular solution
has a very special form~\eqref{eq:part}. To find it, we have to solve a system of
polynomial equations $V'(\vd)=\vd$.  In our considerations it is important to take into account all
particular solutions. An  analysis of this problem leads us to a
definition of a geometrical notion of a Darboux point of a homogeneous
potential. It was introduced in~\cite{mp:05::c} for systems with two degrees of
freedom.  However, the passage from two to three and more degrees of freedom
introduces essential difficulties which force us to modify slightly this concept.
 In this section we give basic definitions and prove several general facts
 concerning Darboux points.

Let us remark that the definition of a Darboux point given in this section is closely related to the notion
of the proper direction (radial orbits) introduced by Guillot \cite{Guillot:01::,Guillot:04::} for polynomial
differential equations with homogeneous right hand sides.

\subsection{Basic notions of algebraic geometry}
\label{ssec:alg}
At first we recall very basic notions from the algebraic geometry which are
necessary for our further considerations. The main purpose is just to fix the
notation. We refer the reader to \cite{Shafarevich:77::}  for a
more general and formal exposition.

  The set of common zeros of polynomials
$f_1,\ldots, f_s\in\C[\vx]:=\C[x_1,\ldots,x_m]$ is denoted by $\scV(f_1,\ldots,f_s)$.
It is called the affine algebraic set. An algebraic set is a finite sum of
disjoint connected components.
We say that
point $\vp\in X:=\scV(f_1,\ldots,f_s)$ is \bfi{a simple point} of $X$,  iff
\begin{equation}
\rank \begin{bmatrix}
       \pder{f_1}{x_1}(\vp) & \cdots &\pder{f_1}{x_m}(\vp)\\
\hdotsfor{3} \\
    \pder{f_s}{x_1}(\vp) & \cdots &\pder{f_s}{x_m}(\vp)
      \end{bmatrix} = m - \dim X_{\vp},
\label{eq:rrank}
\end{equation}
where $X_{\vp}$ is the component of $X$ containing $\vp$. In particular, if $X$
is a finite set, then $\vp\in X$ is a simple point, iff the
above Jacobi matrix has the maximal rank.

A polynomial $f\in\C[\vx]$ can be uniquely written as a sum of homogeneous
terms. The highest degree homogeneous term of $f$ is denoted by $f^+$.

We say that $f_1,\ldots, f_m\in\C[\vx]$  do not intersect at the infinity, iff
the only solution of $f_1^+=\cdots=f_m^+=0$ is $\vx=\vzero$. The   following fact is well known, see e.g.
\cite{Cox:05::,Kozlowski:78::,Ploski:07::}.
\begin{proposition}
\label{pro:inf}
If $f_1,\ldots, f_m\in\C[\vx]$ do not intersect at the infinity, then set
$\scV(f_1,\ldots,f_m)$ is finite.
\end{proposition}

A point in the $m$ dimensional complex projective space $\CP^m$ is specified by
its homogeneous coordinates $[\vz]=[z_0: \cdots:z_m]$, where
$\vz=(z_0,\ldots,z_m)\in\C^{m+1}$. If
\begin{equation}
 U_i:=\defset{[z_0: \cdots:z_m]\in\CP^m}{z_i\neq 0}\mtext{for}i=0,\ldots,m,
\end{equation}
 then
\begin{equation}
 \CP^m=\bigcup_{i=0}^m U_i,
\end{equation}
and we have natural coordinate maps
\begin{equation*}
 \theta_i:\CP^m \supset U_i\rightarrow \C^m ,\qquad
  \theta_i([\vz])=(x_1,\ldots,x_m),
\end{equation*}
where
\begin{equation}
 (x_1,\ldots,x_m)=\left(\frac{z_1}{z_i}, \ldots,
 \frac{z_{i-1}}{z_i}, \frac{z_{i+1}}{z_i}, \ldots, \frac{z_{m}}{z_i}\right).
\end{equation}
Each $U_i$ is homeomorphic to $\C^m$. It is easy to check that charts
$(U_i,\theta_i)$, $i=0,\ldots, m$ form an atlas which makes $\CP^m$ an
holomorphic $m$-dimensional manifold.
It is customary to choose one $U_i$, e.g., $U_0$, and call it the affine part
of $\CP^m$. Then we define the hyperplane at the infinity
\begin{equation}
 H_\infty:=\defset{[z_0: \cdots:z_m]\in\CP^m}{z_0= 0}.
\end{equation}
It is clear that $\CP^m=U_0\cup H_\infty$ and $U_0\cap H_\infty=\emptyset$. The
following map
\begin{equation}
 \theta_\infty:H_\infty\rightarrow\CP^{m-1}, \quad
  \theta_\infty([0:z_1:\cdots:z_m])=[z_1:\cdots:z_m],
\end{equation}
is the bijection, so $\theta_\infty(H_\infty)=\CP^{m-1}$. The above consideration
shows that  $\CP^m$ is homeomorphic to
$\C^m\cup\CP^{m-1}$.

Let $ F_1, \ldots, F_s\in \C[\vz]$ be homogeneous polynomials.   Then  their common zero locus in $\CP^m$, i.e.,

\begin{equation}
 \scV(F_1, \ldots,F_s):=\defset{[\vz]\in\CP^{m}}{F_i(\vz)=0, \quad
i=1,\ldots,s},
\end{equation}
is called the projective algebraic set. The affine part of $X:= \scV(F_1,
\ldots,F_s)$ is, by definition $X\cap U_0$. It is homeomorphic to the algebraic set
$\scV(f_1,\ldots,f_s)$, where $f_i$ is a dehomogenisation of $F_i$, that is
\begin{equation}
 f_i(x_1,\ldots,x_m)=F_i(1,x_1,\ldots,x_m) \mtext{for} i=1, \ldots,s.
\end{equation}
Let us show this. Notice that if $F\in\C[\vz]$ is a homogeneous polynomial, then
$F(\vz)$ is not
a function of point
$[\vz]\in\CP^m$. In order to describe the hypersurface $\scV(F)\in\CP^m$ as a zero set
of a certain function defined on  a subset of $\CP^m$, we define on $U_0$ the following function
\begin{equation}
 \widetilde F:U_0\rightarrow \C, \qquad [z_0:\cdots:z_m]\mapsto \widetilde
 F([\vz]):= z_0^{-deg F}F(\vz),
\end{equation}
and then
\[
 \scV(F)\cap U_0=\defset{[\vz]\in U_0}{ \widetilde F([\vz]) =0}.
\]
The function  $\widetilde F$ is called the function defining $\scV(F)$ on $U_0$. If we set
\[
f=\widetilde F\circ \theta_0^{-1}:\C^m\rightarrow\C,
\quad
f(x_1,\ldots,x_n):=F(1,x_1,\ldots, x_m),
\]
then
\begin{equation*}
\theta_0( \scV(F)\cap U_0)=\scV(f).
\end{equation*}
Polynomial $f$ is called the dehomogenisation of $F$ with respect to the first variable.
The above shows that
\[
   \theta_0\left( \scV(F_1,\ldots, F_s)\cap U_0 \right)=\scV(f_1,\ldots,f_s)
   \]
as we wanted to show.

Let $[\vz]\in U_i$ for a certain $0\leq i\leq n$. We say that $[\vz]$ is a simple
point of $\scV(F_1,\ldots,F_s)$ iff
$\theta_i([\vz])$ is a simple point of $\scV(f_1,\ldots,f_s)$, where $f_j$ is
the dehomogenisation of $F_j$ with respect to the $i$-th variable, for $j=1,\ldots,s$.

For the later use we will need explicit relations between representations of
$\scV(F_1,\ldots,F_s)$ on two different charts. Let
\begin{equation*}
  \theta_1\left( \scV(F_1,\ldots, F_s)\cap U_1\right)=\scV(h_1,\ldots,h_s),
\end{equation*}
where $h_i$ is the dehomogenisation of $F_i$ with respect to the second variable. A relation
between polynomials $h_i$ and $f_i$ can be found with a help of the transition map
$\psi=\theta_0\circ\theta_1^{-1}$. Namely, if
\begin{equation}
 \theta_0([z_0:\cdots:z_m])=(x_1, \ldots, x_m), \mtext{for}[z_0:\cdots:z_m]\in U_0,
\end{equation}
and
\begin{equation}
 \theta_1([z_0:\cdots:z_m])=(y_1, \ldots, y_m), \mtext{for}[z_0:\cdots:z_m]\in U_1,
\end{equation}
then
\begin{equation}
 (x_1,\ldots, x_m)=\psi(y_1,\ldots,y_m)= \left( \frac{1}{y_1},
 \frac{y_2}{y_1}, \ldots,  \frac{y_m}{y_1}\right).
\end{equation}
Then, one can check that
\begin{equation}
\label{eq:hi}
 h_i(\vy):=y_1^{\deg f_i}f_i(\psi(\vy))= y_1^{\deg f_i}f_i\left( \frac{1}{y_1},
 \frac{y_2}{y_1}, \ldots,  \frac{y_m}{y_1}\right),
\end{equation}
for $i=1,\ldots, s$.

In our considerations we invoke several times the B\'ezout Theorem.   Let us recall its formulation, as  the same name have  several different theorems.
\begin{theorem}[B\'ezout]
\label{thm:bezout}
The number of solutions $N$  of a system of $m$ homogeneous equations
\begin{equation*}
 F_i(\vz)=0, \quad i =1,\ldots, m,
\end{equation*}
in $m+1$ unknowns $\vz:=(z_0, \ldots, z_m)$   is either infinite, or is equal
 \begin{equation*}
  N=\prod_{i=1}^m \deg F_i,
 \end{equation*}
provided that solutions considered as points in $\CP^m$ are counted with their multiplicities.
\end{theorem}
A precise definition   of the multiplicity notion  used in this theorem is highly technical and needs involved language of algebraic geometry.  Here we refer the reader to  Chapter~4 in \cite{Cox:05::}. For the purpose of this paper it is sufficient to know that a simple point  has multiplicity one.

\subsection{General properties of Darboux points}

Let $V$ be a homogeneous polynomial potential of degree $k>2$, i.e.,
$V\in\C_k[\vq]$. A direction, i.e., a non-zero
$\vd\in\C^n$, is called  \bfi{a Darboux point} of $V$ iff the gradient $V'(\vd)$  of
$V$ at $\vd$ is parallel to $\vd$. Hence, $\vd$ is a Darboux point of
$V$ iff
\begin{equation}
 \vd\wedge V'(\vd)=\vzero , \qquad \vd\neq \vzero,
\end{equation}
or
\begin{equation}
 V'(\vd)=\gamma \vd, \qquad \vd\neq \vzero,
\end{equation}
for a certain $\gamma\in\C$. Obviously, if $\vd$ satisfies one of the above
conditions, then $\widetilde \vd=\alpha\vd $ for any $\alpha\in\C^\star$ satisfies
them. However, we do not want to distinguish between $\vd$ and $\widetilde\vd$. Hence we
consider a Darboux point $\vd=(d_1,\ldots, d_n)\in\C^n$ as a point
$[\vd]:=[d_1:\cdots :d_n]$ in the projective space $\CP^{n-1}$.

The set ${\scD}(V) \subset \CP^{n-1}$ of all Darboux points of a potential $V$
is a projective algebraic set. In fact, ${\scD}(V)$ is the zero locus in $\CP^{n-1}$ of
homogeneous polynomials $R_{i,j}\in\C_k[\vq]$ which are components of $\vq\wedge
V'(\vq)$, i.e.
\begin{equation}
\label{eq:Rij}
 R_{i,j}:= q_i\pder{V}{q_j}-q_j \pder{V}{q_i}, \mtext{where} 1\leq i<j\leq n.
\end{equation}

We say that a Darboux point $[\vd]\in {\scD}(V)$ is \bfi{a proper Darboux point} of $V$, iff
$V'(\vd)\neq \vzero$. The set of all proper Darboux points of $V$ is denoted by
${\scD}^\star(V)$. If  $[\vd]\in {\scD}(V)\setminus {\scD}^\star(V)$, then  $[\vd]$ is called \bfi{an improper
Darboux point} of potential $V$.

We say that $[\vd]$ is an \bfi{isotropic Darboux point}, iff
 \begin{equation}
 \label{eq:iso}
 d_1^2+\cdots+d_n^2=0.
 \end{equation}
 The set of all isotropic Darboux points of potential $V$ is denoted by $\scD_0(V)$.

We consider also
two additional subsets of  $\CP^{n-1}$
\begin{equation}
\scZ(V):=\scV(V), \qquad \scS(V):= \scV\left( \pder{V}{q_1}, \ldots,
\pder{V}{q_n}\right),
\end{equation}
and we can prove the following relations.
\begin{proposition}
 For a non-zero $V\in\C_k[\vq]$ we have
\begin{enumerate}
 \item ${\scD}^\star(V)={\scD}(V)\setminus \scS(V) $,
\item $\scS(V)\subset\scZ(V)$.
\end{enumerate}
\end{proposition}
The first of the above statements follows just from the definition of
${\scD}^\star(V)$, and the second from the Euler identity for the homogeneous
polynomial $V$.

To describe the set $\scD(V)$ in the affine
coordinates we put, according to our convention,  $U_1:=\CP^{n-1}\setminus
\defset{[\vq]\in\CP^{n-1}}{q_1 =0}$, and
\begin{equation}
\label{eq:th1}
\theta_1:U_1\rightarrow \C^{n-1}, \quad \widetilde\vx:=(x_1,\ldots,x_{n-1})= \theta_1([\vq]),
\end{equation}
 where
\begin{equation}
\label{eq:xi}
x_i=\frac{q_{i+1}}{q_1}, \mtext{for}i=1,\ldots, n-1.
\end{equation}
\begin{lemma}
\label{lem:aff}
On the affine chart $(U_1, \theta_1)$ we have
\begin{equation}
\label{eq:dvu1}
\theta_1(\scD(V)\cap U_1)=\scV(g_1,\dots,g_{n-1}),
\end{equation}
where polynomials $g_1,\ldots,g_{n-1}\in\C[\widetilde \vx]$ are given by
\begin{equation}
\label{eq:g0}
v(\widetilde\vx):=V(1,x_1,\ldots, x_{n-1}), \quad g_0:= k v -\sum_{i=1}^{n-1}
 x_i\pder{v}{x_i},
\end{equation}
and
\begin{equation}
\label{eq:gi}
g_i := \pder{v}{x_i}-x_i g_0, \mtext{for}i=1, \ldots, n-1.
\end{equation}
Moreover, $[\vd]\in \scD(V)\cap U_1$ is an improper Darboux iff its affine
coordinates $\widetilde \va := \theta_1([\vd])$ satisfy $g_0( \widetilde
\va)=0$.
\end{lemma}
\begin{proof}
We know that
\begin{equation}
\label{eq:dvRij}
\scD(V)=\scV(R_{1,2},\ldots, R_{n-1,n}),
\end{equation}
where $R_{i,j},$ for $1\leq i<j\leq n$ are given by~\eqref{eq:Rij}. Hence,
as it was explained in the previous subsection, we have
\begin{equation}
\label{eq:af}
\theta_1(\scD(V)\cap U_1)=\scV(r_{1,2},\ldots, r_{n-1,n}),
\end{equation}
where
\begin{equation*}
r_{i,j}(x_1,\ldots,x_{n-1}):=R_{i,j}(1,x_1,\ldots,x_{n-1}), \mtext{for}1\leq i<j\leq n.
\end{equation*}
Direct calculations give
\begin{equation}
\label{eq:v111}
 \pder{V}{q_1}(\vq)=q_1^{k-1}\left[ kv(\widetilde\vx) -\sum_{i=1}^{n-1}
 x_i\pder{v}{x_i}( \widetilde\vx)\right]=q_1^{k-1}g_0(\widetilde\vx),
\end{equation}
and
\begin{equation}
\label{eq:vi}
  \pder{V}{q_{i}}(\vq)=q_1^{k-1}\pder{v}{x_{i-1}}(\widetilde\vx)=
  q_1^{k-1}(g_{i-1}(\widetilde\vx)+x_{i-1}g_0(\widetilde\vx)), \mtext{for} i=2,\ldots, n.
\end{equation}
Using the above formulae we easily find that
\begin{equation*}
r_{1,i+1}= g_i, \mtext{for} i=1, \ldots,n-1,
\end{equation*}
and
\begin{equation*}
r_{i+1,j+1}=x_ig_j-x_jg_i, \mtext{for} 1\leq i<j\leq n-1.
\end{equation*}
It shows that
\begin{equation}
\label{eq:xx}
\scV(r_{1,2},\ldots, r_{n-1,n})=\scV(g_1,\dots,g_{n-1}),
\end{equation}
and thus
\begin{equation}
\label{eq:dvu11}
\theta_1(\scD(V)\cap U_1)=\scV(g_1,\dots,g_{n-1}).
\end{equation}

To prove the last statement of the lemma let us assume that $[\vd]$ is a Darboux
point such that $[\vd]\in U_1$ with affine coordinates $\widetilde\va=\theta_1([\vd])$.
Then, $g_i(\widetilde{\va})=0$, for $i=1,\ldots, n-1$. If $[\vd ]$ is an
improper Darboux point, then all  partial derivatives of $V$ vanish at $\vd$.
From~\eqref{eq:v111} it follows that if
\[ \pder{V}{q_1}(\vd)=d_1^{k-1}g_0
(\widetilde \va)=0,
\]
 then necessarily $g_0( \widetilde \va)=0$, because $d_1\neq0$.
On the other hand, if $g_i(\widetilde{\va})=0$, for $i=0,\ldots, n-1$, and
$\theta_1^{-1}(\widetilde{\va})=[\va]$, with $\va:=(1,a_1,\ldots, a_{n-1})$, then
from~\eqref{eq:v111} and~\eqref{eq:vi} we have $V'(\va)=\vzero$. Since $[\va]=[\vd]$,
this shows that $[\vd]$ is an improper Darboux point.
\end{proof}

The construction which we are going to describe now plays a very important role
in our considerations. Roughly speaking, the idea is to associate with a Darboux
point which is located in $\CP^{n-1}$, a finite set of points in $\CP^n$. This
procedure is a kind of blowup.

Let us define the following $n$ homogeneous polynomials of $n+1$
variables $\widehat\vq:=(q_0,q_1,\ldots,q_n)$
\begin{equation}
F_i:= \pder{V}{q_i}-q_0^{k-2}q_i, \qquad i=1,\ldots, n,
\end{equation}
and an algebraic set $\widehat{{\scD}}(V)=\scV(F_1, \ldots, F_n)  \subset\CP^n$.

Assume that $[\vd]\in\scD^\star(V)$. Then there exists $\gamma\in\C^\star$, such that
$V'(\vd)=\gamma \vd$,
so $k-2$ points $[\sqrt[k-2]{\gamma}:d_1:\cdots:d_n]\in\CP^n$ belong to
$\widehat{{\scD}}(V)$. These points are well defined as they do not depend on a
representative for $[\vd]$. If $[\vd]$ is an improper Darboux point, then it
defines just one point $[0:d_1:\cdots:d_n]\in\CP^n$ which is a point of
$\widehat{{\scD}}(V)$.

Set $\widehat{{\scD}}(V)$ is not empty because it contains point
$[\vd_0]:=[1:0:\cdots:0]$. If
$[\widehat\vd]=[d_0:d_1:\cdots:d_n]\in\widehat{{\scD}}(V)\setminus\{[\vd_0]\}$,
then $[\vd]=[d_1:\cdots:d_n]$ is a Darboux point of $V$. Moreover, if $d_0\neq
0$, then $[\vd]$ is a proper Darboux point.

\begin{figure}[h]
\centering \includegraphics*[bb=70.0 442.0 595.0 722.0,scale=0.8]{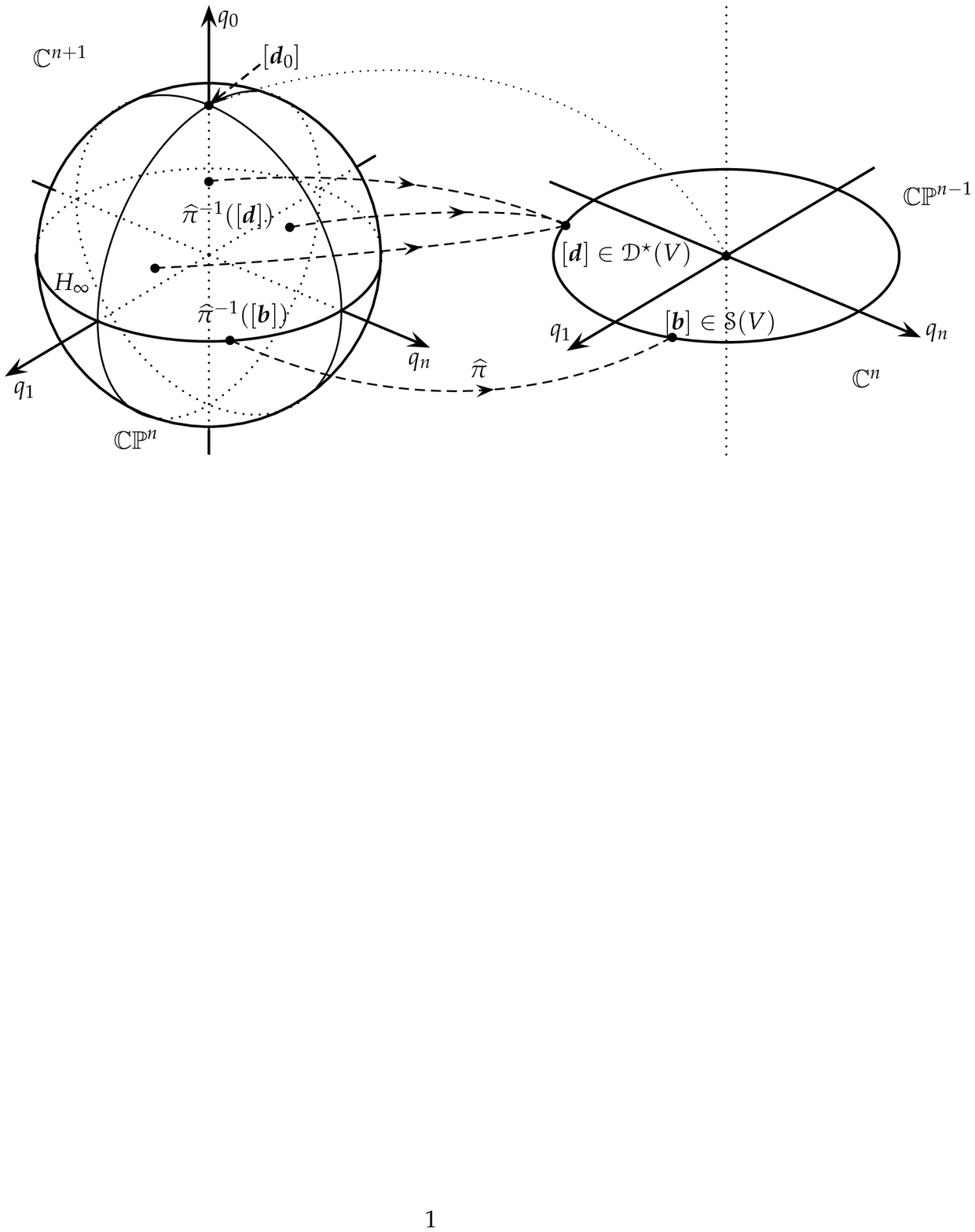}\\[1em]
\caption{Sets $\widehat\scD(V)\subset\CP^n$ and $\scD(V)\subset\CP^{n-1}$}
\label{fig:1}
\end{figure}

The natural projection
\begin{equation}
\label{eq:p}
\pi:\CP^n\setminus\{[\vd_0]\}\rightarrow\CP^{n-1}, \quad
\pi([q_0:q_1:\cdots:q_n])=[q_1:\cdots:q_n],
\end{equation}
maps $\widehat{\scD}(V)\setminus\{[\vd_0]\}$ onto $\scD(V)$, and the intersection
of the inverse image $\pi^{-1}([\vd])$ of a Darboux point $[\vd]\in\scD(V)$ with
$\widehat{\scD}(V)$ is a finite set.  We define also
\begin{equation}
\label{eq:rpi}
\widehat\pi:\widehat{\scD}(V)\setminus\{[\vd_0]\} \rightarrow \scD(V),
\end{equation}
putting  $\widehat{\pi}([\widehat{\vd}]):= \pi([\widehat{\vd}])$  for
$[\widehat{\vd}]\in \widehat{\scD}(V)\setminus\{[\vd_0]\}$. That is,
$\widehat{\pi}$ is
the restriction of $\pi$ to $\widehat{\scD}(V)\setminus\{[\vd_0]\}$. This construction is illustrated in
the Figure~\ref{fig:1}.
\begin{example}
\label{ex:1}
 For potential
\begin{equation}
 V=\frac{1}{k}\sum_{i=1}^n q_i^k, \qquad k>2,
\end{equation}
 $\scS(V)=\emptyset$, so ${\scD}^\star(V)={\scD}(V)$.
Let us calculate the number of points of  $\widehat{{\scD}}(V)$.
Since $F_i= q_i^{k-1}-q_0^{k-2}q_i$ for $i=1, \ldots, n$, it is easy to find that
$\card \widehat{{\scD}}(V)=(k-1)^n$. Hence, $\card{\scD}(V)=
\card{\scD}^\star(V):=D(n,k)$  where
\begin{equation}
 D(n,k)= \frac{(k-1)^n - 1}{k-2}.
\end{equation}
\end{example}
Next lemmas show how to use the above construction for proving several facts
which are not so obvious.

At first we distinguish generic potentials.
We say that \bfi{potential $V$ is generic}  iff all its Darboux points are proper and simple.  This definition is justified be the following  lemma.

\begin{lemma}
 The set of  generic potentials  $\scG_{n,k}\subset\C_k[\vq]$ of  degree $k$ is a non-empty open set in $\C_k[\vq]$.  A generic $V\in\C_k[\vq]$ has $D(n,k)$  proper Darboux points.
\end{lemma}
\begin{proof}
 We proceed as in the proof of Proposition~4 in \cite{Guillot:04::}, see also \cite{Guillot:01::}.
Except for point $[\vd_0]$, every point of  $\widehat{{\scD}}(V)$   is a Darboux point.
Moreover, every proper Darboux point gives  $k-2$ points of $\widehat{{\scD}}(V)$. Hence, if
hypersurfaces   $ \scV(F_i)\subset \CP^n$
intersect at a finite number of points in $\CP^n$  and all these intersections
are transversal, then by the projective B\'ezout Theorem~\ref{thm:bezout}  we have $(k-1)^n$ of such
intersections which give $D(n,k)$ Darboux points.
Example~\ref{ex:1} shows that for $n\geq 2$ and $k>2$ set $\scG_{n,k}$ is not empty.  Fact that $\scG_{n,k}$ is open in $\C_k[\vq]$ considered as a finite dimensional $\C$-vector space with the standard topology, was proved in \cite{mp:07::a}, Lemma~2.
\end{proof}
Let us note that this lemma is a generalisation of Corollary 2.1 in \cite{mp:05::c}, when the case $n=2$ was considered and  then a generic potential has $D(2,k)=k$ proper Darboux points.

\begin{lemma}
 For an arbitrary $V\in\C_k[\vq]$ the set of Darboux points ${\scD}(V)$ is not empty.
\end{lemma}
\begin{proof}
 If ${\scD}(V)=\emptyset$, then $\widehat{{\scD}}(V)=\{[\vd_0]\}$. Then, by the B\'ezout
theorem the multiplicity of $[\vd_0]$ is greater than one. However, as it is
easy to notice, $[\vd_0]$ is a simple point of $\widehat{{\scD}}(V)$. A
contradiction finishes the proof.
\end{proof}
On the affine chart $(U_0,\theta_0)$, the affine part of $\widehat{{\scD}}(V)$
is an algebraic set $\scV(f_1, \ldots,f_n)$, where
$f_i$ is the dehomogenisation of $F_i$, i.e.,
\begin{equation}
\label{eq:fi0}
f_i(q_1,\ldots, q_n):=F_i(1,q_1,\ldots, q_n)=\pder{V}{q_i}-q_i, \mtext{for} i=1,\ldots, n.
\end{equation}
Notice that the polynomials $f_1,\ldots, f_n$ intersect at the infinity iff the
gradient $V'$ vanishes at a certain $\vd\neq \vzero$. But such $\vd$ gives
an improper Darboux point $[\vd]$. On the other hand, the points of intersections of
$\widehat{{\scD}}(V)$ with the hypersurface at the infinity $H_\infty$
correspond exactly to the improper Darboux points, more precisely
\begin{equation*}
\theta_\infty(\widehat{\scD}(V)\cap H_{\infty} )=\scD(V)\setminus\scD^\star(V ).
\end{equation*}
The lemma below gives a necessary condition for  the finiteness of set $\scD(V)$.
\begin{lemma}
If a potential $V$ does not have the improper Darboux points, then it has a finite
number of Darboux points.
\end{lemma}
\begin{proof}
Under assumption of the lemma we have $\widehat{{\scD}}(V)\cap H_\infty
=\emptyset$. So, all points of $\widehat{{\scD}}(V)$ are located in the
affine part of $\CP^n$, and their affine coordinates are points of
$\scV(f_1,\ldots,f_n)$. Since, by the assumption,
$f_1,\ldots,f_n $ do not intersect at the infinity, by Proposition~\ref{pro:inf},
set $\scV(f_1,\ldots,f_n)$ is finite. Hence, $\widehat{{\scD}}(V)$ as well as
$\scD(V)$ are finite.
\end{proof}
\begin{remark}
\label{rem:mv}
Let $[\vd]\in\scD(V)$ be a Darboux point of potential $V$. Then for each
$\vA\in\mathrm{PO}(n,\C)$, the  potential $V_{\vA}$ equivalent to $V$ has Darboux point
$[\vA\vd]$. This fact allows us to fix coordinates of  one Darboux point. Namely, if $[\vd]$ is
not isotropic, then we can assume that $\vd=(0,\ldots, 0,1)$. If $[\vd]$ is
isotropic, then we can assume that
$\vd=(0,\ldots,0,\rmi, 1)$.
\end{remark}
For a Darboux point $[\vd]\in\scD(V)$ we can calculate eigenvalues
$\lambda_1(\vd), \ldots, \lambda_n(\vd)$ of the Hessian matrix $V''(\vd)$.
However, numbers $\lambda_i(\vd)$ are not well defined, as they depend on a
representative $\vd$ of the Darboux point $[\vd]$. There are several
possibilities to define properly the quantities related to the eigenvalues of
$V''(\vd)$ which do not depend on a choice of a representative of $[\vd]$.
However, because of some historical reasons and the convention widely accepted in the
literature, we choose the one which is a simple normalisation. Namely, if it is not
otherwise stated, we always
assume the following
\begin{assumption}
\label{as:1}
 If
$[\vd]$ is a proper Darboux point, then the chosen representative is such that it satisfies
$V'(\vd)=\vd$.
\end{assumption}
If
$[\vd]$ is an improper Darboux point, then the representative of
$[\vd]$ can be chosen arbitrarily.

Before we formulate our next lemma let us introduce    the following auxiliary system
\begin{equation}
 \label{eq:auxgen}
 \Dt \vq= \vf(\vq):=V'(\vq)-\gamma\vq,
\end{equation}
where $\gamma\in\C$. If $[\vd]$ is a Darboux point, then $\vf(\vd)=\vzero$ for a
certain $\gamma$, i.e., $\vd$ is an equilibrium of~\eqref{eq:auxgen}. If
$\gamma\neq0$, then $[\vd]$ is a proper Darboux point and for $\gamma=0$ is an improper one. On the other hand,
 if
$\vd$ is an equilibrium of~\eqref{eq:auxgen} for a certain $\gamma$, then
$[\vd]$ is a Darboux point. Under Assumption~\ref{as:1}, if $[\vd]$ is a proper
Darboux point, then $\vd$ is the equilibrium of~\eqref{eq:auxgen} with
$\gamma=1$, and the Jacobi matrix of the right hand sides of
~\eqref{eq:auxgen} at  $\vd$ is $\vf'(\vd)=V''(\vd)-\vE_n$, so
its eigenvalues are $\Lambda_i(\vd):=\lambda_i(\vd)-1$, for $i=1,\ldots,n$.

We make a change of coordinates in the system~\eqref{eq:auxgen}. Namely,
assuming that $q_1\neq 0$, we put
\begin{equation}
\label{eq:x0xi}
x_0 = q_1, \quad  x_i = \frac{ q_{i+1}}{q_1} \mtext{for} i=1,\ldots, n-1.
\end{equation}
Then $\vx:=(x_0, x_1, \ldots, x_{n-1})$ are
 coordinates on $\C^n\setminus\defset{\vq\in\C^n}{q_1=0}$. Notice that
 $\widetilde{\vx}:=(x_1,\ldots, x_{n-1})=\theta_1([\vq])$.
In these coordinates
system~\eqref{eq:auxgen} reads
\begin{equation}
  \label{eq:auxzog}
\begin{split}
  \dot x_0 &=w_0(\vx):= - \gamma x_0 + x_0^{k-1}g_0,  \\
 \dot x_i &=w_i(\vx):=
x_0^{k-2} g_i, \mtext{for} i=1, \ldots, n-1,
\end{split}
\end{equation}
where $g_0, \ldots, g_{n-1}\in\C[\widetilde\vx]$ are given by~\eqref{eq:g0}
and~\eqref{eq:gi}. It is clear that the Jacobi matrix $\vf'(\vx)$ has the same
spectral properties as
\begin{equation}
\label{eq:jac}
 J(\vx) =\frac{\partial(w_0,\ldots, w_{n-1})}{\partial(x_0,\ldots, x_{n-1})}(\vx).
\end{equation}
 However, it appears that using $J(\vx)$  we obtain simpler proofs of some facts.

Under Assumption~\ref{as:1} we have the following simple but important consequences.
\begin{lemma}
\label{lem:sls}
 If $[\vd]$ is a proper Darboux point of homogeneous potential $V$ of degree $k>2$, then
\begin{equation}
\label{eq:slsp}
 \vq(t):=\varphi(t) \vd,  \quad  \vp(t):=\dot\varphi(t) \vd,
\end{equation}
is a solution of Hamilton's equation~\eqref{eq:eqham} provided  $\ddot
\varphi = -\varphi^{k-1}$. Moreover,
$V''(\vd)\cdot\vd =\lambda_n \vd$ with $\lambda_n=k-1$, and if additionally
 $[\vd]$ is isotropic, then $\lambda_n$ is a multiple eigenvalue of $V''(\vd)$.

If $[\vd]$ is an improper Darboux point, then
\begin{equation}
\label{eq:slsn}
 \vq(t):=t \vd,  \quad  \vp(t):= \vd,
\end{equation}
is a solution of Hamilton's equations~\eqref{eq:eqham}. Moreover,
$V''(\vd)\cdot\vd =\lambda_n\vd$, with $\lambda_n=0$, and if additionally
 $[\vd]$ is isotropic, then $\lambda_n$ is a multiple eigenvalue of $V''(\vd)$.
\end{lemma}
\begin{proof}
 The fact that~\eqref{eq:slsp} and~\eqref{eq:slsn} are solutions of~\eqref{eq:eqham}
 can be checked directly. Moreover, the fact that $\vd$ is an eigenvector of
 $V''(\vd)$ with the prescribed eigenvalue follows directly from the Euler
 identity. Hence, we have only to show that if $[\vd]$ is isotropic, then
 $\lambda_n$ is a multiple eigenvalue of $V''(\vd)$.

Let us assume that $[\vd]$ is a proper and isotropic Darboux point.
We show that matrix $V''(\vd) -\vE_n$ has the multiple eigenvalue $\Lambda_n=k-2$.
Here we notice that this matrix is, under Assumption~\ref{as:1}, exactly the
Jacobi matrix of the right hand sides of the auxiliary
system~\eqref{eq:auxgen} calculated at equilibrium point $\vd$.
Hence eigenvalues of $V''(\vd) -\vE_n$ and their multiplicities are the same as
eigenvalues of the Jacobi matrix of system~\eqref{eq:auxzog} calculated at
an appropriate point.
Direct calculations give
\begin{equation}
\label{eq:Jx}
 J(\vx)= \begin{bmatrix}
-\gamma+(k-1)x_0^{k-2}g_0&x_0^{k-1}\partial_{1}g_0&\cdots&x_0^{k-1}\partial_{{n-1
}}g_0\\
(k-2)x_0^{k-3}g_1&x_0^{k-2}\partial_{1}g_1&\cdots&x_0^{k-2}\partial_{{n-1}}
g_1\\
\vdots&\vdots&\vdots&\vdots\\
(k-2)x_0^{k-3}g_{n-1}&x_0^{k-2}\partial_{1}g_{n-1}&\cdots&x_0^{k-2}\partial_{
{n-1}}g_{n-1}
 \end{bmatrix}(\vx).
\end{equation}
At first we assume that $[\vd]:=[1:\rmi:0:\cdots:0]$ is an isotropic Darboux
point, so $x_0=\gamma=1$ and $\widetilde\vd=\theta_1([\vd])=(\rmi,0,\dots,0)$.
Thus we have
\begin{equation}
 g_0(\widetilde\vd) = kv(\widetilde\vd) -\rmi \pder{v}{x_1}(\widetilde\vd)=1, \quad
g_1( \widetilde\vd)=-\rmi k v(\widetilde\vd),
\end{equation}
and
\begin{equation}
   g_j=\dfrac{\partial v}{\partial x_j}(\widetilde\vd), \mtext{for} j=2,\ldots,n-1.
\end{equation}
Since  at a proper Darboux point we have $g_i( \widetilde\vd)= 0$ for $i=1,\ldots,
n-1$, we obtain also the following conditions
\begin{equation}
\label{eq:condy}
 v(\widetilde\vd)=0,\qquad \dfrac{\partial v}{\partial x_j}(\widetilde\vd)=0,\mtext{ for }  j=2,\ldots,n-1,
\end{equation}
as well as
\begin{equation}
 g_0(\widetilde\vd)=-\rmi \dfrac{\partial v}{\partial
x_1}( \widetilde\vd)=1,
\end{equation}
and finally
\begin{equation}
 J(\vd)= \begin{bmatrix}
  k-2&\partial_{1}g_0&\cdots&\partial_{{n-1}}g_0\\
0&\partial_{1}g_1&\cdots&\partial_{{n-1}}g_1\\
\vdots&\vdots&\vdots&\vdots\\
0&\partial_{1}g_{n-1}&\cdots&\partial_{{n-1}}g_{n-1}
 \end{bmatrix}(\vd).
\label{eq:jjacob}
\end{equation}
But
from
conditions \eqref{eq:condy} it follows that
\[
\dfrac{\partial g_1}{\partial x_1}(\widetilde\vd)=-(k-2)\rmi \dfrac{\partial v}{\partial
x_1}(\widetilde\vd)=k-2,
\]
and
\begin{equation}
 \dfrac{\partial g_1}{\partial x_j}(\widetilde\vd)=0,\mtext{for} j=2,\ldots,n-1.
\end{equation}
Thus, in the second row of $J(\vd)$ all but the second element are zero, and the
second element is $k-2$, so $k-2$ is an eigenvalue of $J(\vd)$ with the multiplicity
at least two.

If $[\vd]$ is an arbitrary proper and isotropic Darboux point, then we can always
change the reference frame with the help of an element of $\mathrm{PO}(n,\C)$ in
such a way that in the new frame $\vd=(1,\rmi,0,\ldots,0)\in\C^{n+1}$.

If $[\vd]$ is an improper and isotropic Darboux point, then we can assume also that it
is $[\vd]=[1:\rmi:0:\cdots:0]$, and then we proceed as in the case of the proper
Darboux point.  For an improper Darboux point $\gamma=0$, and
$g_i(\widetilde\vd)=0$ for $i=0,\ldots,n-1$, where
$\widetilde\vd=\theta_1([\vd])=(\rmi,0,\dots,0)$. Thus we have
\begin{equation}
\label{eq:co1}
 v(\widetilde\vd)=0, \mtext{and} \pder{v}{x_i}(\widetilde\vd) =0, \mtext{for}
 i=1,\ldots,n-1,
\end{equation}
and
\begin{equation}
 J(\vd)= \begin{bmatrix}
  0&\partial_{1}g_0&\cdots&\partial_{{n-1}}g_0\\
0&\partial_{1}g_1&\cdots&\partial_{{n-1}}g_1\\
\vdots&\vdots&\vdots&\vdots\\
0&\partial_{1}g_{n-1}&\cdots&\partial_{{n-1}}g_{n-1}
 \end{bmatrix}(\vd),
\label{eq:jjacob1}
\end{equation}
as $x_0=1$.
But  using~\eqref{eq:co1} and definition of $g_1$, one can easily find out that
\[
\dfrac{\partial g_1}{\partial x_i}(\widetilde\vd)=0,\mtext{for} i=1,\ldots,n-1,
\]
so indeed $0$ is a multiple eigenvalue of $J(\vd)$.
\end{proof}
We underline that the presence of more than one eigenvalue equal to $k-2$ does
not imply automatically that the Darboux point is isotropic.

Next we explain relations between the properties of a Darboux point
$[\vd]\in\scD(V)$, and properties of points
$\widehat{\pi}^{-1}([\vd])\subset\widehat{\scD}(V)$.
\begin{lemma}
\label{lem:sprop}
Assume that set $\scD(V)$ is finite. Then point $[\vd]\in \scD^\star(V)$ is a
simple point of $\scD(V)$ iff each point
$[\widehat\vd]\in\widehat{\pi}^{-1}([\vd])$ is a simple point of
$\widehat{\scD}(V)$.
\end{lemma}
\begin{proof}
Let $[\vd]\in \scD^\star(V)$ be a proper Darboux point satisfying $V'(\vd)=\vd$.
Without loss of
generality we can assume that $d_1\neq 0$. Then the affine coordinates of $[\vd]$
on the chart $(U_1,\theta_1)$ are
$\widetilde{\va}:=(a_1,\ldots,a_{n-1})=\theta_1([\vd])\in\C^{n-1}$, where
\begin{equation}
\label{eq:ai}
a_i=\frac{d_{i+1}}{d_1}, \mtext{for} i=1,\ldots,n-1,
\end{equation}
see formulae~\eqref{eq:th1} and ~\eqref{eq:xi}. Thus, $[\vd]$ is simple iff
\begin{equation*}
\det\vg'(\widetilde\va):=
\det\frac{\partial(g_1,\ldots, g_{n-1})}{\partial(x_1,\ldots,x_{n-1})}(\widetilde\va)
\neq 0.
\end{equation*}

Now, let us take point $[\widehat{\vd}]:=[1:d_1:\cdots:d_n]\in\CP^n$. Of course,
$[\widehat\vd]\in\widehat{\pi}^{-1}([\vd])$. Moreover, the affine
coordinates of $[\widehat\vd]$ on the chart $(U_0,\theta_0)$ in $\CP^n$ are
$\vd$. On this chart $\widehat{\scD}(V)$ is given by $\scV(f_1,\ldots, f_n)$,
see~\eqref{eq:fi0}. Thus $[\widehat{\vd}]$ is simple iff
\begin{equation}
\label{eq:bdim}
\det \vf'(\vd)= \det(V''(\vd)-\vE_n)\neq 0.
\end{equation}
Notice, that $\vf(\vq)$ coincides with the right hand sides of
system~\eqref{eq:auxgen} with $\gamma=1$. Thus, matrix $\vf'(\vq)$ is similar to
the matrix $J(\vx)$ defined by~\eqref{eq:jac}, with $\vx=(x_0, \ldots, x_{n-1})$
given by~\eqref{eq:x0xi}. For $\vq=\vd$, we have $\vx=\va:=(a_0,a_1,\ldots,
a_{n-1})$, where $a_0=d_1$ and $a_i$ for $i>0$ are given by~\eqref{eq:ai}. Matrix $J(\vx)$ is
determined by~\eqref{eq:Jx} (with $\gamma=1$), so we have
\begin{equation}
\label{eq:Ja}
 J(\va)= \begin{bmatrix}
-1+(k-1)a_0^{k-2}g_0&a_0^{k-1}\partial_{1}g_0&\cdots&a_0^{k-1}\partial_{{n-1
}}g_0\\
(k-2)a_0^{k-3}g_1&a_0^{k-2}\partial_{1}g_1&\cdots&a_0^{k-2}\partial_{{n-1}}
g_1\\
\vdots&\vdots&\vdots&\vdots\\
(k-2)a_0^{k-3}g_{n-1}&a_0^{k-2}\partial_{1}g_{n-1}&\cdots&a_0^{k-2}\partial_{
{n-1}}g_{n-1}
 \end{bmatrix},
 \end{equation}
 where $g_i=g_i(\widetilde{\va})$, for $i=0,\ldots,n-1)$, see~\eqref{eq:auxzog} and below.
 But because $[\vd]$ is a Darboux point, $\va$ is an equilibrium of system~\eqref{eq:auxzog},
 so $g_1(\widetilde{\va})=\cdots=g_{n-1}(\widetilde{\va})=0$, and
 $a_0^{k-2}g_0(\widetilde{\va})=1$. Thus we obtain
 \begin{equation}
\label{eq:Jaa}
 J(\va)= \begin{bmatrix}
(k-2)&a_0^{k-1}\partial_{1}g_0&\cdots&a_0^{k-1}\partial_{{n-1
}}g_0\\
0&a_0^{k-2}\partial_{1}g_1&\cdots&a_0^{k-2}\partial_{{n-1}}
g_1\\
\vdots&\vdots&\vdots&\vdots\\
0&a_0^{k-2}\partial_{1}g_{n-1}&\cdots&a_0^{k-2}\partial_{
{n-1}}g_{n-1}
 \end{bmatrix},
 \end{equation}
 and
 \begin{equation}
 \label{eq:detfin}
 \det \vf'(\vd)=\det J(\va)=\frac{k-2}{a_0^{(k-1)(n-1)}}
 \det \vg'(\widetilde\va).
 \end{equation}
 Thus we show that $[\vd]$ is simple if and only if its one preimage, namely
 $[\widehat{\vd}]$, is simple. Let
 \begin{equation*}
  [\widehat{\vd}_j]:=[\varepsilon^j:d_1:\cdots:d_n], \mtext{for}j=0,\ldots, k-3,
 \end{equation*}
 where $\varepsilon$ is a primitive $(k-2)$-root of the unity. Then
 \begin{equation*}
 \widehat{\pi}^{-1}([\vd])=\{   [\widehat{\vd}_0],\ldots,   [\widehat{\vd}_{k-3}] \}.
 \end{equation*}
 Because the affine coordinates of $ [\widehat{\vd}_j] $ are
 $\vd_j:=\varepsilon^{-j}\vd$, we have
 \begin{equation*}
 \vf'(\vd_j)=V''(\varepsilon^{-j}\vd)-\vE_n=\varepsilon^{-j(k-2)}V''(\vd)-\vE_n=\vf'(\vd).
 \end{equation*}
 Hence our conclusion is valid for an arbitrary element of the preimage of
 $[\vd]$, and this finishes the proof.
\end{proof}
If $[\vd]$ is an improper Darboux point, then the situation is more complicated.
We can always assume that $d_1\neq 0$. Then the affine coordinates of $[\vd]$ are
$\widetilde{\va}=\theta_1([\vd])$, where $\widetilde{\va}$ is given
by~\eqref{eq:ai}. Since $[\vd]$ is improper Darboux point $g_i(\widetilde{\va})=0$ for $i=0,\ldots, n-1$, and
thus, as  direct calculations show, we have
\begin{equation}
\label{eq:gp}
\vg'(\widetilde\va)= v''(\widetilde\va)\cdot\left(\vE_{n-1}+ \widetilde\va\cdot\widetilde\va^T\right),
\end{equation}
where we consider $\widetilde\va$ as a one column matrix. One can show that matrix
$\widetilde\va\cdot\widetilde\va^T$ is diagonalisable, its one eigenvalue is
$\lambda=a_1^2+\cdots+a_{n-1}^2$, and the remaining ones are zero. Thus
\begin{equation}
\label{eq:dgp}
\det \vg'(\widetilde\va )=(1+a_1^2+\cdots+a_{n-1}^2)\det v''(\widetilde\va).
\end{equation}
Note that $1+a_1^2+\cdots+a_{n-1}^2=0$ iff $[\vd]$ is  isotropic. Hence, we
have the following.
\begin{proposition}
Assume that set $\scD(V)$ is finite, and let $[\vd]\in \scS(V)$ be an
improper Darboux point. Then it is a simple point of  $\scD(V)$ iff it is not isotropic and
$\det v''(\widetilde\va)\neq 0$.
\end{proposition}
The preimage $\widehat{\pi}^{-1}([\vd])$ of an improper Darboux point
$[\vd]=[d_1:\cdots:d_n]$ is just one point, namely
$[\widehat{\vd}]:=[0:d_1:\cdots:d_n]$. Again we assume that $d_1\neq 0$.
Now, we cannot use chart $(U_0,\theta_0)$ in order to check whether $[\widehat{\vd}]$ is simple.
We will use chart $(U_1,\theta_1)$.
Hence, on the chart $(U_1,\theta_1)$, set $\widehat{\scD}(V)$ is $\scV(h_1,\ldots, h_n)$
where polynomials $\vh=(h_1,\ldots, h_n)$, according to~\eqref{eq:hi}, are
\begin{equation}
\label{eq:hif}
 h_i(\vy):=y_1^{\deg f_i}f_i(\psi(\vy))= y_1^{\deg f_i}f_i\left( \frac{1}{y_1},  \frac{y_2}{y_1}, \ldots,  \frac{y_n}{y_1}\right),
\end{equation}
for $i=1,\ldots, n$.
 Direct calculations show that
\begin{equation}
\label{eq:hiv}
 h_1(\vy):=kv(\widetilde\vy) -\sum_{i=2}^n y_i\pder{v}{y_i}(\widetilde\vy) -y_1^{k-2}, \quad h_i(\vy):= \pder{v}{y_i}(\widetilde\vy) -y_iy_1^{k-2},
\end{equation}
for $ 2\leq i\leq n$, where $\widetilde{\vy}:=(y_2,\ldots,y_n)$, and
\begin{equation*}
v(\widetilde\vy):=V(1,y_2,\ldots,y_n).
\end{equation*}
With the above formulae, we can calculate the Jacobi matrix $\vh'(\vy)$. It has
the following form
\begin{equation}
\label{eq:hj}
  \vh'(\vy)=
\begin{bmatrix}
 -(k-2)y_1^{k-3} & \pder{h_1}{y_2}(\vy)& \pder{h_1}{y_3}(\vy) &  \cdots& \pder{h_1}{y_n}(\vy) \\
-(k-2)y_2 y_1^{k-3} &\dfrac{\partial^2 v}{\partial y_2^2}(\vy) -y_1^{k-2} & \dfrac{\partial^2 v}{\partial y_2\partial y_3}(\vy)& \cdots& \dfrac{\partial^2 v}{\partial y_2\partial y_n}(\vy) \\
\hdotsfor{5}\\
-(k-2)y_n y_1^{k-3} & \dfrac{\partial^2 v}{\partial y_n\partial y_2}(\vy)&   \dfrac{\partial^2 v}{\partial y_n\partial y_3}(\vy)& \cdots& \dfrac{\partial^2 v}{\partial y_n^2}(\vy) -y_1^{k-2} \\
\end{bmatrix},
\end{equation}
where
\begin{equation}
 \pder{h_1}{y_j}(\vy)=(k-1) \pder{v}{y_j}-\sum_{i=2}^ny_i\dfrac{\partial^2 v}{\partial y_i\partial y_j}\mtext{for} 2\leq j\leq n.
\end{equation}
Notice that $\theta_1([\widehat{\vd}])=\va:=(0,a_2,\ldots, a_n)$, where
\begin{equation*}
a_{i}=\frac{d_i}{d_1}, \mtext{for} i=2,\ldots,n.
\end{equation*}
Moreover, because $[\vd]$ is an improper Darboux point, we have
\begin{equation*}
\pder{v}{y_j}(\widetilde{\va})=0,\mtext{for} j=2,\ldots, n,
\end{equation*}
where we denoted $\widetilde{\va}:=(a_2,\ldots, a_n)$.
  This gives
\begin{equation*}
\pder{h_1}{y_j}(\va)=-\sum_{i=2}^n
a_i\dfrac{\partial^2 v}{\partial y_i\partial y_j}(\widetilde\va),\mtext{for} 2\leq j\leq n.
\end{equation*}
Let $k=3$. Using the above formulae we find that
\begin{equation}
\label{eq:hpa}
  \vh'(\va)=
\begin{bmatrix}
 - 1& -v''(\widetilde\va)\cdot \widetilde\va &   \cdots&   -v''(\widetilde\va)\cdot \widetilde\va \\
-a_2  &\dfrac{\partial^2 v}{\partial y_2^2}(\widetilde{\va})  &  \cdots& \dfrac{\partial^2 v}{\partial y_2\partial y_n}(\widetilde{\va}) \\
\hdotsfor{4}\\
-a_n  & \dfrac{\partial^2 v}{\partial y_n\partial y_2}(\widetilde\va)&   \cdots& \dfrac{\partial^2 v}{\partial y_n^2} (\widetilde{\va})
\end{bmatrix}.
\end{equation}
Hence, as it is easy to see, we have
\begin{equation}
\label{eq:dha}
\det \vh'(\va)=-(1+ a_2^2+\cdots+a_n^2)\det v'' (\widetilde{\va}).
\end{equation}
The above calculations show the following.
\begin{proposition}
\label{prop:imk3}
Assume that $k=3$. Then the preimage $\widehat\vd:=\widehat{\pi}^{-1}([\vd])$ of
an improper Darboux point
$[\vd]\in\scD(V)$ is a simple point of $\widehat{\scD}(V)$ iff $[\vd]$ is a simple point of
$\scD(V)$.
\end{proposition}
For $k>3$, the point $\widehat\vd$ is not a simple point of $\widehat{\scD}(V)$,
as all elements of the first column of matrix $\vh'(\va)$ vanish. For such a
case we have the following.
\begin{proposition}
\label{prop:k>3}
Assume that $k>3$. Then
\begin{equation*}
\rank \vh'(\va)=\rank v''(\widetilde{\va})\leq n-1.
\end{equation*}
\end{proposition}
\begin{proof}
For $k>3$ we have
\begin{equation}
\label{eq:hpak}
  \vh'(\va)=
\begin{bmatrix}
 0& -v''(\widetilde\va)\cdot \widetilde\va &   \cdots&   -v''(\widetilde\va)\cdot \widetilde\va \\
0 &\dfrac{\partial^2 v}{\partial y_2^2}(\widetilde{\va})  &  \cdots& \dfrac{\partial^2 v}{\partial y_2\partial y_n}(\widetilde{\va}) \\
\hdotsfor{4}\\
0 & \dfrac{\partial^2 v}{\partial y_n\partial y_2}(\widetilde\va)&   \cdots& \dfrac{\partial^2 v}{\partial y_n^2} (\widetilde{\va})
\end{bmatrix}.
\end{equation}
The first row of this matrix is a  linear combination of the remaining ones, so our claim easily follows.
\end{proof}
With the above notation, we say that $[\vd]\in\scS(V)=\scD(V)\setminus
\scD^\star(V)$ is \bfi{minimally degenerated} iff
$\rank v''(\widetilde{\va})=n-1$.

\subsection{Obstruction for the integrability due to improper Darboux point}
As we have already explained, the proper Darboux points are important because they
give particular solutions of the considered canonical equations~\eqref{eq:eqham}
and, thanks to this, we have the necessary conditions for the integrability given
by Theorem~\ref{thm:MoRa}. The question is if we can obtain any obstruction for
the integrability from the existence of an improper Darboux point.
Lemma~\ref{lem:sls} shows that the improper Darboux points give a  particular
solution~\eqref{eq:slsn} of the considered canonical
equations~\eqref{eq:eqham}. However, this solution has an extremely simple form and
one can doubt if using it we can obtain any obstruction for the integrability.
In fact, it is easy to notice that the monodromy group of the variational
equations along solution~\eqref{eq:slsn} is trivial. Thus, in the frame of the
Ziglin theory we do not obtain any obstacles for the integrability. However,
situation is different if we are using the Morales-Ramis theory. The following
theorem demonstrates the power of the Morales-Ramis theory, and it amazingly shows
that the improper Darboux points give, in some sense, stronger obstacles for the
integrability than the proper ones.
\begin{theorem}
\label{thm:im}
Assume that a homogeneous potential $V\in\C_k[\vq]$ of degree $k>2$ admits an
improper Darboux point $[\vd]\in\CP^{n-1}$. If $V$ is integrable with rational
first integrals, then matrix $V''(\vd)$ is nilpotent, i.e., all its eigenvalues
vanish.
\end{theorem}
\begin{proof}
We prove the theorem by contradiction. We assume that there exists a non-zero
$\lambda\in\spectr V''(\vd)$. Thus, in an appropriate base, the variational
equations along solution~\eqref{eq:slsn} contain equation $\ddot \eta=-\lambda
t^{k-2}\eta$. As it was shown in~\cite{mp:05::c}
the differential Galois group of this equation is $\mathrm{SL}(2,\C)$ and hence
 the identity component of the differential Galois
group of all variational equations is not Abelian. Thus, by Morales-Ramis
Theorem~\ref{thm:mo}, the system is not integrable. A contradiction finishes
the proof.
\end{proof}
The above theorem gives a very strong generalisation of Theorem~2.4 in~\cite{mp:05::c} for
systems with an arbitrary number  of  degrees of freedom.

\section{Basic Theorems}

\subsection{Formulation}
Let $[\vd]$ be a proper Darboux point of potential $V$. Then, thanks to
Assumption~\ref{as:1} we have well defined eigenvalues $\lambda_1(\vd), \ldots,
\lambda_n(\vd)$ of the Hessian matrix $V''(\vd)$. According to our convention
$\lambda_n(\vd)=k-1$ is the trivial eigenvalue. Let
$\vlambda(\vd)=(\lambda_1(\vd), \ldots, \lambda_{n-1}(\vd))$. Hence we have the
following mapping
\begin{equation}
 \scD^\star(V)\ni [\vd]\longmapsto \vlambda(\vd)\in\C^{n-1}.
\end{equation}
Assume that $\scD^\star(V)$ is finite. Then the image of $\scD^\star(V)$ under
the above map is a finite subset of $\C^{n-1}$. The question is if we can find a
potential $V$ of degree $k$ such that the elements in the image have values
prescribed in advance. We show that the answer to this question is
negative. More precisely, we prove that among $\vlambda(\vd)$ taken at all
proper Darboux points $[\vd]\in\scD^\star(V)$  a certain number of
universal relations exists. These relations play the fundamental and central role in our
considerations.

To formulate our first theorem we define  $\vLambda(\vd)=(\Lambda_1(\vd),\ldots,
\Lambda_{n-1}(\vd))$, where $\Lambda_i(\vd):= \lambda_i(\vd)-1$ for $i=1,\ldots,n-1$.
By  $\tau_r$ for $0\leq r \leq n-1$, we denote
the elementary symmetric polynomials in $(n-1)$ variables  of degree $r$, i.e.,
\[
\tau_r(\vx):=\tau_r(x_1,\ldots,x_{n-1})=\sum_{1\leq i_1<\cdots<i_r\leq
n-1}\prod_{s=1}^r x_{i_s}, \qquad 1\leq r\leq n-1,
\]
and $\tau_0(\vx):=1$.

Our first theorem gives the explicit form of the above mentioned relations among
$\vLambda(\vd)$, $[\vd]\in\scD(V)$ for a generic potential $V$.
\begin{theorem}
\label{thm:1}
 Let $V\in\C_k[\vq]$ be a homogeneous potential of degree $k>2$ and let all its
Darboux points be proper and simple.
Then
\begin{equation}
 \sum_{[\vd]\in \mathscr{D}^\star(V)}
\frac{\tau_1(\vLambda(\vd))^r}{\tau_{n-1}(\vLambda(\vd)
)}=(-1)^{n-1}(-n-(k-2))^r,
\label{eq:rkoj}
\end{equation}
and
\begin{equation}
 \sum_{[\vd]\in \mathscr{D}^\star(V)}
\frac{\tau_r(\vLambda(\vd))}{\tau_{n-1}(\vLambda(\vd)
)}= (-1)^{r+n-1}\sum_{i=0}^{r}\binom{n-i-1}{r-i}(k-1)^{i},
\label{eq:rtau}
\end{equation}
for $r=0,\ldots,n-1$.
\end{theorem}
The proof of the above theorem is given in the next subsection. We prove at first the existence of
   relations~\eqref{eq:rkoj}
and later we show that from them
relations~\eqref{eq:rtau} follow immediately. In effect Theorem~\ref{thm:1} gives $n$
independent relations.

Let us explain the importance of Theorem~\ref{thm:1}. To do this we need more
definitions.

 Let $\scC_m$ denote the set of all  unordered tuples  $\vLambda=(\Lambda_1, \ldots,
\Lambda_{m})$, where $\Lambda_i\in\C$ for $i=1,\ldots, m$. For $M>0$, the symbol
$\scC_m^M$ denotes the set of all unordered tuples
$(\vLambda_1,\ldots,\vLambda_M)$, where $\vLambda_i\in\scC_m$, for
$i=1,\ldots,M$.

 We fix $k>2$ and $n\geq 2$, and say that
 a
tuple  $\vLambda\in\scC_{n-1}$ is \bfi{admissible} iff
$\lambda_i=\Lambda_i+1\in\scM_k$ for $i=1,\ldots, n-1$. In other words, $\vLambda_i$ is admissible
iff $\Lambda_i+1$  belongs to
items, appropriate for a given $k$,  in the table of the Morales-Ramis
Theorem~\ref{thm:MoRa}, for $i=1,\ldots, n-1$.
 We denote the set of all
admissible tuples by $\scA_{n,k}$.
If the potential $V$ is
integrable, then for each $ [\vd]\in\scD^\star(V)$, the tuple  $\vLambda(\vd)$ is
admissible.  The set of all admissible elements $\scA_{n,k}$ is countable but infinite.

 If
the set of proper Darboux points of a potential $V$ is non-empty, and $N=\card
\scD^\star(V)$, then the $N$-tuples
\begin{equation}
\label{eq:SV}
\scL(V):=\deftuple{\vLambda(\vd)}{[\vd]\in\scD^\star(V)}\in\scC^N_{n-1},
\end{equation}
is called \bfi{the spectrum of} $V$. Let $\scA^N_{n,k}$ be the subset of
$\scC^N_{n-1}$ consisting of $N$-tuples $(\vLambda_1,\ldots,\vLambda_N)$, such that
$\Lambda_i$ is admissible, i.e., $\vLambda_i\in\scA_{n,k}$, for $i=1,\ldots, N$.
We say that the spectrum $\scL(V)$ of a potential $V$ is admissible iff
$\scL(V)\in\scA^N_{n,k}$. The Morales-Ramis Theorem~\ref{thm:MoRa} says that if
potential $V$ is integrable, then its spectrum $\scL(V)$ is admissible.  However, the problem is that
the set of admissible spectra $\scA_{n,k}^N$ is infinite. We show that from
Theorem~\ref{thm:1} it follows that, in fact, if $V$ is integrable, then its
spectrum $\scL(V)$ belongs to a certain \bfi{finite} subset $\scI^N_{n,k}$ of
$\scA_{n,k}^N$. We call this set  \bfi{distinguished one}, and its elements
\bfi{distinguished spectra}.
\begin{theorem}
\label{thm:ff}
Let potential $V$ satisfy assumptions of Theorem~\ref{thm:1}. If $V$ is
integrable, then there exists a finite subset $\scI_{n,k}^N\subset \scA_{n,k}^N$, where
$N=\card \scD^\star(V)$, such that $\scL(V)\in \scI_{n,k}^N$.
\end{theorem}
Informally speaking, for fixed $k$ and $n$, we restrict the infinite number of
possibilities in each line of the Morales-Ramis table to a finite set of choices.

The importance of the above theorem immediately forces us to ask  if it is
possible to relax assumptions of Theorem~\ref{thm:1} and preserves its
conclusions. Below we give two generalisations of Theorem~\ref{thm:1}. In the
first one we restrict ourselves  to the case $k=3$, but we do not assume that all
Darboux points are proper. However, for this generalisation we obtain a smaller number of relations.
\begin{theorem}
\label{thm:2}
  Let $V\in\C_3[\vq]$ be a homogeneous potential of degree $k=3$ and let all its
Darboux points be simple.
Then relations~\eqref{eq:rkoj} and \eqref{eq:rtau} with $r=0,\ldots,n-2$, are satisfied.
\end{theorem}
In the second generalisation of Theorem~\ref{thm:1} we do not assume that all
Darboux points are proper and $k>3$ can be arbitrary.
\begin{theorem}
\label{thm:3}
  Let $V\in\C_k[\vq]$,   be a homogeneous potential of degree $k>3$ which
satisfies the following conditions
\begin{enumerate}
 \item if $[\vd]\in\mathscr{D}^\star(V)$, then $[\vd]$ is simple,
\item if $[\vd]\in \mathscr{D}(V)\setminus \mathscr{D}^\star(V) $ then $[\vd]$
is minimally degenerated.
\end{enumerate}
Then relations~\eqref{eq:rkoj} and \eqref{eq:rtau} with $r=0,\ldots,n-2$, are satisfied.
\end{theorem}
The fact that Theorems~\ref{thm:2} and~\ref{thm:3} give a smaller number of
relations than Theorem~\ref{thm:1} does not weaken their strength. Namely, we
show that the following result holds true.
\begin{theorem}
\label{thm:ff1}
Let potential $V$ satisfy assumptions of  either Theorem~\ref{thm:2} or
Theorem~\ref{thm:3}. If $V$ is
integrable, then there exists a finite subset $\scI_{n,k}^N\subset \scA_{n,k}^N$, where
$N=\card \scD^\star(V)$, such that $\scL(V)\in \scI_{n,k}^N$.
\end{theorem}
\subsection{The Euler-Jacobi-Kronecker formula and multi-dimensional residues}
\label{ssec:ejk}
In proofs of Theorems~\ref{thm:1}, \ref{thm:2} and \ref{thm:3}  formulated in the previous subsection, we use the
classical Euler-Jacobi-Kronecker formula and its known generalisations. In these
generalisations the multi-dimensional residues are used. We need to calculate
these residues effectively. In this subsection
we recall basic facts about the multi-dimensional residues and the
Euler-Jacobi-Kronecker formula. For details the reader is refered to
\cite{Aizenberg:83::,Griffiths:76::,Griffiths:78::,Tsikh:92::,Khimshiashvili:06::}.

Let $f_i:\C^n\supset U \rightarrow \C$, where $U$ is an open neighbourhood of
the origin, be holomorphic functions for $i=1, \ldots, n$, and $\vx=\vzero$ be
an isolated common zero of $f_i$. We consider differential $n$-form
\begin{equation}
 \omega:= \frac{p(\vx)}{f_1(\vx)\cdots f_n(\vx)} \,
 \rmd x_1\wedge\cdots\wedge \rmd x_n,
\end{equation}
where $p:U\rightarrow\C$ is a holomorphic function. The residue of the form
$\omega$ at $\vx=\vzero$ can be defined as
\begin{equation}
\label{eq:defres}
 \res(\omega,\vzero):=\frac{1}{(2\pi \rmi)^n}\int_\Gamma \omega,
\end{equation}
where
\begin{equation}
 \Gamma:=\defset{\vx\in U}{ \abs{f_1(\vx)}=
 \varepsilon_1, \ldots,  \abs{f_n(\vx)}=\varepsilon_n},
\end{equation}
and $\varepsilon_1, \ldots, \varepsilon_n$ are sufficiently small positive numbers.
The orientation of $\Gamma$ is fixed by
\begin{equation}
 \rmd(\arg f_1)\wedge\cdots \wedge\rmd(\arg f_n)\geq 0.
\end{equation}
Let us denote $\vf:=(f_1,\ldots, f_n)$. It can be shown that if the Jacobian
$\det \vf'(\vzero)\neq 0$, then
\begin{equation}
\label{eq:0res}
  \res(\omega,\vzero)= \frac{p(\vzero)}{\det \vf'(\vzero)}.
\end{equation}
 The following theorem gives the classical
Euler-Jacobi-Kronecker formula, see e.g. \cite{Griffiths:78::}.
\begin{theorem}[Euler-Jacobi-Kronecker]
\label{thm:ejk}
Let $f_1,\ldots, f_n\in\C[\vx]$ be non-constant polynomials such that
$\scV(\vf):=\scV(f_1,\ldots, f_n)$ is finite and all points of this set are
simple. If $f_1,\ldots, f_n$ do not intersect at infinity, then for each
$p\in\C[\vz]$ such that
\begin{equation}
\label{eq:degp}
 \deg p\leq \sum_{i=1}^n \deg f_i -(n+1),
\end{equation}
we have
\begin{equation}
\label{eq:ejk}
 \sum_{\vd \in \scV(\vf)}  \res(\omega,\vd)= \sum_{\vd \in \scV(\vf)}  \frac{p(\vd)}{\det \vf'(\vd)} =0.
\end{equation}
\end{theorem}
The above theorem is not sufficient for our investigations. We have to consider
cases when $f_1, \ldots, f_n$ have intersections at the infinity as well as cases when intersections
 of $f_1, \ldots, f_n$ are not simple.

The  homogenisations of $f_i$ are given by
\begin{equation}
 F_i(z_0,z_1,\ldots, z_n):=z_0^{\deg f_i}f_i\left(\frac{z_1}{z_0},\ldots, \frac{z_n}{z_0}\right), \mtext{for}i=1,\ldots, n.
\end{equation}
They define the projective algebraic set $\scV(\vF):=\scV(F_1,\ldots,F_n)\subset
\CP^n$ whose affine part is homeomorphic to $ \scV(\vf)$. Next we extend the
form $\omega$ to a rational form $\Omega$ defined on $\CP^n$. To this end we
consider $\omega$ as the expression of  $\Omega$ on the chart
$(U_0,\theta_0)$. In order to express $\Omega$ on other charts we use the
standard coordinate transformations. For example, if we are using formulae from Section~\ref{ssec:alg},
on $(U_1,\theta_1)$, the form $\Omega$ is given by the pullback
 $\widetilde \omega:=\psi^\star \omega$.
To write down $\widetilde \omega$ explicitly we put
\begin{equation}
 h_i(\vy):=y_1^{\deg f_i}f_i(\psi(\vy))= y_1^{\deg f_i}f_i\left( \frac{1}{y_1},  \frac{y_2}{y_1}, \ldots,  \frac{y_n}{y_1}\right),
\end{equation}
for $i=1,\ldots, n$, and
\begin{equation}
\label{eq:hy}
 r(\vy):= y_1^{\deg p}p(\psi(\vy))=  y_1^{\deg p}p\left( \frac{1}{y_1},  \frac{y_2}{y_1}, \ldots,  \frac{y_n}{y_1}\right).
\end{equation}
Then we obtain
\begin{equation}
\label{eq:to}
 \widetilde\omega = - \frac{r(\vy)y_1^s}{h_1(\vy)\cdots h_n(\vy)}\rmd y_1\wedge \cdots \wedge \rmd y_n,
\end{equation}
where
\begin{equation}
 s := \sum_{i=1}^n\deg f_i -\deg p -(n+1).
\end{equation}
Notice that $r, h_1,\ldots, h_n\in \C[\vy]$.

Let $[ \vp]=[p_0:\cdots:p_n]\in U_i\cap\scV(F_1,\ldots,F_n)$. We can define
the residue of the form $\Omega$ at point $[\vp]$ as
\begin{equation}
 \res(\Omega, [\vp]):=\res(\widetilde\omega , \theta_i(  [ \vp] )),
\end{equation}
 where $\widetilde\omega$ denotes form $\Omega$ expressed in the chart
 $(U_i,\theta_i)$.

The form $\Omega$ is defined by homogeneous polynomials $F_1,\ldots,F_m$ and
\begin{equation}
 P(z_0,z_1,\ldots, z_n):=z_0^{\deg p}p\left(\frac{z_1}{z_0},\ldots,
 \frac{z_n}{z_0}\right).
\end{equation}
To underline the explicit dependence of $\Omega$ on $F_i$ and $P$ we write
symbolically $\Omega=P/\vF$. The following theorem is a special version of the
global residue theorem.
\begin{theorem}
\label{thm:glo}
 Let $\scV(\vF):=\scV(F_1,\ldots,F_n)$ be a finite set. Then for each polynomial $P$ such that
\begin{equation}
 \deg P\leq \sum_{i=1}^n \deg F_i-(n+1),
\end{equation}
we have
\begin{equation}
 \sum_{[\vs]\in\scV(\vF)} \res(P/\vF, [\vs])=0.
\end{equation}
\end{theorem}
For the proof and the more detailed exposition we refer the reader to \cite{Griffiths:78::,Biernat:92::}.

If $\vzero\in\scV(f)$ is an isolated but not simple point, then we cannot use
formula~\eqref{eq:0res} to calculate the residue of the form $\omega$ at this
point. In such a case we can apply a very nice method developed by Biernat in
\cite{Biernat:89::,Biernat:91::} that reduces the calculation of
multi-dimensional residue to a one dimensional case. We describe it shortly
below.

Let us consider the following analytic set
\begin{equation}
 \scA:=\defset{\vx\in U}{ f_2(\vx)=\cdots=f_n(\vx)=0},
\end{equation}
where $U\subset\C^n$ is a neighbourhood of the origin. Set $\scA$ is a sum of
irreducible one dimensional components $\scA=\scA_1\cup\cdots\cup \scA_m$.
Let $t\mapsto\vvarphi_i(t)\in\scA_i$, $ \vvarphi_i(0)=\vzero$, be an injective
parametrisation of $\scA_i$. Then we define the following forms
\begin{equation}
\label{eq:boi}
 \omega_i = \frac{p(\vvarphi_i(t))}{ \vf'(\vvarphi_i(t))} \frac{f_1'(\vvarphi_i(t))\cdot \dot\vvarphi_i(t)}{f_1(\vvarphi_i(t))}
\rmd t.
\end{equation}
As it was shown in \cite{Biernat:91::} we have
\begin{equation}
 \res(\omega,\vzero)=\sum_{i=1}^m \res(\omega_i, 0).
\end{equation}
\subsection{Proofs}
Let $V\in\C_k[\vx]$, where $\vx=(x_1,\ldots,x_n)$,  be a homogeneous polynomial of degree $k>2$.
We define $n$ polynomials $\vf=(f_1,\ldots, f_n)$ putting
\begin{equation}
 \label{eq:fi}
f_i = \pder{V}{x_i} - x_i, \mtext{for} i=1,\ldots,n,
\end{equation}
i.e., $\vf(\vx)=V'(\vx)-\vx$.  In our considerations the basic role plays the following
$n$ form

\begin{equation}
\label{eq:om}
 \omega= \frac{p(\vx)}{f_1(\vx)\cdots f_n(\vx)}
\,\rmd x_1\wedge \cdots \wedge \rmd x_n,
\end{equation}
where $p$ is a polynomial. The polar loci of this form are points of
$\scV(\vf):=\scV(f_1,\ldots,f_n)$. Let us notice the following facts.
\begin{proposition}
\label{pro:ba}
 Point $\vzero\in\scV(\vf)$ is a simple point and $\vf'(\vzero)=-\vE_n$. Thus we have
\begin{equation}
 \res(\omega,\vzero)=(-1)^n p(\vzero).
\end{equation}
If $\vd\in\scV(\vf)$ and $\vd\neq\vzero$, then
\begin{enumerate}
 \item point $[\vd]\in\CP^{n-1}$ is a proper Darboux point of $V$, i.e.,  $[\vd]\in \scD^\star(V)$,
\item the Jacobi matrix $\vf'(\vd)$ has eigenvalues $\Lambda_1(\vd),
\ldots, \Lambda_{n-1}(\vd),\Lambda_n(\vd)=k-2$,
\item if $\det \vf'(\vd)\neq 0$, then
\begin{equation}
 \res(\omega,\vd)= \frac{p(\vd)}{(k-2)\Lambda_1(\vd)\cdots\Lambda_{n-1}(\vd)},
\end{equation}
\item points  $\vd_j:=\varepsilon^j\vd\in\mathscr{V}(\vf)$,  where $\varepsilon$ is a primitive $(k-2)$-root
of the unity, satisfy  $\vf'(\vd_j)=\vf'(\vd)$, for $j=0, \ldots, k-3$.
\end{enumerate}
\end{proposition}
An easy proof of the above proposition we left to the reader.

Let us notice that $f_i^+=\partial_i V$, so $f_i$ intersect at the infinity at a
point $\vd\neq\vzero$ iff $V'(\vd)=\vzero$. For such $\vd$, point
$[\vd]\in\CP^{n-1}$ is an improper Darboux point. Hence, in order to
investigate the intersection of $\vf$ at the infinity we extend the form $\omega$ into a
rational form $\Omega$ on $\CP^{n}$. We proceed exactly as in the previous
subsection. Namely,  let
\begin{equation}
 F_i(z_0,z_1,\ldots, z_n):=z_0^{\deg f_i}f_i\left(\frac{z_1}{z_0},\ldots, \frac{z_n}{z_0}\right)=\pder{V}{z_i}-z_0^{k-2}z_i,
\end{equation}
be the homogenisation of $f_i$, for $i=1,\ldots, n$, and
\begin{equation*}
 P(z_0,z_1,\ldots, z_n):=z_0^{\deg p}p\left(\frac{z_1}{z_0},\ldots, \frac{z_n}{z_0}\right),
\end{equation*}
the homogenisation of $p$. The affine part of the projective algebraic set
$\scV(\vF):=\scV(F_1,\ldots,F_n)$ is homeomorphic to $\scV(\vf)$.
Notice that $\scV(\vF)=\widehat{\scD}(V)$.
Hence, the form $\Omega=P/\vF$ on the chart $(U_0,\theta_0)$ is given
by~\eqref{eq:om}. According to the previous subsection on the chart
$(U_1,\theta_1)$ it reads
\begin{equation}
\label{eq:om1}
 \widetilde\omega = - \frac{r(\vy)y_1^s}{h_1(\vy)\cdots h_n(\vy)}\rmd y_1\wedge \cdots \wedge \rmd y_n,
\end{equation}
where
\begin{equation}
\label{eq:s}
 s := n(k-1)-\deg p -(n+1),
\end{equation}
and polynomials $h_i$ are defined by~\eqref{eq:hif} and \eqref{eq:hiv},
and $r(\vy)$ is given by~\eqref{eq:hy}.

In the proofs of Theorems~\ref{thm:1}, \ref{thm:2} and
\ref{thm:3} we use the same idea: we calculate residues of the form
$\Omega=P/\vF$ with appropriately chosen $P$ at all points of
$\scV(\vF)=\widehat{\scD}(V)$.

\subsubsection{Proof of Theorem~\ref{thm:1}}
At first we prove the following lemma.
\begin{lemma}
 \label{lem:1}
If assumptions of Theorem~\ref{thm:1} are fulfilled, then for a polynomial $P\in\C[\vq]$, with degree $\deg P=m\leq n(k-2)-1$, satisfying
\begin{equation}
 P(\vx)=P(\varepsilon^i\vx) \mtext{for all} \vx\in\C^n \mtext{and} i=0,\ldots, k-3,
\end{equation}
where $\varepsilon$ is a primitive $(k-2)$-root of unity, we have
\begin{equation}
 \sum_{[\vd]\in\scV} \frac{P(\vd)}{\tau_{n-1}(\vLambda(\vd) )}=(-1)^{n-1}P(\vzero).
\end{equation}
\end{lemma}
\begin{proof}
We take the form  $\Omega=P/\vF$ and we calculate residues of this form at all
points of $\scV(\vF)=\widehat{\scD}(V)$. In the affine part of $\CP^n$, form
$\Omega$ is given by~\eqref{eq:om}.

 Assumptions of Theorem~\ref{thm:1} imply that
$V'(\vx)=\vzero$ only for $\vx=\vzero$. Thus, polynomials $f_1,\ldots, f_n$ do
not intersect at infinity. So all points of $\scV(\vF)$ lie in the affine part of $\CP^n$.

If $\vd\in\scV(f_1, \ldots, f_n)$, and $\vd\neq \vzero$, then $[\vd]$ is a
proper Darboux point of $V$.
By our assumption, $[\vd]$  is a simple point of $\scD^\star(V)$. By Lemma~\ref{lem:sprop},
all points of the preimage $\widehat{\pi}^{-1}([\vd])$ are simple. This is why the inequality holds
\[
 \det
\vf'(\vd)=(k-2)\prod_{i=1}^{n-1}\Lambda_i(\vd)\neq 0.
\]
Moreover, by Proposition~\ref{pro:ba}, $\vd_j:=\varepsilon^j\vd\in\scV(f_1,
\ldots, f_n)$, for $j=0, \ldots,
k-3$, where $\varepsilon$ is a primitive $(k-2)$-root of the unity, and
$\vf'(\vd_j)=\vf'(\vd)$ for $j=0,\ldots, k-3$, so all Jacobi matrices
$\vf'(\vd_j)$ have the same eigenvalues $\Lambda_1(\vd), \ldots,\Lambda_{n-1}(\vd),
\Lambda_n(\vd)=k-2$. Additionally, polynomial $p$ satisfies
restriction~\eqref{eq:degp} of Theorem~\ref{thm:ejk} and $p(\vd_j)=p(\vd)$ for
$j=0,\ldots, k-3$.

The above considerations show that all assumptions of Theorem~\ref{thm:ejk} are satisfied.
For a given  $\vd\in\scV(f_1, \ldots, f_n)$, $\vd\neq \vzero$, we have
\begin{equation}
\label{eq:rj}
 \sum_{j=0}^{k-3}\res( \omega, \vd_j)=
 \frac{p(\vd)}{\Lambda_1(\vd)\cdots \Lambda_{n-1}(\vd) },
\end{equation}
and for $\vzero\in\scV(f_1, \ldots, f_n)$  we obtain
\begin{equation}
\label{eq:r0}
 \res( \omega, \vzero)=(-1)^np(\vzero),
\end{equation}
as $\vf'(\vzero)=-\vE_n$.
Taking the sum of~\eqref{eq:rj}  over all Darboux points and adding \eqref{eq:r0}  we obtain zero.
\end{proof}

The elements of matrix $\vf'(\vx)$ are  polynomials of degree $k-2$. Thus, we can take
\begin{equation}
 p(\vx)=(\tr \vf'(\vx) -(k-2))^r, \mtext{with}  r\in\{0,\ldots, n-1\},
\end{equation}
and apply Lemma~\ref{lem:1}. For this choice of $p(\vx)$ we have
\begin{equation}
 p(\vd)= \tau_1(\vLambda(\vd))^r \mtext{for} \vd\in{\scD}^\star(V),
\end{equation}
and
\begin{equation}
 p(\vzero)= (-n -(k-2))^r.
\end{equation}
Thus, making the above choice for $p(\vx)$ we obtain relations~\eqref{eq:rkoj}
in Theorem~\ref{thm:1}.

In order to prove that relations~\eqref{eq:rkoj} are valid we define
polynomials $ p_i(\vx)$
which are coefficients of the characteristic polynomial of matrix $\vf'(\vx)$,
namely
\begin{equation}
 \det(\vf'(\vx)-\lambda \vE_n)=
 (-1)^n\left(\lambda^n - p_1(\vx)\lambda^{n-1}+\cdots+(-1)^n p_n(\vx)\right).
\end{equation}
Obviously $\deg  p_i=i(k-2)$, and $p_i(\varepsilon^j\vx)=p_i(\vx)$ for all $\vx\in\C^n$, $j=0, \ldots, k-3$,
and $i=0,\ldots,n$.
For $\vd\in{\scD}^\star(V)$ we have  $ p_i(\vd)=\tau_i(\widehat\vLambda(\vd))$, where
\[
 \widehat\vLambda(\vd)=(\Lambda_1(\vd), \ldots, \Lambda_{n-1}(\vd), k-2),
\]
and
\[
 p_i(\vzero)=(-1)^{i}\binom{n}{i}.
\]
Applying Lemma~\ref{lem:1} for  $p(\vx)=p_i(\vx)$ with $ 0\leq i\leq n-1$, we
obtain the following relations
\begin{equation}
 \sum_{[\vd]\in{\scD}^\star(V)}
\frac{\tau_r(\widehat\vLambda(\vd))}{\tau_{n-1}(\vLambda(\vd) )}=(-1)^{n+r-1}\binom{n}{r}.
\end{equation}
Since
\begin{equation}
  \tau_i(\widehat\vLambda(\vd)) =   \tau_i(\vLambda(\vd)) + (k-2)\tau_{i-1}(\vLambda(\vd)),  \quad i=0,\ldots, n-1,
\end{equation}
where $\tau_{-1}(\vLambda(\vd))=0$,  we have the following system of linear equations
\begin{equation}
\label{eq:rtr}
 T_r + (k-2)T_{r-1} = (-1)^{n+r-1}\binom{n}{r}, \quad T_0= (-1)^{n-1},
\end{equation}
where $ r= 1, \ldots, n-1$, and
\begin{equation}
\label{eq:tr}
 T_r =  \sum_{[\vd]\in{\scD}^\star(V)}
\frac{\tau_r(\vLambda(\vd))}{\tau_{n-1}(\vLambda(\vd) )}.
\end{equation}
Now using induction  with respect to $r$ and the relation
\[
 \binom{n-i}{r-i+1}=\binom{n-i-1}{r-i}+\binom{n-i-1}{r-i+1},
\]
it is easy to show that
\begin{equation}
 T_r=(-1)^{r+n-1}\sum_{i=0}^{r}\binom{n-i-1}{r-i}(k-1)^{i}, \mtext{for} r=0.\ldots,n-1.
\end{equation}
This shows that relations~\eqref{eq:rkoj} are satisfied and the proof is  finished.
\subsubsection{Proof of Theorem~\ref{thm:2}}
Our proof is based on the following lemma.
\begin{lemma}
 \label{lem:2}
If assumptions of Theorem~\ref{thm:2} are satisfied, then for a polynomial
$p\in\C[\vq]$, with degree $\deg p=m\leq n(k-2)-2=n-2$, we have
\begin{equation}
 \sum_{[\vd]\in\scV^\star(V)} \frac{p(\vd)}{\tau_{n-1}(\vLambda(\vd) )}=
 (-1)^{n-1}p(\vzero).
\end{equation}
\end{lemma}
\begin{proof}
We proceed in a similar way as in the proof of Theorem~\ref{thm:1}. However, now
polynomials $f_1, \ldots, f_n$ can have intersections at the infinity, and
because of this we have to apply Theorem~\ref{thm:glo}. Thus, we have to
calculate residues of the form $\Omega=P/\vF$ at all points of
$\scV(\vF)=\widehat{\scD}(V)$. Points of $ \widehat{\scD}(V)$ which lie in the
affine part of $\CP^n$ are given by $\scV(\vf)$. Since, by assumption,
$k=3$, there is one to one correspondence between points of $\scD^\star(V)$ and
$\scV(\vd)\setminus\{\vzero\}$. Thus for each $[\vd]\in \scD^\star(V)$,
by assumption, $[\vd]$ is a simple point of $\scD(V)$, so by Lemma~\ref{lem:sprop}
$\widehat{\pi}^{-1}([\vd])$ is a simple point of $\scD(V)$. As a result for each
non-zero $\vd\in\scV(\vf)$ we have
\begin{equation*}
\det \vf'(\vd) =\Lambda_1(\vd)\cdots \Lambda_{n-1}(\vd)\neq 0,
\end{equation*}
and
\begin{equation}
\label{eq:r1}
 \res( \omega, \vd)=\frac{p(\vd)}{\Lambda_1(\vd)\cdots \Lambda_{n-1}(\vd) }.
\end{equation}
Moreover, the local residue  of the form $\omega$ at $\vx=\vzero$ is
\begin{equation}
\label{eq:r00}
 \res( \omega, \vzero)=(-1)^nP(\vzero).
\end{equation}

Let us assume that $[\widehat{\vd}]\in\scV(\vF)$ lies in the hyperplane at the
infinity. Without any loss of the generality we can assume that
$[\widehat{\vd}]=[0:d_1:\cdots:d_n]$ with $d_1= 1$. We know that
$[\vd]=[d_1:\cdots:d_n]$ is an improper Darboux point which is, by assumption, a
simple point of $\scD(V)$, and thus, by Proposition~\ref{prop:imk3},
$[\widehat{\vd}]$ is a simple point of $\widehat{\scD}(V)$. It turns out that we
can calculate the residue of the form $\Omega$ using the local residue formula on
the chart $(U_1,\theta_1)$, where form $\Omega$ is given by~\eqref{eq:om1}.
Since coordinates of $[\widehat{\vd}]$ on $(U_1,\theta_1)$  are
$\va=(0,d_2,\ldots,d_n)=\theta_1([\widehat{\vd}])$, and $s$ given by
formula~\eqref{eq:s} is greater than zero, we have
\[
\res(\widetilde{\omega}, \va)=0.
\]
The above shows that points of $\scV(\vF)$ which lie at infinity do not
enter in the total sum of the residues.
\end{proof}
To prove the theorem we apply Lemma~\ref{lem:2} taking $p(\vx)=\left(\tr
\vf'(\vx)-1\right)^r$ and $p(\vx)=p_r(\vx)$ with $r=0,\ldots, n-2$. Then
repeating arguments used in the proof of Theorem~\ref{thm:1} we show that
relations~\eqref{eq:rkoj} and~\eqref{eq:rtau} are satisfied.

\subsubsection{Proof of Theorem~\ref{thm:3}}
The proof will follow directly from the following lemma.
\begin{lemma}
 \label{lem:3}
  If assumptions of Theorem~\ref{thm:3} are satisfied, then for a polynomial
  $p\in\C[\vq]$, with degree $\deg p=m\leq (n-1)(k-2)-1$, satisfying
\begin{equation}
 P(\vx)=P(\varepsilon^i\vx) \mtext{for all} \vx\in\C^n \mtext{and} i=0,\ldots, k-3,
\end{equation}
where $\varepsilon$ is a primitive $(k-2)$-root of unity, we have
\begin{equation}
 \sum_{[\vd]\in\scD^\star(V)} \frac{P(\vd)}{\tau_{n-1}(\vLambda(\vd) )}=
 (-1)^{n-1}P(\vzero).
\end{equation}
\end{lemma}
\begin{proof}
As in the proof of Lemma~\ref{lem:2} we have to show that residues of the form
$\Omega=P/\vF$ at points $[\widehat\vd]\in\scV(\vF)$ which lie on the
hypersurface at the infinity, vanish. The difficulty that we meet here is that now these
points, by Proposition~\ref{prop:k>3}, are not simple and we cannot use the
local residue formula.

Let us assume that $[\widehat{\vd}]\in\scV(\vF)\cap H_\infty$. Without any loss
of generality we can assume that
$[\widehat{\vd}]=[0:1:d_2:\cdots:d_n]$. The coordinates of $[\widehat{\vd}]$ on
the chart $(U_1,\theta_1)$ are
$\va=(0,d_2,\ldots,d_n)=\theta_1([\widehat{\vd}])$. We have to calculate the
residue of the form $\widetilde{\omega}$ given by~\eqref{eq:om1} at point $\va$.
We know that $[1:d_2:\cdots:d_n]$ is an improper Darboux point. By assumption it
is weakly degenerated. To calculate $\res(\widehat{\omega},\va)$ we apply the
Biernat formula, see the end of Section~\ref{ssec:ejk}. To this end we need to
determine all the branches of the analytic set
\begin{equation}
\label{eq:a}
\scA:=\defset{\vy\in U_{\va}}{h_2(\vy)=\cdots=h_n(\vy)=0},
\end{equation}
 where $U_{\va}$ is a neighbourhood of $\va$. Since $\va$ are the coordinates of an
 improper Darboux point that is minimally degenerated, by
 Proposition~\ref{prop:k>3}, we have
 \begin{equation*}
 \rank\frac{\partial(h_2, \ldots,h_n)}{\partial(y_1,\ldots,y_n)}(\va)=n-1,
 \end{equation*}
 and moreover
 \begin{equation*}
 \det\frac{\partial(h_2, \ldots,h_n)}{\partial(y_2,\ldots,y_n)}(\va)\neq 0.
 \end{equation*}
 Hence, by the implicit function theorem there is only one analytic branch
 passing by point $\va$. We can always choose coordinates such that $\va=\vzero$. Then, because
 \begin{equation*}
 h_i(\vy)=\pder{v}{y_i}-y_iy_1^{k-2}, \mtext{for} i=2, \ldots, n,
 \end{equation*}
 and
 \begin{equation*}
 \pder{v}{y_i}(\vzero)=0,  \mtext{for} i=2, \ldots, n,
 \end{equation*}
we can put
 \begin{equation}
 y_1(t)=t, \mtext{and}  y_i(t)=0, \mtext{for} 2\leq i\leq n,
\end{equation}
as the parametrisation of the unique branch of $\scA$. Then, the Biernat
formula~\eqref{eq:boi}, applied to the form $\widetilde{\omega}$ gives  the
differential form $w(t)\rmd t$, where
\begin{equation}
w(t)= -\dfrac{ r(\vy(t))t^s}{h_1(\vy(t))\det \vh'(\vy(t))}\Dt h_1(\vy(t)).
\end{equation}
Calculations show that
\begin{equation}
\det \vh'(\vy(t))=-(k-2)t^{k-3} h(t) ,
\end{equation}
where
\begin{equation*}
h(t)=\det\left[\dfrac{\partial^2 v}{\partial y_i\partial y_j}(\vzero )-t^{k-2}\delta_{i,j}\right]_{2\leq i, j\leq n}.
\end{equation*}
 As an effect we have
\begin{equation}
 w(t)= \dfrac{ \widetilde{r} (t)  t^{ s-(k-2) }  }{h(t)},
\end{equation}
where $\widetilde{r} (t)=r(y(t))$ is a polynomial and moreover
\begin{equation}
 h(0) = \det\left[\dfrac{\partial^2 v}{\partial y_i\partial y_j}(\vzero)\right]_{2\leq i, j\leq n}\neq 0,
\end{equation}
 by assumption that the considered point is weakly degenerated. Thus,
$w(t)$ is regular at $t=0$, iff $s\geq k-2$. Thus,  if $m=\deg p\leq
(n-1)(k-2)-1$, then $w(t)$ is regular at $t=0$. But this is the assumption of our lemma,
so all the residues at points vanish
at infinity.

We calculate the residues at points in the affine part of $\scV(\vF)$  as in the
proof of~Lemma~\ref{lem:1}.
\end{proof}
To prove the theorem we have only to check if we can apply Lemma~\ref{lem:3}
for $p(\vx)=(\tr \vf'(\vx) -(k-2))^r$ with $r=0,\ldots, n-2$. But $\deg (\tr
\vf'(\vx) -(k-2))^r=r(k-2)$, so
$\deg p \leq (n-1)(k-2)-1 $ for $r=0,\ldots, n-2$.
\subsubsection{Proof of Theorem~\ref{thm:ff} and Theorem~\ref{thm:ff1}}
% If, additonally, the assumptions of Theorem~\ref{thm:1}
% are satisfied, then  $N=D(n,k)$ and  relations~\eqref{eq:rkoj} and \eqref{eq:rtau} must be
% fullfield,  $\scL(V):=\deftuple{\vLambda(\vd)}{[\vd]\in\scD^\star(V)}$

Let us recall that  $\scA_{n,k}$ denotes the set of all admissible $(n-1)$-tuples for a given $k$.
We define the set of distinguished spectra $\scI_{n,k}^N$, which appears in
Theorem~\ref{thm:ff} and~\ref{thm:ff1} as  the set of admissible
$(\vLambda_1,\ldots,\vLambda_N)\in \scA_{n,k}^N$ satisfying all
relations~\eqref{eq:rkoj} and~\eqref{eq:rtau}. In particular,
$(\vLambda_1,\ldots,\vLambda_N)\in \scJ_{n,k}^N$, satisfies
relation~\eqref{eq:rtau} with $r=n-2$, i.e.,
\begin{equation}
 \sum_{i=1}^N
\frac{\tau_{n-2}(\vLambda_i)}{\tau_{n-1}(\vLambda_i
)}= -\frac{(k-1)^n-n(k-2)-1}{(k-2)^2}.
\label{eq:taun-2}
\end{equation}
If we denote $(\vLambda_1,\ldots,\vLambda_N)=(\Lambda_{1},\ldots, \Lambda_{M})$,
where $M=(n-1)N$, then \eqref{eq:taun-2} reads
\begin{equation}
\label{eq:mm}
\sum_{i=1}^M
\frac{1}{\Lambda_i}= -\frac{(k-1)^n-n(k-2)-1}{(k-2)^2}.
\end{equation}
We need the following technical lemma.
\begin{lemma}
\label{lem:jnkN}
For $k>2$, $n\geq 2$ and $N\geq 1$, the set of solutions of \eqref{eq:taun-2}
in $\scA_{n,k}^N$ is finite.
\end{lemma}
\begin{proof}
We rewrite relation~\eqref{eq:mm} in the form
\begin{equation}
\label{eq:x}
\sum_{i=1}^M X_i= -c,\mtext{where} c>0.
\end{equation}
As it is easy to check, the admissible set $\scX_k\subset \Q$ of values of $X_i$
for a given  $k>2$, has the
following properties: $\scX_k=\scX_k^-\cup \scX_k^+ $, where $\scX_k^-$ is
finite and its elements are negative; set $\scX_k^+ $ is infinite and its
elements are positive. From~\eqref{eq:x} it follows that a certain number of
$X_i$ must be negative, and we have
only a finite number of choices of these negative $X_i$. For each of them  we have to
investigate equation
\begin{equation}
\label{eq:xp}
\sum_{i=1}^{p} X_i= c',\mtext{where} c'>0,
\end{equation}
where $X_i\in \scX_k^+ $, for $i=1,\ldots,p<M$. But the set $\scX_k^+$ has an
important property. Namely, its only accumulation point is zero, i.e.,
$\overline{\scX_k^+ }\setminus \scX_k^+=\{0\}$. Hence, applying Lemma~B.1 from
\cite{mp:05::c} we obtain that the set of solutions $(X_1,\ldots,X_{p})\in
(\scX_k^+)^{p}$ is finite.
\end{proof}
Now, we pass to the proofs of Theorems~\ref{thm:ff} and \ref{thm:ff1}. If
an integrable potential $V$ satisfies assumptions of  one of Theorems~\ref{thm:1},
\ref{thm:2}
or \ref{thm:2}, then
\begin{equation}
\label{eq:LV}
\scL(V):=\deftuple{\vLambda(\vd)}{[\vd]\in\scD^\star(V)}\in\scA^N_{n,k},
\mtext{where} N=\card \scD^\star(V),
\end{equation}
 fulfils, among others, relation~\eqref{eq:rtau} with
$r=n-2$. By Lemma~\ref{lem:jnkN}, this relation is fulfilled only by a
finite number of admissible elements.

\subsection{Discussion}

 We know that, for a given  $k>2$, and $n\geq 2$, the set of distinguished spectra  $\scI_{n,k}^N$  is finite
 but we do not know how many elements it has.  We show that this
set is  not empty.

\begin{proposition}
For each $k>2$,  $n\geq 2$,  and each $N>1$,  set $\scI_{n,k}^N$ is not empty and it contains
at least two elements.
\end{proposition}
\begin{proof}
We will use the following convention. If a collection $\scI\in\scC_{n,k}^N$
contains $n_1$ copies of the element $\ve_1$, \ldots, and $n_j$ copies of the element
$\ve_j$, then we write
\begin{equation*}
\scJ=(n_1\times \ve_1, \ldots, n_j\times\ve_j)=(n_i\times\ve_i\,|\,i=1,\ldots,j),
\end{equation*}
 instead of
 \begin{equation*}
 \scJ=(\underbrace{\ve_1, \ldots, \ve_1}_{n_1\ \text{times}},\ldots, \underbrace{\ve_j, \ldots, \ve_j}_{n_j\ \text{times}}).
 \end{equation*}
For given $k$ and $n$ we define
%
%\begin{gather*}
\begin{equation*}
\begin{split}
 &\va_i:=(\underbrace{-1,\ldots,-1}_{n-1-i\
\text{times}},\underbrace{k-2,\ldots,k-2}_{i\ \text{times}}),
 \mtext{for}i=0,\ldots, n-1,\\
 &\vb_i:=\Big(\underbrace{-1,\ldots,-1}_{i\
\text{times}},\underbrace{k-2,\ldots,k-2}_{n-2-i\
\text{times}},-\dfrac{k+1}{2k}\Big), \mtext{for}i=0, \ldots, n-2,\\
&\vc_i:=(\underbrace{-1,\ldots,-1}_{i\
\text{times}},\underbrace{k-2,\ldots,k-2}_{n-i-2\ \text{times}},k+1), \mtext{for}i=0, \ldots, n-2,
\end{split}
\end{equation*}
%\end{gather*}
%
and
\begin{gather}
\label{eq:I1}
\scJ_1:=\deftuple{(k-2)^i\binom{n}{i+1}\times \va_i}{i=0,\ldots, n-1},\\
\label{eq:I2}
\scJ_{2}:=\deftuple{\alpha_l\times \va_l,
\beta_i\times\vb_i,
\gamma_i\times \vc_i}{l=1,\ldots,n-2;\, i=0,\ldots,n-2},
\end{gather}
 where
\begin{equation*}
\begin{split}
&\alpha_i :=(k-2)^{n-l-2}\binom{n-2}{l-1}, \quad \beta_i:=(k-2)^{n-2-i}\binom{n-2}{i},\\
&\gamma_i = (k-1)(k-2)^{n-i-2}\binom{n-2}{i}.
\end{split}
\end{equation*}
Now, it is only the matter of simple but lengthy calculations to check that $
\scJ_1$  and    $\scJ_2$  are elements of $\scI_{n,k}^N$ with $N=D(n,k)$.
\end{proof}
One can wonder whether there exist potentials $V$ with spectrum given by $\scJ_1$ or $\scJ_2$.
Below we give examples of potentials satisfying this condition.

Let us take
\begin{equation}
\label{eq:v1}
V_1=\sum_{i=1}^n v_i q_i^k,  \mtext{where} (v_1,\ldots,v_n)\in \C^n.
\end{equation}
For generic values of parameters $v_i$ this potential possesses $D(n,k)$
Darboux points and one can check that
$\deftuple{\vLambda(\vd)}{[\vd]\in\scD(V_1)}=\scJ_1$.

Potential
\begin{equation}
\label{eq:v2}
V_2=\sum_{i=0}^{[k/2]}2^{-2i}\binom{k-i}{i}q_1^{2i}q_2^{k-2i}+\sum_{i=3}^n
c_iq_i^{k}, \mtext{where} (c_3, \ldots, c_n)\in\C^{n-3},
\end{equation}
possesses also $D(n,k)$ Darboux points (for generic values of parameters). Moreover  now
$\deftuple{\vLambda(\vd)}{[\vd]\in\scD(V_2)}=\scJ_2$.

 It is obvious that one would like to have theorems
more general than those
proved in the previous subsection. Having relations of the form given in the
theorems, we have to keep the assumption that all proper Darboux points are simple.
Hence the only possibility is to weaken assumptions concerning improper Darboux
points. However, due to theoretical difficulties, we were not able to find
good generalisations of the theorems.
For example, we tried to relax the second assumption of Theorem~\ref{thm:3} and
proceed in the same
way as in its proof. However we meet a serious difficulty. We did not find a
general enough method to distinguish and parametrise branches of the analytic
set~\eqref{eq:a}.

Examples suggest that it is possible to find the desired generalisation.
\begin{example}
Potential
\begin{equation}
\label{eq:e1}
V=q_1(aq_1^2+bq_2q_3), \mtext{with}a\neq 0,\mtext{and}b\neq 0,
\end{equation}
has five proper Darboux points $[\vd_i]$, $i=1,\ldots 5$ and two improper
Darboux points. All of them are simple, and
\begin{gather*}
\vLambda(\vd_1)=\left(-\dfrac{3a+b}{3a},-\dfrac{3a-b}{3a}\right), \quad
\vLambda(\vd_2)=\vLambda(\vd_3)=\left(-2,-\dfrac{2(3a-b)}{b}\right),\\
\vLambda(\vd_4)=\vLambda(\vd_5)=\left(-2,-\dfrac{2(3a+b)}{b}\right).
\end{gather*}
Although $V$ does not satisfy assumptions of Theorem~\ref{thm:1}, all
relations~\eqref{eq:rkoj} are fulfilled. Of course~\eqref{eq:e1} satisfies
assumptions of Theorem~\ref{thm:2}.
\end{example}
\begin{example}
Potential
\begin{equation*}
V=q_1(q_1^2 + \rmi q_1q_2 + a q_3^2),
\end{equation*}
has, for  generic values of parameter $a$,  four simple proper Darboux points
and one improper Darboux point which has multiplicity 3. Although assumptions of
Theorem~\ref{thm:2} are not satisfied, relations~\eqref{eq:rkoj} and
\eqref{eq:rtau} for $r=0,1$ are fulfilled. But
\begin{equation*}
\sum_{[\vd]\in\scD^\star(V)}\frac{(\Lambda_1(\vd) +\Lambda_2(\vd) )^2}{\Lambda_1(\vd) \Lambda_2(\vd)}=\dfrac{1 + a(28a-9)}{2a^2}.
\end{equation*}
\end{example}
\begin{example}
Potential
\begin{equation*}
V=(q_2-\rmi q_1)^2(b_1 q_1^2+b_2 q_2^2+b_3 q_3^2+b_4q_1q_2),
\end{equation*}
has, for generic values of parameters, three simple proper Darboux points and a
curve of improper Darboux points $[1:\rmi:\alpha]$, where $\alpha\in\C$. None of
our theorems can be applied for this potential. In fact, we have
\begin{gather*}
\sum_{[\vd]\in\scD^\star(V)}\frac{1}{\Lambda_1(\vd) \Lambda_2(\vd)}=
-\dfrac{4b_1b_2+2(b_1+b_2)b_3-b_4^2}{2(b_1+b_2)^2}, \\
\sum_{[\vd]\in\scD^\star(V)}\frac{\Lambda_1(\vd) +\Lambda_2(\vd) }{\Lambda_1(\vd) \Lambda_2(\vd)}=-\dfrac{2(b_1^2+b_2^2)+b_4^2}{2(b_1+b_2)^2},\\
\sum_{[\vd]\in\scD^\star(V)}\frac{(\Lambda_1(\vd) +\Lambda_2(\vd) )^2}{\Lambda_1(\vd) \Lambda_2(\vd)}=\dfrac{4(b_1^2+b_1b_2+b_2^2)+2(b_1+b_2)b_3+b_4^2}{2(b_1+b_2)^2},
\end{gather*}
so none of the relations of the form~\eqref{eq:rkoj} is satisfied. Nevertheless,
 we have one universal relation for the considered potential. Namely combining
the above formulae we obtain the equality
\begin{equation*}
\sum_{[\vd]\in\scD^\star(V)}\frac{(1+\Lambda_1(\vd) +\Lambda_2(\vd) )^2}{\Lambda_1(\vd) \Lambda_2(\vd)}=0.
\end{equation*}
\end{example}

\section{Three degrees of freedom and potential of third degree}
\label{sec:3d}
In this section we consider systems with three degrees of freedom with a
homogeneous potential of the third degree. The aim is to show how to apply our
theorems to distinguish integrable potentials. In fact, we consider here only
generic potentials having a maximal number of simple Darboux points.
Investigations of non-generic potentials will be presented in a separate
publication.

The whole analysis can be divided roughly into two steps:
\begin{enumerate}
\item determination of the set of distinguished spectra $\scI^N_{n,k}$,
\item reconstruction of the potential. Knowing a distinguished spectrum
$(\vLambda_1,\ldots, \vLambda_N)\in\scI^N_{n,k}$ we have to find all potentials
$V$ with $N$ proper Darboux points $[\vd_i]$, such that
$\vLambda_i=\vLambda(\vd_i)$ for $i=1,\ldots,N$.
\end{enumerate}

\subsection{Distinguished spectra}
For $n=k=3$ we have $N=D(n,k)=7$, so we look for seven admissible pairs
$\vLambda_i=(\Lambda_1^{(i)}, \Lambda_2^{(i)})\in\scA_{3,3}$,  $i=1,\ldots, 7$,
 which satisfy the following three relations
 \begin{equation}
 \left.
 \begin{split}
 &\sum_{i=1}^7\frac{1}{\Lambda_1^{(i)}\Lambda_2^{(i)}}=1,\\
&\sum_{i=1}^7\frac{\Lambda_1^{(i)}+\Lambda_2^{(i)}}{\Lambda_1^{(i)}\Lambda_2^{(i)}}= -4,\\
 &\sum_{i=1}^7\frac{(\Lambda_1^{(i)}+\Lambda_2^{(i)})^2}{\Lambda_1^{(i)}\Lambda_2^{(i)}}=16.
 \end{split}
 \quad\right\}
\label{eq:rele3}
\end{equation}
Numbers $\Lambda_j^{(i)}+1$ belong to the set $\scM_3$ which corresponds to items 1--6 in the table of
the Morales-Ramis Theorem~\ref{thm:MoRa}, i.e.,
\begin{multline*}
\scM_3=\defset{ p + \dfrac{3}{2}p(p-1)}{ p\in\Z}\cup
\defset{\dfrac 1
{2}\left[\dfrac {2} {3}+3p(p+1)\right]}{p\in\Z} \\
\cup\defset{-\dfrac 1 {24}+\dfrac 1 {6}\left( 1 +3p\right)^2 }{ p\in\Z}\cup
\defset{-\dfrac 1 {24}+\dfrac 3 {32}\left(  1  +4p\right)^2 }{ p\in\Z} \\
\cup\defset{-\dfrac 1 {24}+\dfrac 3 {50}\left(  1  +5p\right)^2 }{ p\in\Z}\cup \defset{-\dfrac 1 {24}+\dfrac{3}{50}\left(2 +5p\right)^2 }{ p\in\Z}.
\end{multline*}
The first step in solving~\eqref{eq:rele3} is to find all solutions of the
second relation which we write in the form
\begin{equation}
\label{eq:rx}
\sum_{l=1}^{14}X_l = -4,
\end{equation}
where $X_l=1/\Lambda_i^{(j)}$. Knowing $\scM_3$ we easily determine the set
$\scX_3$ of admissible values of $X_l$. It has, as it is easy to check, the following properties:
\begin{enumerate}
 \item $\scX_3= \scX_3^-\cup \scX_3^+$, and $\scX_3^-$ is finite,
\item if $x\in \scX_3^-$, then $x\leq-1$,
\item if $x \in \scX_3^+$, then $x\leq 1$.
\end{enumerate}
It follows that among $X_l$ satisfying~\eqref{eq:rx}, there is a certain number of negative ones.
Moreover, at most nine of $X_l$ are negative and we have
a finite number of choices of
them.
\begin{example}
\label{ex:gen}
 Assume that $X_1, \ldots, X_7$ are
negative.  Taking $X_1, \ldots, X_7\in \scX_3^-$ we obtain $2426$ different choices satisfying
$
 X_1+\cdots+ X_7<-4
$.
\end{example}
Thus we reduce the problem to finding solutions of
\begin{equation}
\label{eq:pp}
 \sum_{i=1}^{p} X_i = c, \mtext{where} c>0, \quad p<14,\qquad X_i\in \scX_3^+.
\end{equation}
If $(X_1, \ldots, X_p)$ is a solution of the above equation, then we can assume
that $X_1\leq X_2\leq \cdots \leq X_p$. Hence $X_p\geq c/p$, and we have only a
finite number of choices for $X_p$. For each of this choices we have to find
the solution of equation~\eqref{eq:pp} where we replace  $p$ by $p-1$.  Repeating
this reasoning we  end up with equation
\begin{equation}
\label{eq:xeps}
 X_1 + X_2 = \varepsilon,
\end{equation}
so, $\varepsilon >X_2\geq \varepsilon/2$, and we have only a finite number of
choices for $X_2$. This procedure proves that we have only a finite number of
solutions of~\eqref{eq:pp}, however a serious practical problem appears because we
 do not know \emph{a priori}
the lower bound for $X_1$.
We can order elements of $X_3^+$ into decreasing sequence $\{x_i\}$. Then
$x_i$ decreases with $i$ as $i^{-2}$. For a small $\varepsilon$ interval
$[\varepsilon/2,\varepsilon )$ contains approximately $[1/\varepsilon]$
admissible elements $X_2$. For each of them we have to check if
$X_1=\varepsilon-X_2$ is admissible, i.e., if $X_1\in\scX_3^+$. Examples show
that $ \varepsilon$ can be of order
 $10^{-15}$ and smaller. Very small values of $\varepsilon$ in \eqref{eq:xeps}
 appear when the number of negative elements is smaller than seven. For such
 cases, even using a dedicated software running on the quickest accessible
 computers, we were unable to perform the computations up to the end in a
 reasonable period of the time (one month). This is a reason  why we restrict our searches
 giving \emph{a priori} lower bound for the smallest element $X_1$.

A solution of~\eqref{eq:rx} gives 14 numbers $(L_1,\ldots, L_{14})$, where $L_i=1/X_i$.
We divide them into all possible seven unordered pairs $\vLambda_i=(\Lambda_1^{(i)},
\Lambda_2^{(i)})$ and check if these pairs satisfy the remaining two
relations~\eqref{eq:rele3}.

Using the described algorithm we have found the distinguished spectra listed in
Table~\ref{tab:ds}.
\begin{table}
\label{tab:ds}
\caption{The distinguished spectra for $n=k=3$}
\begin{tabular}{ll}
1.& $\left(3\times (-1,-1),3\times (-1,1),(1,1)\right)$,\\[0.2em]
2.& $\left(
(-1,-1),\left(-1,-\dfrac{2}{3}\right),\left(1,-\dfrac{2}{3}\right),2\times
(-1,4),2\times (1,4)\right)$,\\[1em]
3.& $\left(
(-1,-1),\left(-1,-\dfrac{7}{8}\right),\left(1,-\dfrac{7}{8}\right),2\times
(-1,14),2\times (1,14)\right)$,\\[1em]
4.& $\left(
(-1,-1),\left(-1,-\dfrac{2}{3}\right),\left(1,-\dfrac{2}{3}\right),\left(
-1,\dfrac{7}{3}\right),\left(1,\dfrac{7}{3}\right),(-1,14),(1,14)\right)
$,\\[1em]
 5.&$\left(\left(-\dfrac{2}{3},-\dfrac{2}{3}\right),2\times
\left(-\dfrac{2}{3},\dfrac{7}{3}\right),
(-1,4),(1,4),2\times(4,14)\right)$,\\[1em]
6.&$\left(\left(-\dfrac{2}{3},-\dfrac{7}{8}\right),
\left(-\dfrac{7}{8},4\right),
 \left(-\dfrac{3}{8},4\right),2\times \left(\dfrac{7}{3},4\right)
2\times (4,21)\right)$,\\[1em]
7.&$\left(2\times (-1,6),\left(-1,\dfrac{7}{3}\right),
\left(-\dfrac{7}{8},-\dfrac{2}{3}\right),
 \left(\dfrac{13}{8},14\right),
2\times (14.39)\right)$,\\[1em]
8.&$\left(\left(-\dfrac{2}{3},-\dfrac{2}{3}\right),3\times
\left(-1,\dfrac{7}{3}\right),
3\times (6,14)\right)$,\\[1em]
9.&$\left(\left(-\dfrac{2}{3},-\dfrac{7}{8}\right),\left(-\dfrac{2}{3},
\dfrac{7}{3}\right),
\left(-\dfrac{3}{8},14\right), 2\times \left(1,\dfrac{52}{3}\right), 2\times
(14,39)\right)$,\\[1em]
10.&$\left((-1,1),\left(-1,-\dfrac{3}{8}\right),2\times
\left(-\dfrac{2}{3},\dfrac{7}{3}\right),\left(1,\dfrac{13}{8}\right),
 2\times(14,39)\right)$.
\end{tabular}
\end{table}
\begin{remark}
The described algorithm has an obvious generalisation for arbitrary $k$ and $n$.
\end{remark}
\subsection{Normalisation of the potential}
The $\C$-linear space $\C_k[\vq]$ of homogeneous polynomials of degree 3 in three variables has
dimension $10$. Starting from the general potential
 of degree three
\begin{equation}
\begin{split}
 V=& a_1q_1^3+a_2q_1^2q_2+a_3q_1^2q_3+a_4q_1q_2^2+a_5q_2^3+a_6q_2^2q_3+a_7q_3^3+ a_8q_1q_3^2+a_9q_2q_3^2\\
&+a_{10}q_1q_2q_3,
\end{split}
\label{eq:pot3gen}
\end{equation}
we have to select its `good' equivalent representative with the minimal number
of free parameters. The group
$\mathrm{PO}(3,\C)$ is four dimensional, so a typical orbit of its action on
$\C_3[\vq]$ has dimension $6$. However, the problem is that there is no good
parametrisation of these orbits.

Here we consider only a generic case and this simplifies the analysis a lot. In
our further considerations we will need the following fact.

\begin{proposition}
\label{prop:nor}
Assume that a potential $V\in\C_3[\vq]$ has a proper non-isotropic Darboux point
$[\vd]$ such that
matrix $V''(\vd)$ is semi-simple. Then it is equivalent to
 \begin{equation}
 V=a_1q_1^3+a_2q_1^2q_2+a_3q_1^2q_3+a_4q_1q_2^2+a_5q_2^3+a_6q_2^2q_3+\dfrac{1}{3
}q_3^3.
\label{eq:pot3}
\end{equation}
Moreover, $\vd=(0,0,1)$ and matrix $V''(\vd)$ has eigenvalues $\lambda_1=2a_3$,
$\lambda_2=2a_6$ and $\lambda_3=2$.
\end{proposition}
\begin{proof}
By Remark~\ref{rem:mv} we can assume that $\vd=(0,0,1)$.
  This implies that  in~\eqref{eq:pot3gen} we have $a_8=a_9=0$
and $a_7=1/3$. Vector $\vd$  is an eigenvector of  $V''(\vd)$ with
eigenvalue $\lambda_3=k-1=2$. Hence $V''(\vd)$ has the following form
\begin{equation}
 V''(\vd)=\begin{bmatrix}
2a_3&a_{10}&0\\
a_{10}&2a_6&0\\
0&0&2
          \end{bmatrix},
\label{eq:niediag}
\end{equation}
and its remaining eigenvalues  are
$\lambda_{1,2}=a_3+a_6\pm\sqrt{a_{10}+(a_3-a_6)^2}$.
If
\begin{equation}
 a_{10}\neq\pm\rmi (a_3-a_6),
\label{eq:noediag}
\end{equation}
then we can make a rotation around $\vd$ such that in the new frame
$a_{10}=0$. Under this assumption  the normalised potential reads
\begin{equation}
 V=a_1q_1^3+a_2q_1^2q_2+a_3q_1^2q_3+a_4q_1q_2^2+a_5q_2^3+a_6q_2^2q_3+\dfrac{1}{3
}q_3^3.
\label{eq:pot3f}
\end{equation}
If condition~\eqref{eq:noediag} is not fulfilled, then $V''(\vd)$ has a double
eigenvalue $\lambda_{1,2}=a_3+a_6$.  In this case, because  $V''(\vd)$ is semi-simple, we have
\begin{equation}
\label{eq:sesi}
\rank (V''(\vd) - (a_3+a_6)\vE_3)<2.
\end{equation}
This condition gives
\begin{equation}
\label{eq:c}
(a_3+a_6-2)(a_3-a_6)=0.
\end{equation}
If $a_3+a_6-2=0$, then all eigenvalues of $V''(\vd)$ are 2. But now,
if $a_3\neq 1$, then $\rank(V''(\vd)-2\vE_3)=1$. Hence, $a_3=a_6=1$, and so
$a_{10}=0$.
 In the alternative subcase $a_3=a_6$, so again $a_{10}=0$, and this finishes the proof.
\end{proof}
Notice that for the normalised potential~\eqref{eq:pot3} the proper Darboux point $[\vd]=[0:0:1]$ lies in the
line at infinity $H_\infty$.  We called it the Darboux point at the infinity, although, as the following proposition
 shows, we can have more Darboux points  in this line.
\begin{proposition}
\label{prop:3i}
  Potential~\eqref{eq:pot3} has at most 3 proper Darboux points in the line at  infinity  $H_\infty$.
If $a_4\neq 0$, then it has exactly one proper Darboux point at infinity.
\end{proposition}
\begin{proof}
If $[\vq]$ is a proper Darboux point of $V$ which lies in $H_\infty$, then $V'(\vq)=\vq$ and $q_1=0$.
 For the potential~\eqref{eq:pot3} this gives the following system of equations
\begin{equation}
 a_4 q_2^2 = 0, \quad q_2(-1 + 3 a_5 q_2 + 2a_6q_3)=0,\quad
 a_6 q_2^2 - q_3 + q_3^2 =0.
\end{equation}
It is easy to see that if $a_4\neq 0$, then the above system has only one non-zero solution corresponding
to $[\vd]=[0:0:1]$. Thus, $a_4=0$. Now, we have equations
\begin{equation}
 -1 + 3 a_5 q_2 + 2a_6q_3=0,\quad  a_6 q_2^2 - q_3 + q_3^2 =0,
\end{equation}
that admit at most 2 solutions.
\end{proof}

\subsection{Reconstruction of potential}
At this point the problem is following. For each distinguished spectrum given in Table~\ref{tab:ds}
we have to find all non-equivalent potentials having such a spectrum.  This problem is difficult.
We have to determine the coefficients of the potential knowing only the eigenvalues of its Hessian matrix
calculated at certain points with unknown coordinates.

For each distinguished spectrum from Table~\ref{tab:ds} we can assume that the potential has the
form~\eqref{eq:pot3}. So,  we assign a pair $(\Lambda_1,\Lambda_2)$ from the spectrum to the Darboux point at
infinity.  By Proposition~\ref{prop:nor},  we have  $\Lambda_1=2a_3 -1$ and $\Lambda_2=2a_6 -1$. In this way,
 we fix values of two unknown parameters $a_3$ and $a_6$.  Hence, the unknown parameters are
 $\va=(a_1,a_2,a_4,a_5)\in\C^4$.

We do not know the coordinates of the remaining Darboux points.  So we have the following problem.  Let us assume
 that we know  eigenvalues $\Lambda_1$, $\Lambda_2$ and $\Lambda_3=k-2$  of  matrix $\vf'(\vq)=V''(\vq)-\vE_3$,
 at an unknown point $\vq$ satisfying $\vf(\vq):=V'(\vq)-\vq=\vzero$.  We have to derive   restrictions on the
 coefficients of the potential.  A direct approach to this problem is based on a simple idea.
On the one hand, we know numerical values of symmetric polynomials $s_i=\tau_i(\Lambda_1,\Lambda_2,\Lambda_3)$
for $i=1,2,3$.
On the other hand,  we can express these values as coefficients $\tilde p_i(\vq)$ of the characteristic
polynomial of matrix $\vf'(\vq)$, namely
\begin{equation*}
 \det (\vf'(\vq)-\lambda \vE_3)= -(\lambda^3 -\lambda^2\tilde p_1(\vq) + \lambda \tilde p_2(\vq)-p_3(\vq)).
\end{equation*}
As a result we obtain the following system of polynomial equations
\begin{equation}
 f_i(\vq)=0, \quad  p_i(\vq):=\tilde p_i(\vq)-s_i=0, \quad i=1,2,3,
\end{equation}
where polynomials $f_i$ and   $p_i$ are considered as elements of the ring $\C[\va][\vq]$.
An elimination of variables $\vq$ from the above system gives rise to a set of polynomial equations for the
coefficients of potential $V$.
We repeat the calculations  for each Darboux point and then find solutions of all obtained equations.
\begin{remark}
If the considered distinguished spectrum contains $l\leq7$ different admissible pairs, then,
because one of them is attached to the Darboux point at the infinity, we have at our disposal $l-1$ pairs.
Hence, if $l<7$,  we have to attach the same pair to two or more Darboux points.
As an effect, we have to apply the algorithm only for $l-1$ Darboux points with different attached pairs.
\end{remark}

The elimination step in the described procedure consists of determination of the so called  elimination ideal
 of a given set of polynomials, see \cite{Cox:97::,Cox:05::}.
It  can be performed only with the help of a computer algebra system.  However, in practice, the known algorithms
 work effectively only in cases with a relatively small number of variables. Hence, in order to have a chance to
 obtain the desired result, we have to minimalise the number  of unknowns.
In the considered problem we can do this introducing the affine coordinates $x_1$ and $x_2$ for the unknown
 Darboux point.
They satisfy the following polynomial equations
\begin{equation}
\label{eq:g12}
 g_1(x_1,x_2) = 0 \mtext{and} g_2(x_1,x_2) = 0,
\end{equation}
where
\begin{equation*}
 g_i := \pder{v}{x_i}-x_i g_0, \mtext{for}i=1, 2,
\end{equation*}
and
\begin{equation*}
 g_0=kv-x_1\frac{\partial v}{\partial
x_1}-x_2\frac{\partial v}{\partial x_2}, \quad v(x_1,x_2)=V(1,x_1,x_2),
\end{equation*}
see Lemma~\ref{lem:aff} and its proof. Moreover, we take values of two symmetric  polynomials
 $S_1=\tau_1(\Lambda_1,\Lambda_2)=\Lambda_1+\Lambda_2$ and $S_2=\tau_2(\Lambda_1,\Lambda_2)=\Lambda_1\Lambda_2$.
Then,  we have to express $S_1$ and $S_2$ as functions of $x_1$ and $x_2$. To this end we recall that
the eigenvalues of $\vf'(\vq)$ coincide with the eigenvalues of Jacobi matrix $J(\vx)$ defined by~\eqref{eq:jac}.  Performing  simple calculations like in the proof of Lemma~\ref{lem:sls} we find that
\begin{equation}
  g_0S_1=\left(\frac{\partial
g_1}{\partial x_1}+\frac{\partial g_2}{\partial x_2}\right)
\mtext{and} g_0^2S_2=\left(\frac{\partial
g_1}{\partial x_1}\frac{\partial g_2}{\partial x_2}-
\frac{\partial g_1}{\partial x_2}\frac{\partial g_2}{\partial
x_1}\right).
\label{eq:lambdusie}
\end{equation}
Now, eliminating $x_1$ and $x_2$ from equations~\eqref{eq:g12} and \eqref{eq:lambdusie} we obtain a set of
equations for unknown coefficients of the potential.
\begin{remark}
 \label{rem:ini}
The drawback of using the affine coordinates is following. We do not know in advance how many points are
located at the infinity and what are their admissible spectra. Thus we have to check all the possibilities.
One can perform another normalisation of the potential with the assumption that all Darboux points are located
in the affine part od $\CP^2$.  However this  causes  other computational problems.
\end{remark}

Using the described algorithm it is always possible to derive a set of polynomial equations  which must be
satisfied by the coefficients of the potential we are looking for. However, generally  the equations are very
complicated, and  it is impossible to find their explicit solutions even with the help of the strongest tools
 of the computer algebra. In order to overcome this problem we  are forced to  apply a different strategy.

In fact we do not  want to know the explicit form of all potentials with a given distinguished spectrum,
only those which are integrable have to be selected.  Our potential satisfies a priori all the necessary
integrability conditions of Theorem~\ref{thm:MoRa}. However, the stronger integrability conditions were
formulated by Morales, Ramis, and Sim\'o,  see  \cite{Morales:99::c,Morales:00::a,Morales:06::}.
\begin{theorem}[Morales, Ramis, Sim\'o, 2006]
\label{thm:mrs}
  If a complex Hamiltonian system is integrable in the Liouville sense
  in a neighbourhood of a particular non-equilibrium solution, then the identity
  component $\scG^0_m$ of the differential Galois group  $\scG_m$ of $m$-th order variational
  equations along this solution is Abelian, for an arbitrary $m\in\N$.
\end{theorem}
Generally it is difficult to apply the above theorem, because higher order variational equations have a big
dimension.
However, if the first variational equations  are a direct product of  Lam\'e equations of the form
\begin{equation*}
 \frac{\rmd^2 x}{\rmd t^2}= [n(n+1)\wp(t) +B]x,\quad n\in\Z, \quad B\in\C,
\end{equation*}
then we can use the following  local criterion. If $\scG_1$ is not a finite group and  a local solution around $t=0$ of $m$-th order variational
equations contains a logarithmic term, then  $\scG^0_m$ is not Abelian, see  Appendix~A in the recent survey of Morales and Ramis~\cite{Morales:07::}.

For details how to apply
this criterion for homogeneous potentials with two degrees of freedom see
\cite{mp:04::d,mp:05::c}.
In the case of homogeneous Hamiltonian systems with three degrees of freedom the
strategy is completely similar to this applied in \cite{mp:04::d,mp:05::c}.
We remark only that the criterion is applicable for a solution generated by such  a
Darboux point $[\vd]$ that its eigenvalues $\lambda_i$ of $V''(\vd)$ are admissible and all are non-negative
 integers. In all cases, when we apply the higher order variational equation, we  take  a particular solution
 generated by the Darboux point at the infinity.

\subsubsection{Families $1$--$4$}

Let us take as the distinguished spectrum the first line in Table~\ref{tab:ds}.
From Lemma~\ref{lem:sls} it follows that  a potential with such a spectrum admits a proper and non-isotropic
Darboux point.   We can assume that this point is $[\vd]=[0:0:1]$ and  assign to it the admissible pair
 $(\Lambda_1,\Lambda_2)=(-1,-1)$. Moreover,  by Remark~\ref{rem:j}, we can assume that matrix
 $V''(\vd)$ is semi-simple. As an effect, we  consider potential of the form~\eqref{eq:pot3}.
 By Proposition~\ref{prop:nor}, we have $\Lambda_1=2a_3 -1$ and  $\Lambda_2=2a_3 -1$, so $a_3=a_6=0$.
  But then in the potential  \eqref{eq:pot3}
 variable $q_3$ is separated from $q_1$ and $q_2$.  This fact, and the assumption that the potential has
seven Darboux points implies that  one can make a
rotation in the plane $(q_1,q_2)$ and achieve that
$a_4=0$.  Then the potential simplifies to the form
\begin{equation}
 V=a_1q_1^3+a_2q_1^2q_2+a_5q_2^3+\dfrac{1}{3
}q_3^3,
\label{eq:pot3ff}
\end{equation}
and in the affine coordinates we have
\begin{equation}
\begin{split}
&g_1=(3a_5-2a_2)x_1^2 -3a_1x_2+a_2,\qquad
g_2=x_2^2 -
 x_2(3a_1+2a_2x_1),\\
&g_0=3a_1 + 2a_2x_1.
\end{split}
\label{eq:ggg}
\end{equation}
By the B\'ezout theorem, the number of solutions of $g_1=g_2=0$ is not bigger than four, so there is at most
four Darboux points  in the affine part of $\CP^2$.  Since by Proposition~\ref{prop:3i}, at most three Darboux
points can lie at infinity, and  we assumed that the potential has seven Darboux points,  four Darboux points
must lie  in the affine part of $\CP^2$ and three at infinity.   This fact implies that
 $a_5\neq 0$ and   $3a_5-2a_2\neq 0$.
For each  Darboux point in the affine part of $\CP^2$  we  attach the admissible pair
   $(\Lambda_1,\Lambda_2)\in\{(-1,-1),(-1,1),(1,1)\}$. Next
we  calculate symmetric functions
 $S_1=S_1(\Lambda_1,\Lambda_2)$,  $S_2=S_2(\Lambda_1,\Lambda_2)$,
and then, using a computer algebra system,  perform an elimination of  $x_1$ and $x_2$ from the corresponding
equations~\eqref{eq:g12} and \eqref{eq:lambdusie}.  This elimination gives
only one restriction
 on parameters $a_i$, namely $a_2=0$.
But then, from \eqref{eq:ggg} we see that $g_0=a_1$. By assumption all Darboux points are proper,
 and we have necessarily
$a_1\neq 0$.
The affine coordinates of four Darboux points  are following
\[
(x_1,x_2)=(0,0),\quad (x_1,x_2)=(0,3a_1),\quad
(x_1,x_2)=\left(\dfrac{a_1}{a_5},0\right),\quad
(x_1,x_2)=\left(\dfrac{a_1}{a_5},3a_1\right).
\]
We can check directly that the non-trivial eigenvalues  $(\Lambda_1,\Lambda_2)$ are as we required
$(-1,-1)$, $(-1,1)$, $(-1,1)$ and $(1,1)$, respectively.

Assumption $a_5\neq 0$ guarantees that   three Darboux points
are localised at infinity, and their coordinates are following
\[
(q_1,q_2,q_3)=(0,0,1),\quad
(q_1,q_2,q_3)=\left(0,\dfrac{1}{3a_5},0\right),\quad
(q_1,q_2,q_3)=\left(0,\dfrac{1}{3a_5},1\right).
\]
The non-trivial  eigenvalues  $(\Lambda_1,\Lambda_2)$ at these points  are  $(-1,-1)$, $(-1,-1)$, and  $(-1,1)$, respectively.

In this way we showed that if a potential admits seven proper Darboux points with a distinguished spectrum
given by the first line in Table~\ref{tab:ds},  then it  is equivalent to the
separable potential
\begin{equation}
 V_1=a_1q_1^3+a_5q_2^3+\dfrac{1}{3}q_3^3,\qquad a_1a_5\neq0,
\label{eq:fam1}
\end{equation}
 with obvious additional first integrals
\[
 I_1=\dfrac{1}{2}p_3^2+\dfrac{1}{3}q_3^3,\qquad I_2=\dfrac{1}{2}p_2^2+a_5q_2^3.
\]

An analysis for the next three distinguished spectra is completely analogous to that described above.
For the spectrum given by the second line  in Table~\ref{tab:ds} we obtained   one integrable potential
\begin{equation}
 V_2=\frac{a_5}{2}q_1^2q_2+a_5q_2^3+\dfrac{1}{3}q_3^3,\qquad a_5\neq0,
\label{eq:v22}
\end{equation}
  with first integrals
\[
 I_1=\dfrac{1}{2}p_3^2+\dfrac{1}{3}q_3^3,\qquad I_2=8p_1(q_1p_2- q_2p_1) +
a_5q_1^2(q_1^2 + 4q_2^2).
\]

The third spectrum gives rise   to another integrable potential
\begin{equation}
 V_3=\dfrac{3a_5}{16}q_1^2q_2+a_5q_2^3+ \dfrac{1}{3}q_3^3,\qquad a_5\neq 0,
\label{eq:v32}
\end{equation}
 with additional first integrals
\[
 I_1=\dfrac{1}{2}p_3^2+\dfrac{1}{3}q_3^3,\qquad I_2=128p_1^4
-a_5q_1^2(32p_1p_2q_1 - 96p_1^2q_2 +
   a_5q_1^2(q_1^2 + 6q_2^2)).
\]

The fourth spectrum  generates also  an integrable potential
\begin{equation}
 V_4=a_5( \rmi\sqrt{3}q_1^3+9q_1^2q_2+18q_2^3)+\dfrac{1}{3}q_3^3,\qquad
a_5\neq0,
\label{eq:v4}
\end{equation}
which  possesses two additional first integrals
\[
 \begin{split}
I_1&= \dfrac{1}{2}p_3^2+\dfrac{1}{3}q_3^3,\\
 I_2&=3p_1^4 + 2\rmi\sqrt{3}p_1^3p_2 -
 54a_5p_1p_2q_1^2(q_1 + \rmi\sqrt{3}q_2) +
 18a_5p_1^2q_1(\rmi\sqrt{3}q_1^2 + 6q_1q_2 +
   6\rmi\sqrt{3}q_2^2) \\
&+ 9\rmi a_5q_1^3
  (2\sqrt{3}p_2^2 + 3a_5(5\rmi q_1^3 +
     6\sqrt{3}q_1^2q_2 + 9\rmi q_1q_2^2 +
     12\sqrt{3}q_2^3)).
 \end{split}
\]

Let us note that potentials $V_1,V_2,V_3$ and $V_4$  have the form
 $V_i(q_1,q_2,q_3)=\widetilde V_i(q_1,q_2) +q_3^3/3$, where $ \widetilde V_i(q_1,q_2) $
are integrable  two dimensional homogeneous potentials.
They have
appeared in the famous Hietarinta table in \cite{Hietarinta:83::} and later were
re-obtained in the differential Galois framework in \cite{mp:04::d} where it was shown that they are the
 only integrable potentials with the maximal number of Darboux points for $k=3$.

The reconstruction of the potentials corresponding to the next families is more
complicated by  reason of the absence of the admissible pair $(-1,-1)$. Still one can
easily decode from the pair of $\Lambda$s at the infinity two coefficients of the
potential but, separation of the part of $V$ depending on $(q_1,q_2)$ and
$q_3$ no more appears, thus we cannot put $a_4=0$. Thus all six Darboux points rest at
the affine part of $\mathbb{CP}^2$ and the level of computational difficulties
increases very quickly.

\subsubsection{Families $5$--$7$}
For the fifth distinguished spectrum in Table~\ref{tab:ds}    we assign the pair
 $(\Lambda_1,\Lambda_2)=(-1,4)$ to the
Darboux point at the  infinity. Then we obtain immediately two coefficients
 $a_3=0$ and $a_6=5/2$ of the potential \eqref{eq:pot3}. Thus, we will analyse
the higher order variational equations for the potential
\begin{equation}
  V=a_1q_1^3 + a_2q_1^2q_2 + a_4q_1q_2^2 + a_5q_2^3 + \dfrac{5}{2}q_2^2q_3 +
\dfrac{1}{3}q_3^3.
\label{eq:poth1}
\end{equation}
 A local solution around $t=0$  of the second order variational equations does not have
logarithmic terms. Such terms appear  in the solutions of
the third order variational equations. Comparing them to zero we obtain the following equations
\[
\begin{split}
 &a_4(35a_2 - 18a_5) = 0, \quad
 -147 + 28a_4^2 - 27a_5^2 = 0, \quad
 a_2(9a_1 - a_4) = 0, \\
 &84a_2^2 + 168a_1a_4 - 56a_4^2 -
   27a_2a_5 = 0,
\end{split}
\]
with solutions
\begin{eqnarray}
& a_2 = 0,\qquad  a_4 = 0,\qquad a_5=\pm\dfrac{7\rmi}{3}
\label{eq:vv3_1}\\
&a_1=0,\qquad  a_2 = \pm\dfrac{3\rmi}{4},\qquad
  a_5 =\pm\dfrac{7\rmi}{3},\qquad  a_4 =0,
\label{eq:vv3_2}
\\
&a_1 = -\dfrac{\rmi}{2}\sqrt{\dfrac{21}{269}},\quad a_2 =\pm 6\rmi
\sqrt{\dfrac{14}{269}}, \quad
  a_4 = -\dfrac{9\rmi}{2}\sqrt{\dfrac{21}{269}},\quad a_5 =
\pm\dfrac{35\rmi}{3}\sqrt{\dfrac{14}{269}},
\label{eq:vv3_3}\\
&a_1 = \dfrac{\rmi}{2}\sqrt{\dfrac{21}{269}},\quad a_2 =\pm 6\rmi
\sqrt{\dfrac{14}{269}}, \quad
  a_4 = \dfrac{9\rmi}{2}\sqrt{\dfrac{21}{269}},\quad a_5 =
\pm\dfrac{35\rmi}{3}\sqrt{\dfrac{14}{269}},
\label{eq:vv3_4}\\
&a_1 = \dfrac{1}{2}\sqrt{\dfrac{7}{3}},\quad   a_2 = 0, \quad a_4
=\pm\dfrac{1}{2} \sqrt{21},\quad a_5=0.
\label{eq:vv3_5}
\end{eqnarray}
The first solution \eqref{eq:vv3_1}  gives the
 integrable potential
\[
 V=a_1q_1^3\pm\dfrac{7\rmi}{3}q_2^3+\dfrac{5}{2}q_2^2q_3+\dfrac{1}{3}q_3^3,
\]
which is  equivalent to potential $V_2$ defined by~\eqref{eq:v22}.

Solution \eqref{eq:vv3_2}  generates the integrable potential
\[
V_5=\dfrac{3\rmi }{4}q_1^2q_2 + \dfrac{7\rmi}{3}q_2^3 + \dfrac{5}{2}q_2^2q_3 +
\dfrac{1}{3}q_3^3,
\]
with first integrals
\[
 \begin{split}
 & I_1=48p_1(\rmi p_2 + p_3)q_1 - (3q_1^2 + 2(q_2 - 2\rmi q_3)^2)
  (3q_1^2 + 2q_2(5q_2 - 4\rmi q_3)) -
 48p_1^2(\rmi q_2 + q_3) \\
&+ 32(p_2 - 2\rmi p_3)
  (p_2q_3-p_3q_2),\\
& I_2=24p_1^4 - 8p_1^2(12(p_2 - \rmi p_3)(p_2 - 2\rmi p_3) -
   9\rmi q_1^2q_2 + 2(2q_2 - \rmi q_3)(q_2 - 2\rmi q_3)
    (5\rmi q_2 + 4q_3))\\
& +
 3q_1^2(3q_1^4 + 16(p_2 + \rmi p_3)(p_3q_2 - p_2q_3) +
   6q_1^2(3q_2^2 - 4\rmi q_2q_3 + q_3^2) \\
&+
   4q_2(q_2 - 2\rmi q_3)(7q_2^2 - 4\rmi q_2q_3 +
     2q_3^2)) + 24p_1q_1
  (p_3(-3q_1^2 - 22q_2^2 + 20\rmi q_2q_3)\\
& -
   \rmi p_2(3q_1^2 + 6q_2^2 + 4\rmi q_2q_3 + 8q_3^2)).
 \end{split}
\]
Solutions \eqref{eq:vv3_3} --\eqref{eq:vv3_5}  give potentials
\[
\begin{split}
& V=-\dfrac{\rmi}{2}\sqrt{\dfrac{21}{269}}q_1^3 \mp 6\rmi
\sqrt{\dfrac{14}{269}}q_1^2q_2 -
 \dfrac{9\rmi}{2}\sqrt{\dfrac{21}{269}}q_1q_2^2 \mp
\dfrac{35\rmi}{3}\sqrt{\dfrac{14}{269}}q_2^3 +
 \dfrac{5}{2}q_2^2q_3 + \dfrac{1}{3}q_3^3,\\
& V=\dfrac{\rmi}{2}\sqrt{\dfrac{21}{269}}q_1^3 \mp 6\rmi
\sqrt{\dfrac{14}{269}}q_1^2q_2 +
 \dfrac{9\rmi}{2}\sqrt{\dfrac{21}{269}}q_1q_2^2 \mp
\dfrac{35\rmi}{3}\sqrt{\dfrac{14}{269}}q_2^3 +
 \dfrac{5}{2}q_2^2q_3 + \dfrac{1}{3}q_3^3,\\
& V=\pm\dfrac{1}{2}\sqrt{\dfrac{7}{3}}q_1^3 \pm \dfrac{\sqrt{21}}{2}q_1q_2^2 +
\dfrac{5}{2}q_2^2q_3 + \dfrac{1}{3}q_3^3.
\end{split}
\]
All of them are not integrable because the respective solutions of the fifth order variational equations
 have logarithmic terms.

For the sixth distinguished  spectrum  we assign pair $(\Lambda_1,\Lambda_2)=(4,21)$ to the
 Darboux point at the  infinity and we apply the same procedure as in the previous case.
Solving conditions obtained from an analysis of solutions of  the third order variational equations,
 we obtained only one integrable potential
\begin{equation}
\begin{split}
V_6= & 364\sqrt{17}q_1^3 + 2835\rmi\sqrt{17}q_1^2q_2 + 1560\sqrt{17}q_1q_2^2 +
 6552\rmi \sqrt{17}q_2^3 + 4335q_1^2q_3\\ & + 19074q_2^2q_3 + 578q_3^3,
\end{split}
\label{eq:cyc}
\end{equation}
with one first integral of degree two in the momenta
\[
 \begin{split}
&I_1=  34\sqrt{17}p_3^2q_1 +
 8p_2p_3(18\rmi q_1 + q_2) -
 2p_2^2(7\sqrt{17}q_1 + 4q_3) -
 p_1^2(54\rmi \sqrt{17}q_2 + 26q_3) \\
&+
 2p_1(p_2(27\rmi \sqrt{17}q_1 +
     7\sqrt{17}q_2 + 9\rmi q_3) +
   p_3(13q_1 - 81\rmi q_2 - 17\sqrt{17}
      q_3)) \\
&- 51(-2568q_2^4 +
   36\rmi q_2^3(695q_1 + 52\sqrt{17}
      q_3) - 36\rmi q_1q_2(151q_1^2 +
     241\sqrt{17}q_1q_3 + 221q_3^2) \\
&+
   2q_2^2(31351q_1^2 + 5386\sqrt{17}q_1
      q_3 + 2924q_3^2) +
   q_1(12205q_1^3 + 1668\sqrt{17}q_1^2
      q_3 + 3978q_1q_3^2 + 1156\sqrt{17}
      q_3^3)),
 \end{split}
\]
and the second more complicated of degree four in the momenta is given in  Appendix.

For the seventh distinguished spectrum in Table~\ref{tab:ds} we assign pair $(\Lambda_1,\Lambda_2) =(-1,6)$
to the  Darboux point at the infinity.  The conditions that there are no  logarithmic terms in
solutions of variational equations of orders 3, 4 and 7 yield one integrable potential
\[
V_7= 44  \sqrt{7}  q_1^3 + 240\rmi \sqrt{14} q_1^2  q_2   +
 330  \sqrt{7}  q_1    q_2^2 + 935\rmi \sqrt{14}  q_2^3 +
 3087  q_2^2  q_3 + 294  q_3^3.
         \]
One its first integral is of degree four with respect the momenta
\[
 \begin{split}
 &I_1= 19 p_1^4 + 47 p_2^4 - 14 \rmi  \sqrt{14} p_2^3 p_3 +
 16 p_1^3 (8\rmi \sqrt{2} p_2 + 7  \sqrt{7} p_3) +
 294 p_2 p_3 (-436\rmi  \sqrt{2} q_1^3\\
& +
   6 q_1 q_2 (-195\rmi  \sqrt{2} q_2 + 56  \sqrt{7} q_3) +
   6 q_1^2 (17 q_2 - 77\rmi  \sqrt{14} q_3) +
   q_2^2 (827 q_2 - 399\rmi  \sqrt{14} q_3)) \\
&+
 252 p_2^2 (12  \sqrt{7} q_1^3 +
   q_1^2 (145\rmi  \sqrt{14} q_2 - 98 q_3) +
   2  \sqrt{7} q_1 (185 q_2^2 - 98 q_3^2) +
   q_2 (612\rmi  \sqrt{14} q_2^2 \\
&+ 2303 q_2 q_3 -
     49\rmi  \sqrt{14} q_3^2)) +
 18 p_1^2 (-11 p_2^2 + 14\rmi  \sqrt{14} p_2 p_3 +
   7 (7 p_3^2 - 136  \sqrt{7} q_1^3 + 480\rmi  \sqrt{14}
      q_1^2 q_2\\
& - 2756  \sqrt{7} q_1 q_2^2 -
     27\rmi  \sqrt{14} q_2^3 + 6272\rmi  \sqrt{2} q_1 q_2
      q_3 - 9898 q_2^2 q_3 + 784  \sqrt{7} q_1 q_3^2 +
     3430\rmi  \sqrt{14} q_2 q_3^2\\
& + 4116 q_3^3)) +
 p_1 (-65\rmi  \sqrt{2} p_2^3 - 147  \sqrt{7} p_2^2 p_3 -
   147 p_3 (1216 q_1^3 + 24 q_1^2 (-213\rmi  \sqrt{2} q_2 +
       28  \sqrt{7} q_3)\\
& + 96 q_1 q_2 (172 q_2 -
       49\rmi  \sqrt{14} q_3) +
     q_2 (4189\rmi  \sqrt{2} q_2^2 + 5964  \sqrt{7} q_2 q_3 -
       6174\rmi  \sqrt{2} q_3^2)) +
   21 p_2 (14\rmi  \sqrt{2} p_3^2 \\
&- 1664\rmi  \sqrt{14}
      q_1^3 + 6607  \sqrt{7} q_2^3 - 28665\rmi  \sqrt{2}
      q_2^2 q_3 - 3822  \sqrt{7} q_2 q_3^2 -
     6174\rmi  \sqrt{2} q_3^3 \\
&+ 384 q_1^2
      (4  \sqrt{7} q_2 - 49\rmi  \sqrt{2} q_3) +
     6 q_1 (-1555\rmi  \sqrt{14} q_2^2 + 392 q_2 q_3 -
       1666\rmi  \sqrt{14} q_3^2))) +
 294 (27  \sqrt{7} p_3^2 (8 q_1^3 \\
&+ 22\rmi  \sqrt{2} q_1^2
      q_2 - 24 q_1 q_2^2 + 9\rmi  \sqrt{2} q_2^3) +
   112 (434 q_1^6 + 96 q_1^5 (-7\rmi  \sqrt{2} q_2 +
       12  \sqrt{7} q_3) + 6 q_1^4 (119 q_2^2\\
& -
       324\rmi  \sqrt{14} q_2 q_3 + 924 q_3^2) +
     12 q_1 q_2^2 (2044\rmi  \sqrt{2} q_2^3 +
       1782  \sqrt{7} q_2^2 q_3 - 2625\rmi  \sqrt{2} q_2
        q_3^2 - 504  \sqrt{7} q_3^3) \\
&+
     q_1^3 (2233\rmi  \sqrt{2} q_2^3 + 5931  \sqrt{7} q_2^2
        q_3 - 14364\rmi  \sqrt{2} q_2 q_3^2 +
       1134  \sqrt{7} q_3^3) - 8 q_2^3 (6986 q_2^3 -
       3492\rmi  \sqrt{14} q_2^2 q_3\\
& - 7749 q_2 q_3^2 +
       378\rmi  \sqrt{14} q_3^3) - 3 q_1^2 q_2
      (3626 q_2^3 + 48\rmi  \sqrt{14} q_2^2 q_3 -
       3465 q_2 q_3^2 + 1386\rmi  \sqrt{14} q_3^3))),
 \end{split}
\]
and the second of  degree six with respect to the momenta  is  given in Appendix.

\subsubsection{Families $8$--$9$}
For the  eighth distinguished spectrum from Table~\ref{tab:ds} we assign pair $(\Lambda_1,\Lambda_2)=(6,14)$
to the Darboux point at the infinity.  The conditions that there are no  logarithmic terms in
solutions of variational equations of order four are following
\begin{equation}
 \begin{split}
&a_4(-7291848638025 - 195171588288a_2^2 -
   7947435510a_1a_4 + 20220446680a_4^2\\
& +
   315697519440a_2a_5 - 183025477350a_5^2)=0,\\
 &a_4(411968789569575 + 2262392768724a_2^2 +
   1291180817430a_1a_4 + 339419628940a_4^2\\
& +
   1213154206380a_2a_5 + 4209585979050a_5^2)=0,\\
 &2013340175a_1 + 114593850a_1^3 + 9791320a_1a_2^2 -
  6923862a_2^2a_4=0, \\
&729732302775a_1 -
  633757824a_1a_2^2 + 43836570855a_4 +
  8866402020a_1^2a_4 + 4915882928a_2^2a_4 \\
&+
  1092887040a_1a_4^2 + 217032200a_4^3 +
  16783940880a_1a_2a_5 - 5584905348a_2a_4a_5=0.
 \end{split}
\label{eq:fromhove}
\end{equation}
The above system has very complicated solutions. In order to discard those which do not satisfy our assumptions
 about the number of Darboux points and their spectra, we decided to proceed as follows.
At first we assume that there are six Darboux points in the affine part of $\CP^2$, i.e., by
Proposition~\ref{prop:3i}, we assume that $a_4\neq 0$.
For three  of them  we assign  $\left(-\frac{2}{3},-\frac{2}{3}\right)$,
$\left(-1,\frac{7}{3}\right)$ and $(6,14)$, as the respective   admissible pairs
$(\Lambda_1,\Lambda_2)$.  For each of these points   we considered system of equations~\eqref{eq:g12} and~\eqref{eq:lambdusie}.
Then we calculated the  elimination ideal with
respect to variables $(x_1,x_2)$ and its Groebner basis. Next
we join all these three bases and polynomials given by the left hand sides of equations
\eqref{eq:fromhove}. The Groebner basis of all these
polynomials is   very simple. Equating to zero all  polynomials from this basis we obtained equations
\begin{equation}
\label{eq:s8}
\begin{split}
& a_5^3 + 28431/320 a_2 + 729/64 a_5=0,\quad
    a_2^2 - 125/117a_5^2 - 600/13=0,\\
&    a_2a_5 - 145/117 a_5^2 - 540/13=0,\quad
    a_1=a_4=0.
\end{split}
\end{equation}
So we obtained a contradiction with our assumption that $a_4=0$.  Hence, only four points are in the affine
part of $\CP^2$.
Assuming that to three of them the attached admissible pairs are as above, we arrive to the same
 equations~\eqref{eq:s8}.
Their solutions give rise to
 only one  integrable potential
\[
 V_8=\dfrac{7}{2}q_1^2q_3-\dfrac{5\rmi\sqrt{3}}{2}q_1^2q_2 -
\dfrac{9\rmi\sqrt{3}}{2}q_2^3  + \dfrac{15}{2}q_2^2q_3 + \dfrac{1}{3}q_3^3,
\]
that admits the following first integrals
\[
\begin{split}
& I_1=26 \rmi \sqrt{3} p_1^3 + 3 p_1 (6 \rmi \sqrt{3} p_2^2 -
   48 p_2 p_3 - 32 \rmi \sqrt{3} p_3^2 + 27 q_2^3 +
   69 \rmi \sqrt{3} q_2^2 q_3 + 15 q_2 (13 q_1^2 - 8 q_3^2)\\
& +
   \rmi \sqrt{3} q_3 (91 q_1^2 + 16 q_3^2)) +
 3 q_1 (p_2 (-91 q_1^2 + 45 q_2^2 + 42 \rmi \sqrt{3} q_2 q_3 +
     72 q_3^2) - \rmi p_3 (65 \sqrt{3} q_1^2 \\
&-
     99 \sqrt{3} q_2^2 - 480 \rmi q_2 q_3 +
     112 \sqrt{3} q_3^2)),
\end{split}
\]
\[
 \begin{split}
 & I_2=8619 p_1^4 + 459 p_2^4 + 9000 \rmi \sqrt{3} p_2^3 p_3 +
 21519 p_3^4 + 43940 q_1^6 +
 78 p_1^2 (141 p_2^2 + 220 \rmi \sqrt{3} p_2 p_3 + 41 p_3^2\\
& -
   369 \rmi \sqrt{3} q_2^3 + 1547 q_1^2 q_3 +
   2595 q_2^2 q_3 + 214 q_3^3 + 85 \rmi \sqrt{3} q_2
    (-13 q_1^2 + 4 q_3^2)) -
 18 p_2^2 (2649 p_3^2\\
& - 791 \rmi \sqrt{3} q_2^3 -
   6357 q_1^2 q_3 - 765 q_2^2 q_3 + 86 q_3^3 -
   5 \rmi \sqrt{3} q_2 (299 q_1^2 + 100 q_3^2)) +
 20 p_2 p_3 (-1568 \rmi \sqrt{3} p_3^2 \\
&+ 6075 q_2^3 +
   9675 \rmi \sqrt{3} q_2^2 q_3 -
   189 q_2 (65 q_1^2 + 24 q_3^2) + 21 \rmi \sqrt{3} q_3
    (65 q_1^2 + 112 q_3^2)) +
 6 p_3^2 (21573 \rmi \sqrt{3} q_2^3\\
& + 11011 q_1^2 q_3 -
   104085 q_2^2 q_3 + 4782 q_3^3 - 5 \rmi \sqrt{3} q_2
    (3939 q_1^2 + 7840 q_3^2)) +
 780 p_1 q_1 (p_3 (13 q_1^2 + 693 q_2^2 \\
&+
     262 \rmi \sqrt{3} q_2 q_3 - 56 q_3^2) +
   3 \rmi p_2 (13 \sqrt{3} q_1^2 - 35 \sqrt{3} q_2^2 +
     32 \rmi q_2 q_3 - 4 \sqrt{3} q_3^2)) +
 12 (28674 q_2^6 \\
&+ 25110 \rmi \sqrt{3} q_2^5 q_3 +
   30758 q_1^4 q_3^2 + 7917 q_1^2 q_3^4 + 797 q_3^6 +
   405 q_2^4 (-13 q_1^2 + 30 q_3^2) -
   15 q_2^2 q_3^2 (-6500 q_1^2 \\
&+ 381 q_3^2) +
   4 \rmi \sqrt{3} q_2^3 q_3 (-11583 q_1^2 + 5959 q_3^2) +
   10 \rmi \sqrt{3} q_2 q_3 (-2197 q_1^4 + 1118 q_1^2 q_3^2 +
     392 q_3^4)).
 \end{split}
\]
It does not end the analysis, because it is possible that to four Darboux points in the affine part
 of $\CP^2$ only two different admissible pairs are attached. To analyse all these cases we proceed as follows.
 We know that two additional Darboux points are in line at the infinity. Their coordinates and
  spectra $(\Lambda_1^{(i)},\Lambda_2^{(i)})$ can be found explicitly. Since now $a_4=0$, they can  depend
 only on three  parameters $a_1$, $a_2$, and $a_5$. In fact they depend only on $a_2$, and $a_5$.
So $(\Lambda_1^{(1)},\Lambda_2^{(1)})$ and $(\Lambda_1^{(2)},\Lambda_2^{(2)})$ are not independent,
 and the following relations hold
\begin{equation}
\begin{split}
(8+15 \Lambda_2^{(1)}) \Lambda_1^{(2)}&=14(\Lambda_2^{(2)}-\Lambda_2^{(1)}),\\
(8+15 \Lambda_2^{(2)}) \Lambda_1^{(1)}&=-14(\Lambda_2^{(2)}-\Lambda_2^{(1)}),\\
 (14+15\Lambda_2^{(1)})\Lambda_1^{(1)}&=-14\Lambda_1^{(2)}.
\end{split}
\end{equation}
Now, it is easy to check that from the considered distinguished spectrum there is only one choice of two pairs
 satisfying the  above relations, namely
\begin{equation*}
 (\Lambda_1^{(1)},\Lambda_2^{(1)})=\left(-\frac{2}{3},-\frac{2}{3}\right), \qquad (\Lambda_1^{(2)},\Lambda_2^{(2)})=\left(\frac{7}{3},-1\right),
\end{equation*}
and this gives
\begin{equation*}
 a_2= \pm 5\rmi \frac{\sqrt{3}}{2} \mtext{and}a_5=\pm 9\rmi \frac{\sqrt{3}}{2}.
\end{equation*}
For both sign choices equations~\eqref{eq:fromhove} imply that $a_1=0$. In this way,
 we recover a potential equivalent to $V_8$.

For the  ninth distinguished spectrum from Table~\ref{tab:ds} we assign the pair $(\Lambda_1,\Lambda_2)=(14,39)$
to the
Darboux point at the  infinity, and then we proceed in the same way as in the previous case.
Logarithmic terms appear only in solutions of the  seventh order variational equations.
The coefficients of these terms are  four very
complicated polynomials depending on $a_1,a_2,a_4,a_5$.
We  join these polynomials and all  eliminating ideals obtained for five Darboux points in the affine part
 of $\CP^2$.
The Groebner basis of this ideal is very simple and it gives the following equations
\[
 a_4^2 + \dfrac{4609248}{125}=0,\quad a_5^2 - \dfrac{1395372}{125}=0,\quad  a_1
+ \dfrac{9}{152}a_4=0,\quad a_2 + \dfrac{621}{682}a_5=0.
\]
This system yields the following integrable potential
\[
\begin{split}
 V_{9}&=27\rmi \sqrt{3990}q_1^3 + 3726\sqrt{15}q_1^2q_2 -
 456\rmi \sqrt{3990}q_1q_2^2 - 4092\sqrt{15}q_2^3 - 1125q_1^2q_3\\
& -
 3000q_2^2q_3 - 50q_3^3,
\end{split}
\]
with very complicated first integrals of degree 4 and 6 in the momenta which are given in Appendix.

Potential $V_9$ was obtained under assumption that six Darboux points are localised in the affine part
 of $\CP^2$. The cases when a smaller number of points is in the  affine part of $\CP^2$ we analyse in a
 similar way as for the eighth distinguished spectrum. It appears that such cases for an integrable potential
 are impossible.

\subsubsection{Family $10$}
For the  distinguished spectrum from Table~\ref{tab:ds} we  assign the pair $(\Lambda_1,\Lambda_2)=(-1,1)$ to
the Darboux point at  the infinity.
Doing this we additionally assume that the admissible pair $(-1,1)$ is attached to a non-isotropic Darboux point.  The case corresponding to the isotropic Darboux point will be considered separately. The potential has the form
\[
 V=a_1q_1^3 + a_2q_1^2q_2 +
 a_4q_1q_2^2 + a_5q_2^3 +
 q_2^2q_3 + \dfrac{1}{3}q_3^3.
\]
The solutions of the  second  and fourth order variational equations do not have  logarithmic terms iff  $a_2=0$ and
\[
 a_4(-20 - 144a_1a_4 + 160a_4^2 - 45a_5^2) =  0,\qquad  a_4(-100 - 90a_1a_4 +
128a_4^2 -
    225a_5^2)=0.
\]
The above equations have the following solutions
\[
 \{a_4=0\},\qquad \left\{a_1 = \dfrac{16}{15}a_4,
  a_5 =\pm \dfrac{2}{15}\sqrt{-25 + 8a_4^2}\right\}.
\]
The first solution yields the potential
\[
 V=a_1q_1^3+a_5q_2^3+q_2^2q_3+\dfrac{1}{3}q_3^3,
\]
that is equivalent to \eqref{eq:fam1}.

Let us take the second solution with the sign $+$. Then fifth order variational
equations yield
\[
 a_4=0\qquad  \text{or}\qquad a_4=\pm\dfrac{5}{2\sqrt{2}}.
\]
If  $a_4=0$,  then the corresponding potential has only one Darboux point, so we discard this case.
  The second solution yields the following integrable potential
\[
 V_{10}=\dfrac{4\sqrt{2}q_1^3}{3} + \dfrac{5q_1q_2^2}{2\sqrt{2}} + q_2^2q_3 +
\dfrac{1}{3}q_3^3,
\]
with first integrals of degree 4 and 6 in the momenta
\[
 \begin{split}
 & I_1=12 p_2^4 - 27 q_2^6 - 18 q_2^4 (q_1^2 - 4  \sqrt{2} q_1 q_3 + 2 q_3^2) +
 4 (6 p_1^2 - 3 p_3^2 + 16  \sqrt{2} q_1^3 - 2 q_3^3) (3 p_3^2 + 2 q_3^3)\\
& +
 12 q_2^2 (3 p_3^2 ( \sqrt{2} q_1 - 4 q_3) + 12 p_1 p_3 (q_1 +  \sqrt{2} q_3) -
   2 q_3^2 (12 q_1^2 +  \sqrt{2} q_1 q_3 + 2 q_3^2))\\
& -
 12 p_2 q_2 (2 p_3 (16 q_1^2 + 3 q_2^2 + 8  \sqrt{2} q_1 q_3 - 4 q_3^2) +
   3  \sqrt{2} p_1 (q_2^2 + 4 q_3^2))\\
& -
 12 p_2^2 (2 p_3 (2  \sqrt{2} p_1 + p_3) - 4 (q_2 - q_3) q_3 (q_2 + q_3) -
    \sqrt{2} q_1 (5 q_2^2 + 8 q_3^2)),
 \end{split}
\]
\[
 \begin{split}
&I_2=81 q_2^8 (2 \sqrt{2} q_1 + q_3) + 216 p_2 p_3 q_2^5 (\sqrt{2} q_1 + 2 q_3)
+
 54 q_2^6 (p_2^2 - 3 p_3^2 + 4 \sqrt{2} q_1^3 - 24 q_1^2 q_3\\
& -
   6 \sqrt{2} q_1 q_3^2) + 384 p_2 p_3 q_1^2 q_2 (3 p_2^2 + 8 \sqrt{2} q_1^3 +
   8 q_1^2 q_3 - 2 \sqrt{2} q_1 q_3^2) - 72 p_1^4 (3 p_3^2 + 2 q_3^3) \\
&+
 144 p_2 p_3 q_2^3 (p_2^2 + 8 q_1^2 (2 \sqrt{2} q_1 + 3 q_3)) +
 144 p_1^3 (\sqrt{2} p_2^2 p_3 + 3 \sqrt{2} p_2 q_2 q_3^2 -
   3 p_3 q_2^2 (q_1 + \sqrt{2} q_3)) \\
&-
 32 (p_2^6 + 12 p_2^4 q_1^2 q_3 + 12 p_2^2 q_1^3 (\sqrt{2} p_3^2 + 4 q_1 q_3^2)
+
   32 q_1^6 (3 p_3^2 + 2 q_3^3)) -
 12 p_1^2 (4 p_2^4 \\
&- 6 p_2 p_3 q_2 (16 q_1^2 + 9 q_2^2 + 8 \sqrt{2} q_1 q_3 -
     4 q_3^2) + 9 q_2^4 (2 q_1^2 + 4 \sqrt{2} q_1 q_3 + q_3^2) +
   32 \sqrt{2} q_1^3 (3 p_3^2 + 2 q_3^3)\\
& +
   12 p_2^2 (p_3^2 + 4 q_2^2 q_3 - \sqrt{2} q_1 (q_2^2 - 2 q_3^2)) +
   6 q_2^2 (9 \sqrt{2} p_3^2 q_1 + 2 q_3^2 (-6 q_1^2 + 2 \sqrt{2} q_1 q_3 +
       q_3^2))) \\
&- 144 q_2^4 (p_2^2 (7 q_1^2 + 5 \sqrt{2} q_1 q_3 + 2 q_3^2) +
   3 q_1^2 (3 p_3^2 - 2 q_3 (-2 q_1^2 + 2 \sqrt{2} q_1 q_3 + q_3^2)))\\
& -
 48 q_2^2 (p_2^4 (5 \sqrt{2} q_1 + 4 q_3) +
   4 p_2^2 q_1^2 (8 q_1^2 + 2 \sqrt{2} q_1 q_3 + 3 q_3^2) +
   8 q_1^3 (9 p_3^2 q_1 + q_3^2 (-6 \sqrt{2} q_1^2\\
& + 4 q_1 q_3 +
       \sqrt{2} q_3^2))) + 6 p_1 (16 \sqrt{2} p_2^4 p_3 +
   16 p_2^2 p_3 (8 q_1^3 - 6 q_1 q_2^2 + 3 \sqrt{2} q_2^2 q_3) +
   4 p_2^3 q_2 (-16 \sqrt{2} q_1^2 \\
&+ 32 q_1 q_3 +
     \sqrt{2} (3 q_2^2 + 4 q_3^2)) +
   3 p_3 q_2^2 (9 \sqrt{2} q_2^4 - 64 q_1^3 (\sqrt{2} q_1 + 2 q_3) -
     12 q_2^2 (2 \sqrt{2} q_1^2 + 8 q_1 q_3\\
& + \sqrt{2} q_3^2)) +
   12 p_2 q_2 (-3 \sqrt{2} p_3^2 q_2^2 + 9 q_1 q_2^4 + 32 q_1^3 q_3^2 +
     4 q_2^2 (4 q_1^3 + 6 q_1 q_3^2 + \sqrt{2} q_3^3))).
 \end{split}
\]

Let us consider the case when pair $(\Lambda_1,\Lambda_2)=(-1,1)$  is attached to an isotropic Darboux point.
In this case we assign pair $(\Lambda_1,\Lambda_2)=(14,39)$ to the Darboux point at the infinity and this gives
 $a_3=15/2$ and $a_6=20$.
Let $[\vd]$  be  the isotropic   Darboux point corresponding to admissible pair $ (-1,1)$.
We have two possibilities:
\begin{enumerate}
 \item either $d_3\neq 0$, and then using a rotation around the third axis one can
annihilate one component, e.g. $d_1=0$ and
$\vd=(0,s,\rmi s)$, where $s\neq 0$,  or
\item $d_3=0$ and $\vd=(s,\rmi s,0)$, where $s\neq 0$.
\end{enumerate}
In the first case the condition that $\vd=(0,s,\rmi s)$ is a Darboux point
yields
\[
 a_4=0, \quad a_5=-\dfrac{59}{3}\rmi ,\qquad s= \frac{\rmi}{19}.
\]
Then the eigenvalues of  $V''(\vd)$ are $((2\rmi a_2-15)/19,2,2)$ and $V''(\vd)$ is not semi-simple.
Thus by Remark~\ref{rem:j} this potential is not integrable.

In the second case condition that $\vd=(s,\rmi s,0)$ satisfies $V'(\vd)=\vd$ leads to  $s=0$ but it was
assumed that $s\neq 0$.
\section{Final remarks}
Presented calculations show that, formulated at the beginning, classification programme  for homogeneous
potentials is very ambitious and meets a lot of  non-trivial theoretical, as well as computational difficulties.
 Natural question arises  about possible simplifications and improvements allowing  to dream about
effective calculations for $n=3$ and $k>3$, as well as for $n>3$.
One possibility is to  use a full power  of the higher order variational equations techniques.
   Here we used only the local criterion of the existence of logarithmic terms in solutions of higher order
 variational equations, which  implies that their differential Galois groups are not Abelian.
In fact it is the only known way to use the higher variational equations.  However, it seems that,  for
systems with homogeneous potentials,  stronger obstacles for the integrability can be deduced from an analysis
 of global  differential  Galois groups of these equations.

Another question is if we have already detected  all possible relations  between the eigenvalues of
 Hessian $V''(\vd)$ at different Darboux points.  In \cite{Guillot:04::} Guillot  mentioned that other
relations exist, but their form is completely not obvious.  If such additional relations exist, then they give new
 obstacles for the integrability and they limit the number of elements of the distinguished spectra.
 Thus the answer to this question is of great practical importance.

Let us mention also that relations  \eqref{eq:rkoj} and \eqref{eq:rtau}
were already presented in \cite{mp:07::a}. However in  \cite{mp:07::a} only the cases
with maximal number of proper Darboux points were analysed, corresponding here to
Theorem~\ref{thm:1}. Furthermore, the proof of this result was based
on results of Guillot  \cite{Guillot:04::}  obtained with the help of highly advanced techniques \cite{Baum:70::}
for which there are no obvious ways for  generalisations.  In present  paper Theorem~\ref{thm:1} was obtained
 ab initio and
its proof is completely elementary. Additionally,  using the introduced here approach of multidimensional residue
calculus, we were able to analyse also some non-generic   cases.

The first integrals given for the presented integrable potentials were found by means of the direct method
 \cite{Hietarinta:87::}. We have postulated the form of a first integral as a weight-homogeneous polynomial
function with respect to the coordinates and the momenta. We have started from the lowest possible weight degree
 equal to 2, and increased this degree up to finding the first integral. Thus the presented first integrals are
of the lowest possible degrees. All of them are irreducible polynomials. This follows from  the results contained
 in \cite{mp:04::c}.
First integrals for some potentials  are very complicated functions.   We only mention that theses expressions
have already been simplified.

Another problem is that some integrable potentials possess imaginary  coefficients.  In all cases it is possible
 to convert them into real potentials, but together with a simultaneous change of the signature of the kinetic
 energy form.  There is an open problem how to find their equivalent real form, or how to prove that such a
form does not exist.

Also the problems how to improve the calculations of  distinguished spectra that are the starting point to look
 for new integrable potentials and their reconstruction are worth considering.
Our calculations described in the last section show that the one distinguished spectrum gives rise to
a finite number of potentials. However, in some cases (the first four distinguished spectra) we obtained families
of integrable potentials depending on parameters. Let us notice, however, that this occurs only in cases when
  potentials  separate, i.e., if they are  sums of two integrable potentials depending on different coordinates.

\section*{Acknowledgments}
The author is very grateful to Andrzej J. Maciejewski for many helpful comments
and suggestions concerning improvements and simplifications of some results, as well as for his help in
numerical calculations of distinguished spectra.

Special thanks for Alain Albouy, Jean-Pierre Marco and Alexei Tsygvintsev for their stimulative questions
 about the role of improper Darboux points.
The author wishes also to thank Arkadiusz P{\l}oski for his useful comments about multidimensional residues
 and, in particular, for pointing our  attention to papers of G.~Biernat.
The author is greatly indebted to Delphine Boucher for her explanations about the computer algebra system MAGMA.  Without these comments some cases of potential's reconstruction were intractable.

This research has been partially supported  by grant No. N N202 2126 33 of Ministry of Science and Higher Education of Poland, by projet de l'Agence National de la
Recherche
``~Int\'egrabilit\'e r\'eelle et complexe en m\'ecanique hamiltonienne''
N$^\circ$~JC05$_-$41465, and  by  UMK grants 343-A as well as 414-A.

\section*{Appendix}
For potential
\begin{equation*}
\begin{split}
V_6= & 364\sqrt{17}q_1^3 + 2835\rmi\sqrt{17}q_1^2q_2 + 1560\sqrt{17}q_1q_2^2 +
 6552\rmi \sqrt{17}q_2^3 + 4335q_1^2q_3 + 19074q_2^2q_3 + 578q_3^3,
\end{split}
%\label{eq:cyc}
\end{equation*}
the second first integral is following
\[
 \begin{split}
&I_2=20 \rmi p_1^4 (2 \sqrt{17} q_2 + 5 \rmi q_3) + 120 p_2^4 q_3 +
 24 \rmi p_2^3 p_3 (29 q_1 + 5\rmi q_2 - 14 \sqrt{17} q_3) +
 4 p_1^3 (-2\rmi p_2 (5 \sqrt{17} q_1\\
& - 7\rmi \sqrt{17} q_2 - 31 q_3) +
   p_3 (25 q_1 + 247\rmi q_2 + 103 \sqrt{17} q_3)) +
 p_2^2 (682193 q_1^4 - 572220 q_2^4 \\
&+
   8 p_3^2 (134 \sqrt{17} q_1 + 42 \rmi \sqrt{17} q_2 - 713 q_3) +
   537856 \sqrt{17} q_1^3 q_3 - 5080620 q_1^2 q_3^2 +
   1895840 \sqrt{17} q_1 q_3^3\\
& - 1228828 q_3^4 +
   432 \rmi q_2^3 (8279 q_1 + 2566 \sqrt{17} q_3) +
   180 q_2^2 (24259 q_1^2 + 1136 \sqrt{17} q_1 q_3 + 50286 q_3^2)\\
& +
   72\rmi q_2 (17153 q_1^3 - 9978 \sqrt{17} q_1^2 q_3 + 39882 q_1 q_3^2 -
     8092 \sqrt{17} q_3^3)) + p_1^2 (590325 q_1^4 - 11864844 q_2^4 \\
&+
   215380 \sqrt{17} q_1^3 q_3 - 1647300 q_1^2 q_3^2 +
   714408 \sqrt{17} q_1 q_3^3 + 1268132 q_3^4 +
   p_2^2 (56 \sqrt{17} q_1 + 96\rmi \sqrt{17} q_2 \\
&+ 76 q_3) -
   4 p_3^2 (103 \sqrt{17} q_1 + (249 I) \sqrt{17} q_2 + 1600 q_3) +
   72\rmi q_2^3 (60775 q_1 + 28124 \sqrt{17} q_3)\\
& -
   12 q_2^2 (129965 q_1^2 - 18626 \sqrt{17} q_1 q_3 - 22542 q_3^2) +
   36\rmi q_2 (22525 q_1^3 - 22680 \sqrt{17} q_1^2 q_3 + 310386 q_1 q_3^2\\
& -
     50864 \sqrt{17} q_3^3) +
   4 p_2 p_3 (439 q_2 - 3\rmi (103 q_1 + 35 \sqrt{17} q_3))) +
 4 p_2 p_3 (2709\rmi \sqrt{17} q_1^4 + 202986\rmi \sqrt{17} q_2^4 \\
&+
   10 q_1^3 (20909 \sqrt{17} q_2 - 158661\rmi q_3) + 8007612 q_2^3 q_3 -
   4442508\rmi \sqrt{17} q_2^2 q_3^2 -
   3\rmi q_1^2 (92985 \sqrt{17} q_2^2 - \\
&42194\rmi q_2 q_3 -
     345933 \sqrt{17} q_3^2) + 2 q_2 (713 p_3^2 - 1858270 q_3^3) +
   238\rmi \sqrt{17} q_3 (2 p_3^2 - 2601 q_3^3)\\
& +
   2 q_1 (-1327\rmi p_3^2 + 6 (4877 \sqrt{17} q_2^3 + 462111\rmi q_2^2 q_3 +
       18207 \sqrt{17} q_2 q_3^2 - 312987\rmi q_3^3))) \\
&+
 4 p_1 (-24\rmi p_2^3 (\sqrt{17} q_1 - 2 q_3) -
   2 p_2^2 p_3 (229 q_1 + 111\rmi q_2 - 41 \sqrt{17} q_3) +
   p_3 (2765 \sqrt{17} q_1^4\\
& - 723492 \sqrt{17} q_2^4 +
     q_1^3 (-146601\rmi \sqrt{17} q_2 - 533783 q_3) + 9665316\rmi q_2^3 q_3 +
     2909652 \sqrt{17} q_2^2 q_3^2\\
& - 6 q_1^2 (52184 \sqrt{17} q_2^2 -
       1105425\rmi q_2 q_3 - 148835 \sqrt{17} q_3^2) +
     4 q_1 (400 p_3^2 - 55035\rmi \sqrt{17} q_2^3 - 286110 q_2^2 q_3 \\
&-
       873936\rmi \sqrt{17} q_2 q_3^2 - 1704233 q_3^3) +
     544 \sqrt{17} q_3 (p_3^2 + 1156 q_3^3) +
     4\rmi q_2 (620 p_3^2 + 1714059 q_3^3)) \\
&+
   p_2 (-963900\rmi q_2^4 + 6\rmi p_3^2 (59 \sqrt{17} q_1 + 29 q_3) +
     42 q_2^3 (73457 q_1 - 9986 \sqrt{17} q_3) -
     18\rmi q_2^2 (58820 q_1^2\\
& + 36293 \sqrt{17} q_1 q_3 - 121958 q_3^2) -
     7 q_2 (50 \sqrt{17} p_3^2 - 188887 q_1^3 + 24558 \sqrt{17} q_1^2 q_3 -
       296514 q_1 q_3^2\\
& + 47396 \sqrt{17} q_3^3) -
     9\rmi (20825 q_1^4 + 7419 \sqrt{17} q_1^3 q_3 + 95370 q_1^2 q_3^2 -
       30634 \sqrt{17} q_1 q_3^3 + 69360 q_3^4))) \\
&+
 17 (29097600 \sqrt{17} q_1^7 + 90 q_1^6 (2023749\rmi\sqrt{17} q_2 +
     2380085 q_3) + 24 q_1^5 (13365407 \sqrt{17} q_2^2\\
& +
     31554108\rmi q_2 q_3 + 12438441 \sqrt{17} q_3^2) +
   q_1^4 (26857 p_3^2 + 12 (128834505\rmi \sqrt{17} q_2^3 +
       575958164 q_2^2 q_3\\
& - 82635759\rmi \sqrt{17} q_2 q_3^2 -
       217032064 q_3^3)) - 24 q_1^2 (-113786991\rmi \sqrt{17} q_2^5 -
     714952935 q_2^4 q_3 \\
&+ 127808550\rmi \sqrt{17} q_2^3 q_3^2 +
     153 q_3^2 (-33 p_3^2 + 202589 q_3^3) + 9\rmi \sqrt{17} q_2 q_3
      (775 p_3^2 + 999073 q_3^3) \\
&+ 29 q_2^2 (273 p_3^2 + 10760626 q_3^3)) +
   24 q_1^3 (12944100 \sqrt{17} q_2^4 + 174584934\rmi q_2^3 q_3 +
     25566130 \sqrt{17} q_2^2 q_3^2\\
& + \sqrt{17} q_3
      (-1585 p_3^2 + 32136222 q_3^3) +
     6\rmi q_2 (-857 p_3^2 + 39752817 q_3^3)) +
   16 q_2 (-6839721\rmi \sqrt{17} q_2^6 \\
&- 133592868 q_2^5 q_3 +
     85711365\rmi \sqrt{17} q_2^4 q_3^2 - 51 q_2 q_3^2
      (13009 p_3^2 + 4308412 q_3^3) - 9\rmi \sqrt{17} q_2^2 q_3
      (7220 p_3^2
\end{split}
\]
\[
 \begin{split}
&+ 23787879 q_3^3) +
     18 q_2^3 (1669 p_3^2 + 41705012 q_3^3) - 7\rmi \sqrt{17}
      (p_3^4 - 10404 p_3^2 q_3^3 + 2255067 q_3^6))\\
& -
   16 q_1 (8 \sqrt{17} p_3^4 + p_3^2 (-30663\rmi q_2^3 -
       52140 \sqrt{17} q_2^2 q_3 + 15147\rmi q_2 q_3^2 -
       4624 \sqrt{17} q_3^3)\\
& + 6 (10068741 \sqrt{17} q_2^6 -
       84040605\rmi q_2^5 q_3 + 11505600 \sqrt{17} q_2^4 q_3^2 -
       235629792\rmi q_2^3 q_3^3 \\
&- 43242781 \sqrt{17} q_2^2 q_3^4 +
       29227437\rmi q_2 q_3^5 - 2672672 \sqrt{17} q_3^6))) .
 \end{split}
\]
For the next potential
\[
V_7= 44  \sqrt{7}  q_1^3 + 240\rmi \sqrt{14} q_1^2  q_2   +
 330  \sqrt{7}  q_1    q_2^2 + 935\rmi \sqrt{14}  q_2^3 +
 3087  q_2^2  q_3 + 294  q_3^3
         \]
the second first integral has the following form
\[
 \begin{split}
&I_2=3030 p_1^6 + 13140 p_2^6 + 522 \rmi  \sqrt{14} p_2^5 p_3 +
 96 p_1^5 (120\rmi  \sqrt{2} p_2 + 127  \sqrt{7} p_3) +
 21 p_2^4 (2029 p_3^2 \\
&+ 2 (69576  \sqrt{7} q_1^3 + 1775759\rmi  \sqrt{14}
      q_2^3 + 5794740 q_2^2 q_3 + 10962\rmi  \sqrt{14} q_2 q_3^2 +
     880040 q_3^3\\
& + 12 q_1^2 (38327\rmi  \sqrt{14} q_2 + 23618 q_3) +
     24  \sqrt{7} q_1 (24385 q_2^2 - 5257 q_3^2))) +
 14 p_2^3 p_3 (20\rmi  \sqrt{14} p_3^2\\
& +
   21 (37144\rmi  \sqrt{2} q_1^3 - 74257 q_2^3 + 25839\rmi  \sqrt{14} q_2^2
      q_3 - 121506 q_2 q_3^2 - 58730\rmi  \sqrt{14} q_3^3 -
     12 q_1^2 (3897 q_2\\
& + 1903\rmi  \sqrt{14} q_3) +
     6 q_1 (10555\rmi  \sqrt{2} q_2^2 + 6008  \sqrt{7} q_2 q_3 +
       39270\rmi  \sqrt{2} q_3^2))) +
 2058 p_2 p_3 (906048\rmi  \sqrt{14} q_1^6 \\
&-
   864 q_1^5 (6577  \sqrt{7} q_2 + 6237\rmi  \sqrt{2} q_3) +
   576 q_1^4 (8213\rmi  \sqrt{14} q_2^2 + 30646 q_2 q_3 +
     1232\rmi  \sqrt{14} q_3^2) \\
&+ 12 q_1 q_2 (-3186975  \sqrt{7} q_2^4 +
     5829390\rmi  \sqrt{2} q_2^3 q_3 - 5851272  \sqrt{7} q_2^2 q_3^2 +
     45\rmi \sqrt{2} q_2 (79 p_3^2 + 281456 q_3^3) \\
&+
     4  \sqrt{7} q_3 (-757 p_3^2 + 403809 q_3^3)) +
   q_2 (-7918053\rmi  \sqrt{14} q_2^5 - 55828710 q_2^4 q_3 -
     828576\rmi  \sqrt{14} q_2^3 q_3^2 \\
&-
     2 q_2^2 (7519 p_3^2 - 35492856 q_3^3) +
     116424 q_3^2 (p_3^2 - 441 q_3^3) - 30\rmi  \sqrt{14} q_2 q_3
      (809 p_3^2 + 1222746 q_3^3)) \\
&+ 12 q_1^2 (-99678\rmi  \sqrt{14} q_2^4 +
     2562546 q_2^3 q_3 - 248094\rmi  \sqrt{14} q_2^2 q_3^2 +
     11\rmi  \sqrt{14} q_3 (41 p_3^2 - 25284 q_3^3) \\
&+
     q_2 (1313 p_3^2 + 3566220 q_3^3)) -
   8 q_1^3 (157\rmi  \sqrt{2} p_3^2 + 6 (695966  \sqrt{7} q_2^3 -
       296079\rmi  \sqrt{2} q_2^2 q_3 \\
&+ 604296  \sqrt{7} q_2 q_3^2 -
       359758\rmi  \sqrt{2} q_3^3))) +
 147 p_2^2 (264 p_3^4 + 2 p_3^2 (21824  \sqrt{7} q_1^3 +
     530687\rmi  \sqrt{14} q_2^3\\
& + 24\rmi q_1^2 (6465  \sqrt{14} q_2 +
       3703 \rmi q_3) + 1911084 q_2^2 q_3 + 178710\rmi  \sqrt{14} q_2 q_3^2 -
     116424 q_3^3 \\
&+ 12 q_1 (16503  \sqrt{7} q_2^2 - 39270\rmi  \sqrt{2} q_2
        q_3 + 10598  \sqrt{7} q_3^2)) +
   7 (-1915063849 q_2^6 + 6\rmi  \sqrt{2} q_2^5 (107803418 q_1 \\
&+
       149377659  \sqrt{7} q_3) - 6 q_2^4 (150209548 q_1^2 -
       49807524  \sqrt{7} q_1 q_3 - 236319531 q_3^2) \\
&+
     8\rmi  \sqrt{2} q_2^3 (31990100 q_1^3 + 33308910  \sqrt{7} q_1^2 q_3 -
       5276250 q_1 q_3^2 + 15723981  \sqrt{7} q_3^3) -
     12 q_2^2 (9189296 q_1^4\\
& - 3549024  \sqrt{7} q_1^3 q_3 -
       6806016 q_1^2 q_3^2 + 2010792  \sqrt{7} q_1 q_3^3 - 35685867 q_3^4) +
     24\rmi  \sqrt{2} q_2 (898328 q_1^5 \\
&+ 534000  \sqrt{7} q_1^4 q_3 -
       936936 q_1^3 q_3^2 + 1340220  \sqrt{7} q_1^2 q_3^3 + 2474010 q_1 q_3^4 -
       616665  \sqrt{7} q_3^5) \\
&- 8 (128528 q_1^6 - 229392  \sqrt{7} q_1^5 q_3 +
       720720 q_1^4 q_3^2 + 687456  \sqrt{7} q_1^3 q_3^3 -
       1801044 q_1^2 q_3^4\\
& - 633276  \sqrt{7} q_1 q_3^5 - 916839 q_3^6))) +
 686 (-517045760  \sqrt{7} q_1^9 - 218437632\rmi  \sqrt{14} q_1^8 q_2\\
& -
   16128 q_1^7 (1231361  \sqrt{7} q_2^2 - 730296\rmi  \sqrt{2} q_2 q_3 -
     116424  \sqrt{7} q_3^2) - 9408 q_1^6 (269 p_3^2\\
& +
     1626662\rmi  \sqrt{14} q_2^3 + 8138214 q_2^2 q_3 +
     299376\rmi \sqrt{14} q_2 q_3^2 + 690396 q_3^3) -
   4032 q_1^4 (52644172\rmi  \sqrt{14} q_2^5\\
& + 375243372 q_2^4 q_3 -
     18652536\rmi  \sqrt{14} q_2^3 q_3^2 + 12\rmi  \sqrt{14} q_2 q_3
      (227 p_3^2 - 358092 q_3^3)\\
& + 924 q_3^2 (5 p_3^2 - 588 q_3^3) +
     1813 q_2^2 (25 p_3^2 + 38064 q_3^3)) +
   2016 q_1^5 (p_3^2 (27181\rmi  \sqrt{2} q_2 - 1152  \sqrt{7} q_3)\\
& -
     56 (2260588  \sqrt{7} q_2^4 - 2543688\rmi  \sqrt{2} q_2^3 q_3 -
       193725  \sqrt{7} q_2^2 q_3^2 + 344190\rmi  \sqrt{2} q_2 q_3^3 +
       56448  \sqrt{7} q_3^4)) \\
&+ q_2^2 (-1803124370432\rmi  \sqrt{14} q_2^7 -
     17762238821376 q_2^6 q_3 + 4091606410752\rmi  \sqrt{14} q_2^5 q_3^2 \\
&+
     539101332 p_3^2 q_3^4 - 147 q_2^4 (20185577 p_3^2 -
       13484416512 q_3^3) + 18144\rmi  \sqrt{14} q_2^3 q_3
      (74785 p_3^2 \\
&+ 62376608 q_3^3) + 15876 q_2^2 q_3^2
      (99587 p_3^2 + 148533504 q_3^3) + 810\rmi  \sqrt{14} q_2
      (153 p_3^4 + 374752 p_3^2 q_3^3
 \end{split}
\]
\[
 \begin{split}
&- 125389824 q_3^6)) +
   36 q_1^2 q_2 (-31113582752\rmi  \sqrt{14} q_2^6 - 248262862176 q_2^5 q_3 +
     36273632976\rmi  \sqrt{14} q_2^4 q_3^2 \\
&+ 77616 q_2 q_3^2
      (-107 p_3^2 + 112602 q_3^3) - 392 q_2^3 (89717 p_3^2 +
       70945098 q_3^3) + 126\rmi  \sqrt{14} q_2^2 q_3
      (63435 p_3^2\\
& + 86562224 q_3^3) + 99\rmi  \sqrt{14}
      (p_3^4 + 16268 p_3^2 q_3^3 - 460992 q_3^6)) +
   72 q_1 q_2^2 (-26066508160  \sqrt{7} q_2^6 \\
&+ 83309091264\rmi  \sqrt{2} q_2^5
      q_3 + 18122715072  \sqrt{7} q_2^4 q_3^2 + 84  \sqrt{7} q_2^2 q_3
      (148769 p_3^2 - 32104800 q_3^3) \\
&+ 23520\rmi  \sqrt{2} q_2 q_3^2
      (-221 p_3^2 + 326046 q_3^3) + 49\rmi  \sqrt{2} q_2^3
      (385339 p_3^2 + 67039392 q_3^3) +
     9  \sqrt{7} (183 p_3^4 \\
&+ 72520 p_3^2 q_3^3 + 112482048 q_3^6)) +
   96 q_1^3 (108  \sqrt{7} p_3^4 + 7 p_3^2 (767347\rmi  \sqrt{2} q_2^3 +
       241704  \sqrt{7} q_2^2 q_3\\
& - 23436 \rmi  \sqrt{2} q_2 q_3^2 +
       18144  \sqrt{7} q_3^3) - 196 (65790574  \sqrt{7} q_2^6 -
       128347758\rmi  \sqrt{2} q_2^5 q_3 - 11248713  \sqrt{7} q_2^4 q_3^2\\
& +
       7491624\rmi  \sqrt{2} q_2^3 q_3^3 + 6990732  \sqrt{7} q_2^2 q_3^4 -
       3699108\rmi  \sqrt{2} q_2 q_3^5 - 190512  \sqrt{7} q_3^6))) +
 2 p_1^3 (-4843\rmi  \sqrt{2} p_2^3 \\
&- 9825  \sqrt{7} p_2^2 p_3 +
   21 p_2 (634\rmi  \sqrt{2} p_3^2 + 277120\rmi  \sqrt{14} q_1^3 +
     938901  \sqrt{7} q_2^3 - 1818243\rmi  \sqrt{2} q_2^2 q_3 \\
&+
     265062  \sqrt{7} q_2 q_3^2 - 687274\rmi  \sqrt{2} q_3^3 +
     384 q_1^2 (2612  \sqrt{7} q_2 - 2205\rmi  \sqrt{2} q_3) +
     6\rmi q_1 (104991  \sqrt{14} q_2^2 \\
&+ 594664\rmi q_2 q_3 +
       96586  \sqrt{14} q_3^2)) - 7 p_3 (1040  \sqrt{7} p_3^2 +
     21 (-184640 q_1^3 + 804879\rmi  \sqrt{2} q_2^3 \\
&+
       706404  \sqrt{7} q_2^2 q_3 - 469098\rmi  \sqrt{2} q_2 q_3^2 +
       29120  \sqrt{7} q_3^3 + 24 q_1^2 (905\rmi  \sqrt{2} q_2 +
         1524  \sqrt{7} q_3) \\
&- 96 q_1 (22202 q_2^2 - 3229\rmi  \sqrt{14} q_2
          q_3 - 1232 q_3^2)))) +
 6 p_1^4 (12319 p_2^2 - 5284\rmi  \sqrt{14} p_2 p_3 -
   7 (879 p_3^2 \\
&+ 2 (12420  \sqrt{7} q_1^3 - 97200\rmi  \sqrt{14} q_1^2 q_2 -
       1227530\rmi  \sqrt{14} q_2^3 - 6324675 q_2^2 q_3 +
       622650\rmi  \sqrt{14} q_2 q_3^2\\
& + 120442 q_3^3 +
       18 q_1 (13883  \sqrt{7} q_2^2 - 47040\rmi \sqrt{2} q_2 q_3 -
         7112  \sqrt{7} q_3^2))))
+
 6 p_1 (499\rmi  \sqrt{2} p_2^5 \\
&+ 1249  \sqrt{7} p_2^4 p_3 +
   7 p_2^3 (-361\rmi  \sqrt{2} p_3^2 + 65656\rmi  \sqrt{14} q_1^3 -
     421447  \sqrt{7} q_2^3 + 1100295\rmi \sqrt{2} q_2^2 q_3\\
& +
     283542  \sqrt{7} q_2 q_3^2 + 637882\rmi \sqrt{2} q_3^3 -
     48 q_1^2 (1499  \sqrt{7} q_2 - 6615\rmi  \sqrt{2} q_3) +
     6\rmi q_1 (36241  \sqrt{14} q_2^2 \\
&+ 94472 \rmi q_2 q_3 +
       12194  \sqrt{14} q_3^2)) + 7 p_2^2 p_3 (235  \sqrt{7} p_3^2 +
     21 (50840 q_1^3 + 154327\rmi  \sqrt{2} q_2^3 \\
&+ 85406  \sqrt{7} q_2^2 q_3 -
       199990\rmi  \sqrt{2} q_2 q_3^2 + 23828  \sqrt{7} q_3^3 +
       q_1^2 (-5048\rmi  \sqrt{2} q_2 - 53024  \sqrt{7} q_3) \\
&+
       4 q_1 (35881 q_2^2 + 2064\rmi  \sqrt{14} q_2 q_3 + 25256 q_3^2))) +
   343 p_3 (3294720  \sqrt{7} q_1^6 - 1728\rmi q_1^5 (1117  \sqrt{14} q_2\\
& -
       5852\rmi q_3) + 2304 q_1^4 (17302  \sqrt{7} q_2^2 -
       10843\rmi  \sqrt{2} q_2 q_3 - 1232  \sqrt{7} q_3^2) -
     8 q_1^3 (1304 p_3^2\\
& + 3 (31037\rmi  \sqrt{14} q_2^3 + 398244 q_2^2 q_3 -
         774774\rmi  \sqrt{14} q_2 q_3^2 - 365344 q_3^3)) +
     12 q_1 q_2 (6167733\rmi  \sqrt{14} q_2^4 \\
&+ 23478420 q_2^3 q_3 +
       565194\rmi  \sqrt{14} q_2^2 q_3^2 + 8\rmi  \sqrt{14} q_3
        (139 p_3^2 - 122010 q_3^3) + 16 q_2 (247 p_3^2 + 659442 q_3^3)) \\
&+
     q_2 (-54123960  \sqrt{7} q_2^5 + 426439314\rmi  \sqrt{2} q_2^4 q_3 +
       339970176  \sqrt{7} q_2^3 q_3^2 - 57330\rmi \sqrt{2} q_3^2
        (p_3^2 + 588 q_3^3) \\
&+ 12  \sqrt{7} q_2 q_3 (6991 p_3^2 +
         443352 q_3^3) - \rmi  \sqrt{2} q_2^2 (47317 p_3^2 + 325894296 q_3^3))
+
     24 q_1^2 (13 p_3^2 (-39\rmi  \sqrt{2} q_2\\
& + 20  \sqrt{7} q_3) +
       6 (832542  \sqrt{7} q_2^4 - 521255\rmi  \sqrt{2} q_2^3 q_3 +
         315252  \sqrt{7} q_2^2 q_3^2 - 36750\rmi  \sqrt{2} q_2 q_3^3 \\
&+
         25480  \sqrt{7} q_3^4))) + 49 p_2 (-130\rmi  \sqrt{2} p_3^4 +
     3 p_3^2 (6064\rmi  \sqrt{14} q_1^3 + 104229  \sqrt{7} q_2^3 +
       119329\rmi  \sqrt{2} q_2^2 q_3\\
& - 84266  \sqrt{7} q_2 q_3^2 -
       6370\rmi  \sqrt{2} q_3^3 + 64 q_1^2 (1319  \sqrt{7} q_2 +
         1064\rmi  \sqrt{2} q_3) - 2\rmi q_1 (10559  \sqrt{14} q_2^2\\
& -
         71400\rmi q_2 q_3 + 14098  \sqrt{14} q_3^2)) +
     42 (219648\rmi  \sqrt{2} q_1^6 - 6654538\rmi  \sqrt{2} q_2^6 -
       12529491  \sqrt{7} q_2^5 q_3 \\
&+ 23141139\rmi  \sqrt{2} q_2^4 q_3^2 -
       1432060  \sqrt{7} q_2^3 q_3^3 + 11319588\rmi  \sqrt{2} q_2^2 q_3^4 +
       1078980  \sqrt{7} q_2 q_3^5 \\
&+ 802620\rmi  \sqrt{2} q_3^6 -
       6912 q_1^5 (173 q_2 - 22\rmi  \sqrt{14} q_3) +
       24\rmi q_1^4 (1481  \sqrt{2} q_2^2 + 29912\rmi  \sqrt{7} q_2 q_3 \\
&-
         18634  \sqrt{2} q_3^2) - 4 q_1^3 (1134901 q_2^3 -
         733701\rmi  \sqrt{14} q_2^2 q_3 + 410886 q_2 q_3^2 -
         46886\rmi  \sqrt{14} q_3^3)\\
& + 12\rmi q_1^2 (140671  \sqrt{2} q_2^4 +
         914392\rmi  \sqrt{7} q_2^3 q_3 + 710178  \sqrt{2} q_2^2 q_3^2 +
         367136\rmi  \sqrt{7} q_2 q_3^3 + 177968  \sqrt{2} q_3^4)\\
& -
       6 q_1 (2730321 q_2^5 - 367184\rmi  \sqrt{14} q_2^4 q_3 +
         2040738 q_2^3 q_3^2 - 546070\rmi  \sqrt{14} q_2^2 q_3^3 -
         660520 q_2 q_3^4 \\
&+ 16268\rmi  \sqrt{14} q_3^5)))) +
 6 p_1^2 (8163 p_2^4 - 786\rmi  \sqrt{14} p_2^3 p_3 -
   532\rmi  \sqrt{14} p_2 p_3^3 + 294 p_2 p_3 (-26692\rmi \sqrt{2} q_1^3 \\
&+
     49739 q_2^3 - 22723 \rmi  \sqrt{14} q_2^2 q_3 - 9856 q_2 q_3^2 +
     7560\rmi  \sqrt{14} q_3^3 - 2 q_1^2 (61333 q_2 - 6017\rmi  \sqrt{14}
        q_3)
\end{split}
\]
\[
 \begin{split}
&+ q_1 (-64586\rmi  \sqrt{2} q_2^2 + 85296  \sqrt{7} q_2 q_3 -
       2464 \rmi  \sqrt{2} q_3^2)) +
   14 p_2^2 (1132 p_3^2 + 43048  \sqrt{7} q_1^3 + 2048623\rmi  \sqrt{14}
      q_2^3\\
& + 7341474 q_2^2 q_3 - 86100\rmi  \sqrt{14} q_2 q_3^2 +
     381024 q_3^3 + 6 q_1^2 (69267\rmi  \sqrt{14} q_2 + 148666 q_3) \\
&+
     12 q_1 (80973  \sqrt{7} q_2^2 - 26460\rmi  \sqrt{2} q_2 q_3 -
       28238  \sqrt{7} q_3^2)) +
   196 (51 p_3^4 + p_3^2 (-6100  \sqrt{7} q_1^3\\
& + 166937\rmi  \sqrt{14} q_2^3 +
       567546 q_2^2 q_3 - 20055\rmi  \sqrt{14} q_2 q_3^2 + 59976 q_3^3 +
       3 q_1^2 (5361\rmi  \sqrt{14} q_2 + 9856 q_3) \\
&-
       12 q_1 (8877  \sqrt{7} q_2^2 + 7896\rmi  \sqrt{2} q_2 q_3 +
         910  \sqrt{7} q_3^2)) + 7 (-258525485 q_2^6 +
       3\rmi  \sqrt{2} q_2^5 (40352798 q_1\\
& + 43391235  \sqrt{7} q_3) -
       24 q_2^4 (4778174 q_1^2 - 3705267  \sqrt{7} q_1 q_3 - 10673418 q_3^2) +
       4\rmi  \sqrt{2} q_2^3 (11847545 q_1^3 \\
&+ 7974936  \sqrt{7} q_1^2 q_3 -
         15765939 q_1 q_3^2 + 273609  \sqrt{7} q_3^3) -
       12 q_2^2 (1241236 q_1^4 - 1253718  \sqrt{7} q_1^3 q_3 \\
&-
         612486 q_1^2 q_3^2 - 78036  \sqrt{7} q_1 q_3^3 - 1675065 q_3^4) +
       12\rmi  \sqrt{2} q_2 (284864 q_1^5 + 126480  \sqrt{7} q_1^4 q_3\\
& -
         481950 q_1^3 q_3^2 + 159012  \sqrt{7} q_1^2 q_3^3 - 572712 q_1 q_3^4 +
         131271  \sqrt{7} q_3^5) - 16 (54365 q_1^6 - 25920  \sqrt{7} q_1^5
          q_3 \\
&- 11088 q_1^4 q_3^2 + 2457  \sqrt{7} q_1^3 q_3^3 +
         116424 q_1^2 q_3^4 + 57330  \sqrt{7} q_1 q_3^5 - 157437 q_3^6)))).
 \end{split}
\]

Potential
\[
\begin{split}
 V_{9}&=27\rmi \sqrt{3990}q_1^3 + 3726\sqrt{15}q_1^2q_2 -
 456\rmi \sqrt{3990}q_1q_2^2 - 4092\sqrt{15}q_2^3 - 1125q_1^2q_3-
 3000q_2^2q_3 - 50q_3^3,
\end{split}
\]
admits the following first integrals
\[
 \begin{split}
&I_1=621756 p_1^4 - 171\rmi \sqrt{266} p_1^3 (639 p_2 + 82 \sqrt{15} p_3) -
 171 p_1^2 (11366 p_2^2\\
& + 3381 \sqrt{15} p_2 p_3 +
  60 (21 p_3^2 + 5 (1234 \rmi \sqrt{3990} q_1^3 +
       36 q_1^2 (2159 \sqrt{15} q_2 + 1515 q_3) \\
&- \rmi \sqrt{266} q_1
        (4861 \sqrt{15} q_2^2 + 12780 q_2 q_3 + 41 \sqrt{15} q_3^2) -
       6 (3616 \sqrt{15} q_2^3 + 22665 q_2^2 q_3 \\
&+ 966 \sqrt{15} q_2 q_3^2 +
         35 q_3^3)))) + 3 (81377 p_2^4 + 40584 \sqrt{15} p_2^3 p_3\\
& +
   75 p_2 p_3 (558657 \rmi \sqrt{266} q_1^3 + 110124 q_1^2
      (424 q_2 + 3 \sqrt{15} q_3) - 2 \rmi\sqrt{266} q_1
      (1842408 q_2^2\\
& + 162352 \sqrt{15} q_2 q_3 - 65205 q_3^2) -
     304 q_2 (72658 q_2^2 + 15753 \sqrt{15} q_2 q_3 + 910 q_3^2)) \\
&+
   750 (-380805543 q_1^6 + 9234 \rmi \sqrt{266} q_1^5
      (41202 q_2 - 641 \sqrt{15} q_3) + 92340 q_1^4
      (396143 q_2^2\\
& - 5802 \sqrt{15} q_2 q_3 + 235 q_3^2) -
     4560\rmi \sqrt{266} q_1^3 (1322406 q_2^3 + 12177 \sqrt{15} q_2^2 q_3 -
       8505 q_2 q_3^2\\
& + 65 \sqrt{15} q_3^3) -
     3420 q_1^2 (39391416 q_2^4 + 1584720 \sqrt{15} q_2^3 q_3 +
       643960 q_2^2 q_3^2\\
& - 28980 \sqrt{15} q_2 q_3^3 + 3465 q_3^4) +
     304 q_2^2 (80689622 q_2^4 + 9830304 \sqrt{15} q_2^3 q_3\\
& +
       9897000 q_2^2 q_3^2 + 650400 \sqrt{15} q_2 q_3^3 + 206325 q_3^4) +
     24 \rmi \sqrt{266} q_1 (234137952 q_2^5\\
& + 18121160 \sqrt{15} q_2^4 q_3 +
       13755600 q_2^3 q_3^2 + 406400 \sqrt{15} q_2^2 q_3^3 -
       217350 q_2 q_3^4 + 875 \sqrt{15} q_3^5) \\
&+
     p_3^2 (12825 \rmi \sqrt{3990} q_1^3 + 14364 q_1^2 (69 \sqrt{15} q_2 +
         14 q_3) - 5776 q_2^2 (111 \sqrt{15} q_2 + 196 q_3)\\
& -
       2\rmi \sqrt{266} q_1 (44031 \sqrt{15} q_2^2 + 34776 q_2 q_3 -
         350 \sqrt{15} q_3^2))) +
   10 p_2^2 (14497 p_3^2\\
& + 15 (42237\rmi \sqrt{3990} q_1^3 +
       114 q_1^2 (41736 \sqrt{15} q_2 - 6085 q_3) - 2\rmi \sqrt{266} q_1
        (233356 \sqrt{15} q_2^2\\
& + 157860 q_2 q_3 - 4843 \sqrt{15} q_3^2) -
       76 (45750 \sqrt{15} q_2^3 + 85660 q_2^2 q_3 + 534 \sqrt{15} q_2 q_3^2\\
& +
         665 q_3^3)))) + p_1 (63144\rmi \sqrt{266} p_2^3 +
   28596 \rmi \sqrt{3990} p_2^2 p_3 + 135 p_2 (322\rmi \sqrt{266} p_3^2\\
& +
     5 (-1374327 \sqrt{15} q_1^3 + 95\rmi\sqrt{266} q_1^2
        (4706 \sqrt{15} q_2 + 639 q_3) + 38 q_1 (247668 \sqrt{15} q_2^2\\
& +
         365320 q_2 q_3 - 8211 \sqrt{15} q_3^2) - 2 \rmi \sqrt{266}
        (105848 \sqrt{15} q_2^3 + 368340 q_2^2 q_3 + 6406 \sqrt{15} q_2
          q_3^2\\
& - 2415 q_3^3))) + 50 p_3 (-28 \rmi \sqrt{3990} p_3^2 +
     9 (-11433231 q_1^3 + 57 \rmi \sqrt{266} q_1^2 (49734 q_2\\
& +
         1763 \sqrt{15} q_3) + 342 q_1 (157509 q_2^2 + 18676 \sqrt{15} q_2
          q_3 + 210 q_3^2) - 2 \rmi \sqrt{266} (576558 q_2^3\\
& +
         140443 \sqrt{15} q_2^2 q_3 + 57960 q_2 q_3^2 + 350 \sqrt{15} q_3^3)))),
 \end{split}
\]

\[
 \begin{split}
 & I_2=69896714949 p_1^6 - 4410941719 p_2^6 - 12178545120 \sqrt{15} p_2^5 p_3 \\
&-
 5185404 \rmi \sqrt{266} p_1^5 (3321 p_2 + 1130 \sqrt{15} p_3) -
 48735 p_1^4 (9477831 p_2^2 + 6664140 \sqrt{15} p_2 p_3\\
& + 17011399 p_3^2 +
   161435514 \rmi \sqrt{3990} q_1^3 + 342 q_1^2 (31570326 \sqrt{15} q_2 +
     28306975 q_3)\\
& - 912 \rmi\sqrt{266} q_1 (849176 \sqrt{15} q_2^2 +
     2324700 q_2 q_3 + 19775 \sqrt{15} q_3^2) -
   4 (1115132226 \sqrt{15} q_2^3\\
& + 6001374300 q_2^2 q_3 +
     230353200 \sqrt{15} q_2 q_3^2 + 208136575 q_3^3)) -
 15 p_2^4 (11812848011 p_3^2\\
& + 6 (3637289871 \rmi \sqrt{3990} q_1^3 +
     57 q_1^2 (5057639934 \sqrt{15} q_2 + 4023927275 q_3)\\
& -
     8 \rmi \sqrt{266} q_1 (2599271621 \sqrt{15} q_2^2 + 8552117700 q_2 q_3 +
       178766525 \sqrt{15} q_3^2)\\
& - 2 (54592416078 \sqrt{15} q_2^3 +
       441094171900 q_2^2 q_3 + 10148787600 \sqrt{15} q_2 q_3^2\\
& +
       73999814225 q_3^3))) - 420 p_2^3 p_3 (200059064 \sqrt{15} p_3^2 +
   75 (391073751 \rmi \sqrt{266} q_1^3\\
& +
     684 q_1^2 (51068432 q_2 + 871817 \sqrt{15} q_3) -
     2 \rmi \sqrt{266} q_1 (1391808024 q_2^2\\
& + 204067408 \sqrt{15} q_2 q_3 -
       95297139 q_3^2) - 16 (1053113326 q_2^3 + 358832133 \sqrt{15} q_2^2
        q_3 \\
&+ 26186342 q_2 q_3^2 + 7756986 \sqrt{15} q_3^3))) -
 225 p_2^2 (1434130249 p_3^4\\
& + 4 p_3^2 (12485855187 \rmi \sqrt{3990} q_1^3 -
     532583214012 \sqrt{15} q_2^3 - 2347905250680 q_2^2 q_3\\
& -
     28980666960 \sqrt{15} q_2 q_3^2 + 98949864350 q_3^3 +
     171 q_1^2 (6148072746 \sqrt{15} q_2\\
& + 1899245905 q_3) -
     12\rmi \sqrt{266} q_1 (7036440863 \sqrt{15} q_2^2 +
       14574507780 q_2 q_3\\
& - 560965895 \sqrt{15} q_3^2)) -
   60 (841855524804 q_1^6 - 58482 \rmi \sqrt{266} q_1^5
      (3649122 q_2\\
& + 1662263 \sqrt{15} q_3) +
     48735 q_1^4 (171927564 q_2^2 - 272077812 \sqrt{15} q_2 q_3 -
       109530025 q_3^2)\\
& - 1140 \rmi \sqrt{266} q_1^3
      (2587132818 q_2^3 - 1736199048 \sqrt{15} q_2^2 q_3 -
       3090859380 q_2 q_3^2\\
& - 9410749 \sqrt{15} q_3^3) -
     1140 q_1^2 (57246559116 q_2^4 - 28572680070 \sqrt{15} q_2^3 q_3\\
& -
       109359463740 q_2^2 q_3^2 - 7143208482 \sqrt{15} q_2 q_3^3 +
       1499017285 q_3^4)\\
& + 32 \rmi \sqrt{266} q_1 (59926048758 q_2^5 -
       24985763220 \sqrt{15} q_2^4 q_3 - 210263522175 q_2^3 q_3^2\\
& -
       20299324615 \sqrt{15} q_2^2 q_3^3 - 24297828975 q_2 q_3^4 +
       606252325 \sqrt{15} q_3^5)\\
& + 4 (766529106284 q_2^6 -
       264155752944 \sqrt{15} q_2^5 q_3 - 9323232022800 q_2^4 q_3^2\\
& -
       755223062340 \sqrt{15} q_2^3 q_3^3 - 2978323008400 q_2^2 q_3^4 -
       10859780400 \sqrt{15} q_2 q_3^5\\
& - 48468636375 q_3^6))) -
 6300 p_2 p_3 (6889792 \sqrt{15} p_3^4 +
   15 p_3^2 (735779997 \rmi \sqrt{266} q_1^3 \\
&+
     1596 q_1^2 (33673656 q_2 + 1541923 \sqrt{15} q_3) -
     2 \rmi \sqrt{266} q_1 (2058576648 q_2^2\\
& + 331731248 \sqrt{15} q_2 q_3 -
       56134785 q_3^2) - 16 (1603777022 q_2^3 + 482897263 \sqrt{15} q_2^2
        q_3 \\
&+ 304743530 q_2 q_3^2 - 21530600 \sqrt{15} q_3^3)) -
   30 (54812312982 \sqrt{15} q_1^6\\
& - 87723 \rmi \sqrt{266} q_1^5
      (305074 \sqrt{15} q_2+ 740505 q_3) - 1169640 q_1^4
      (1118502 \sqrt{15} q_2^2 \\
&+ 8577605 q_2 q_3 + 98030 \sqrt{15} q_3^2)+
     3420 \rmi \sqrt{266} q_1^3 (12610437 \sqrt{15} q_2^3 +
       705901770 q_2^2 q_3\\
& + 19015055 \sqrt{15} q_2 q_3^2 + 21915 q_3^3) -
     6840 q_1^2 (273535608 \sqrt{15} q_2^4 - 8796044710 q_2^3 q_3\\
& -
       855723590 \sqrt{15} q_2^2 q_3^2 + 131287940 q_2 q_3^3 +
       1126965 \sqrt{15} q_3^4)\\
& + 4 \rmi \sqrt{266} q_1
      (37432776912 \sqrt{15} q_2^5- 569283400800 q_2^4 q_3 -
       128920054050 \sqrt{15} q_2^3 q_3^2\\
& - 26461961100 q_2^2 q_3^3 -
       407403200 \sqrt{15} q_2 q_3^4 + 1772465625 q_3^5) +
     32 (24752235324 \sqrt{15} q_2^6
\end{split}
\]
\[
\begin{split}
& - 229256806470 q_2^5 q_3 -
       110898924000 \sqrt{15} q_2^4 q_3^2 - 65789240150 q_2^3 q_3^3\\
&-
       6222223575 \sqrt{15} q_2^2 q_3^4 + 7013938750 q_2 q_3^5 -
       242219250 \sqrt{15} q_3^6))) -
 1125 (35195551 p_3^6\\
& + 6 p_3^4 (1651614273 \rmi \sqrt{3990} q_1^3 +
     171 q_1^2 (605201734 \sqrt{15} q_2 + 757529455 q_3)\\
& -
     16 \rmi \sqrt{266} q_1 (466133664 \sqrt{15} q_2^2 + 1385148660 q_2 q_3 +
       41393975 \sqrt{15} q_3^2)\\
& - 2 (22781303074 \sqrt{15} q_2^3 +
       109281749540 q_2^2 q_3+ 9645708800 \sqrt{15} q_2 q_3^2\\
& +
       879888775 q_3^3))+ 60 p_3^2 (-532302462216 q_1^6 +
     175446 \rmi \sqrt{266} q_1^5 (2547126 q_2\\
& + 76525 \sqrt{15} q_3) +
     438615 q_1^4 (97795636 q_2^2 + 6459140 \sqrt{15} q_2 q_3 -
       2853339 q_3^2)\\
& - 1140 \rmi \sqrt{266} q_1^3 (5079550338 q_2^3+
       1315963350 \sqrt{15} q_2^2 q_3 - 1009710576 q_2 q_3^2 \\
&-
       7741537 \sqrt{15} q_3^3) - 3420 q_1^2 (28073465772 q_2^4 +
       14499700170 \sqrt{15} q_2^3 q_3 \\
&+ 7760364852 q_2^2 q_3^2 -
       2099095722 \sqrt{15} q_2 q_3^3 + 317762785 q_3^4) \\
&+
     16\rmi \sqrt{266} q_1 (180012985686 q_2^5 + 141678458700 \sqrt{15}
        q_2^4 q_3 + 328269338460 q_2^3 q_3^2\\
& - 21395932365 \sqrt{15} q_2^2
        q_3^3 - 34000372350 q_2 q_3^4 + 289757825 \sqrt{15} q_3^5) \\
&+
     4 (2321951680484 q_2^6+ 2297218340400 \sqrt{15} q_2^5 q_3 +
       10702431842160 q_2^4 q_3^2\\
& + 117217238580 \sqrt{15} q_2^3 q_3^3-
       1202239869200 q_2^2 q_3^4 - 86811379200 \sqrt{15} q_2 q_3^5\\
& +
       4399443875 q_3^6))+ 200 (1364637585276 q_1^8 q_3 +
     45001899 \rmi \sqrt{266} q_1^7 (166420 \sqrt{15} q_2^2\\
& - 796932 q_2 q_3+
       56849 \sqrt{15} q_3^2) + 45001899 q_1^6 (46075240 \sqrt{15} q_2^3 -
       215348732 q_2^2 q_3 \\
&+ 15423138 \sqrt{15} q_2 q_3^2 + 553855 q_3^3) -
     116964 \rmi \sqrt{266} q_1^5 (6854437620 \sqrt{15} q_2^4\\
& -
       29454178554 q_2^3 q_3+ 1471374774 \sqrt{15} q_2^2 q_3^2 +
       1180569330 q_2 q_3^3 - 12999235 \sqrt{15} q_3^4)\\
& -
     292410 q_1^4 (134591782680 \sqrt{15} q_2^5 - 470046837672 q_2^4 q_3 -
       5439697794 \sqrt{15} q_2^3 q_3^2\\
& + 60909452080 q_2^2 q_3^3 +
       105428700 \sqrt{15} q_2 q_3^4 - 214498155 q_3^5) \\
&+
     684 \rmi \sqrt{266} q_1^3 (5729304476700 \sqrt{15} q_2^6 -
       14290700495904 q_2^5 q_3 - 1565486188560 \sqrt{15} q_2^4 q_3^2\\
& +
       1995653240100 q_2^3 q_3^3 + 269861238600 \sqrt{15} q_2^2 q_3^4 -
       89796016800 q_2 q_3^5 + 597920425 \sqrt{15} q_3^6)\\
&+
     2052 q_1^2 (28177936037640 \sqrt{15} q_2^7 - 39924026866124 q_2^6 q_3 -
       11962790190192 \sqrt{15} q_2^5 q_3^2\\
& - 3423685332600 q_2^4 q_3^3 +
       2329388397900 \sqrt{15} q_2^3 q_3^4 + 1510872401600 q_2^2 q_3^5\\
& -
       55513413450 \sqrt{15} q_2 q_3^6+ 8949071875 q_3^7) -
     288 \rmi \sqrt{266} q_1 (5860980890620 \sqrt{15} q_2^8 \\
&-
       1807317655776 q_2^7 q_3 - 2808869611056 \sqrt{15} q_2^6 q_3^2 -
       4526348790060 q_2^5 q_3^3\\
& + 328876708700 \sqrt{15} q_2^4 q_3^4 +
       1155725294850 q_2^3 q_3^5 + 10950315825 \sqrt{15} q_2^2 q_3^6 \\
&+
       1417263750 q_2 q_3^7- 206969875 \sqrt{15} q_3^8) -
     8 (673938162327960 \sqrt{15} q_2^9\\
& + 547277822406288 q_2^8 q_3-
       295571373362208 \sqrt{15} q_2^7 q_3^2 - 943247686228260 q_2^6 q_3^3\\
& -
       37145453338080 \sqrt{15} q_2^5 q_3^4 + 208729566212400 q_2^4 q_3^5 +
       12138937320150 \sqrt{15} q_2^3 q_3^6 \\
&+ 6956841982500 q_2^2 q_3^7 +
       122078502000 \sqrt{15} q_2 q_3^8 + 21997219375 q_3^9))) \\
&-
 7980 p_1^3 (-3038931 \rmi \sqrt{266} p_2^3 - 3361710 \rmi \sqrt{3990} p_2^2
    p_3 + 9 p_2 (-1940193 \rmi \sqrt{266} p_3^2\\
& + 3227221953 \sqrt{15} q_1^3 -
     44403 \rmi \sqrt{266} q_1^2 (19658 \sqrt{15} q_2 + 14175 q_3) \\
&-
     114 q_1 (148858116 \sqrt{15} q_2^2 + 391208200 q_2 q_3 +
       465525 \sqrt{15} q_3^2) \\
&+ 2 \rmi \sqrt{266}
      (181930844 \sqrt{15} q_2^3 + 1018181700 q_2^2 q_3 +
       24910050 \sqrt{15} q_2 q_3^2 + 54061425 q_3^3))\\
& +
   2 p_3 (-990533 \rmi \sqrt{3990} p_3^2 +
     225 (589059945 q_1^3 - 1083 \rmi \sqrt{266} q_1^2
        (136890 q_2
\end{split}
\]
\[
\begin{split}
&+ 8249 \sqrt{15} q_3) -
       342 q_1 (8389805 q_2^2 + 1513572 \sqrt{15} q_2 q_3 + 103404 q_3^2) \\
&+
       2 \rmi \sqrt{266} (32018130 q_2^3+ 10832491 \sqrt{15} q_2^2 q_3 +
         7252524 q_2 q_3^2 + 9322 \sqrt{15} q_3^3)))) \\
&+
 855 p_1^2 (216247837 p_2^4 + 340148340 \sqrt{15} p_2^3 p_3 +
   2 p_2^2 (1362509849 p_3^2\\
& + 6 (927489579 \rmi \sqrt{3990} q_1^3 +
       57 q_1^2 (1233825666 \sqrt{15} q_2 + 815729225 q_3) \\
&-
       4 \rmi \sqrt{266} q_1 (1283526583 \sqrt{15} q_2^2 +
         3508986600 q_2 q_3 + 10864700 \sqrt{15} q_3^2)\\
& -
       2 (14147249322 \sqrt{15} q_2^3 + 88856665600 q_2^2 q_3 +
         1646534400 \sqrt{15} q_2 q_3^2\\
& + 7901396775 q_3^3))) +
   84 p_2 p_3 (7399217 \sqrt{15} p_3^2 + 25 (98234883 \rmi \sqrt{266} q_1^3\\
& +
       342 q_1^2 (21342982 q_2+ 860271 \sqrt{15} q_3)-
       2 \rmi \sqrt{266} q_1 (276720192 q_2^2\\
& + 44322832 \sqrt{15} q_2 q_3 -
         9351261 q_3^2) - 4 (834242922 q_2^3+ 272614248 \sqrt{15} q_2^2
          q_3\\
& + 96947032 q_2 q_3^2 - 1221759 \sqrt{15} q_3^3))) +
   15 (53114167 p_3^4 + 4 p_3^2 (1073099271 \rmi \sqrt{3990} q_1^3\\
&+
       171 q_1^2 (412247118 \sqrt{15} q_2 + 410808415 q_3) -
       4 \rmi \sqrt{266} q_1 (1281131112 \sqrt{15} q_2^2 \\
&+
         3566728620 q_2 q_3 + 33689845 \sqrt{15} q_3^2) -
       2 (15405850098 \sqrt{15} q_2^3 + 76504861120 q_2^2 q_3\\
& +
         4369269240 \sqrt{15} q_2 q_3^2 - 810863525 q_3^3)) -
     20 (262207731366 q_1^6\\
& - 19494 \rmi \sqrt{266} q_1^5
        (20214 q_2 + 2011301 \sqrt{15} q_3) + 48735 q_1^4
        (153003356 q_2^2 \\
&- 79345668 \sqrt{15} q_2 q_3 - 69945825 q_3^2) -
       3420 \rmi \sqrt{266} q_1^3 (499000974 q_2^3 - 161887814 \sqrt{15}
          q_2^2 q_3\\
& - 328039740 q_2 q_3^2 - 10804817 \sqrt{15} q_3^3) -
       3420 q_1^2 (12226096388 q_2^4 - 2645440290 \sqrt{15} q_2^3 q_3 \\
&-
         10784656720 q_2^2 q_3^2 - 786696606 \sqrt{15} q_2 q_3^3-
         218993995 q_3^4) + 48 \rmi \sqrt{266} q_1 (36201520638 q_2^5\\
& -
         4196311620 \sqrt{15} q_2^4 q_3 - 40491476550 q_2^3 q_3^2 -
         4439552965 \sqrt{15} q_2^2 q_3^3 - 4418895600 q_2 q_3^4\\
& -
         815675 \sqrt{15} q_3^5) + 4 (1788936608636 q_2^6 -
         9012180096 \sqrt{15} q_2^5 q_3 - 2441620201200 q_2^4 q_3^2 \\
&-
         339300426660 \sqrt{15} q_2^3 q_3^3 - 656376048600 q_2^2 q_3^4 -
         16245381600 \sqrt{15} q_2 q_3^5\\
& - 3250847375 q_3^6)))) -
 84 p_1 (32579496 \rmi \sqrt{266} p_2^5 + 69384500 \rmi \sqrt{3990} p_2^4 p_3\\
& +
   5 p_2^3 (153859518 \rmi \sqrt{266} p_3^2 - 110609531073 \sqrt{15} q_1^3 +
     171 \rmi \sqrt{266} q_1^2 (187871254 \sqrt{15} q_2\\
& + 137031525 q_3) +
     342 q_1 (1782095052 \sqrt{15} q_2^2 + 5504929400 q_2 q_3 +
       68484675 \sqrt{15} q_3^2)\\
& - 2 \rmi \sqrt{266}
      (5940190504 \sqrt{15} q_2^3 + 45204050700 q_2^2 q_3 +
       887539050 \sqrt{15} q_2 q_3^2 \\
&+ 6030275175 q_3^3)) +
   10 p_2^2 p_3 (26762086 \rmi \sqrt{3990} p_3^2 +
     75 (-20781023121 q_1^3\\
& + 513 \rmi \sqrt{266} q_1^2
        (12502546 q_2 + 325249 \sqrt{15} q_3) +
       342 q_1 (390450499 q_2^2 + 58084012 \sqrt{15} q_2 q_3\\
& -
         22450386 q_3^2) - 2 \rmi \sqrt{266} (1525002138 q_2^3 +
         498142221 \sqrt{15} q_2^2 q_3 + 80902584 q_2 q_3^2\\
& +
         2878382 \sqrt{15} q_3^3))) +
   75 p_2 (9056586 \rmi \sqrt{266} p_3^4 +
     p_3^2 (-127897609527 \sqrt{15} q_1^3\\
& + 171 \rmi \sqrt{266} q_1^2
        (208667026 \sqrt{15} q_2 + 128025855 q_3) +
       342 q_1 (2125927828 \sqrt{15} q_2^2\\
& + 5043793480 q_2 q_3 -
         84854595 \sqrt{15} q_3^2) - 2 \rmi \sqrt{266}
        (8427582736 \sqrt{15} q_2^3\\
& + 39102611940 q_2^2 q_3 +
         1197339030 \sqrt{15} q_2 q_3^2 - 1363176675 q_3^3))\\
& +
     30 (-17935756857\rmi \sqrt{266} q_1^6 - 555579 q_1^5
        (926178 q_2 + 1164607 \sqrt{15} q_3)\\
& - 48735 \rmi \sqrt{266} q_1^4
        (9819012 q_2^2 - 6205816 \sqrt{15} q_2 q_3 - 3190425 q_3^2) \\
&-
       194940 q_1^3 (173644949 q_2^3 - 63035364 \sqrt{15} q_2^2 q_3 -
         107487665 q_2 q_3^2 - 1185437 \sqrt{15} q_3^3)\\
& +
       1140 \rmi\sqrt{266} q_1^2 (2832758478 q_2^4 - 671800025 \sqrt{15}
          q_2^3 q_3 - 2502926370 q_2^2 q_3^2\\
& - 153606221 \sqrt{15} q_2 q_3^3 +
         16305030 q_3^4) + 228 q_1 (156415390456 q_2^5
\end{split}
\]
\[
\begin{split}
& -
         19923354240 \sqrt{15} q_2^4 q_3 - 180078382850 q_2^3 q_3^2 -
         18181769880 \sqrt{15} q_2^2 q_3^3\\
& - 16318512200 q_2 q_3^4 +
         417338775 \sqrt{15} q_3^5) - 4 \rmi \sqrt{266}
        (135040386672 q_2^6\\
& - 752761672 \sqrt{15} q_2^5 q_3 -
         198974124900 q_2^4 q_3^2 - 23736446770 \sqrt{15} q_2^3 q_3^3 \\
&-
         54371797200 q_2^2 q_3^4 - 418389450 \sqrt{15} q_2 q_3^5 -
         172155375 q_3^6))) + 50 p_3 (946148 \rmi \sqrt{3990} p_3^4\\
& +
     5 p_3^2 (-122280569379 q_1^3 + 171 \rmi \sqrt{266} q_1^2
        (172161162 q_2 + 12932861 \sqrt{15} q_3)\\
& +
       342 q_1 (1655383241 q_2^2 + 319116756 \sqrt{15} q_2 q_3 +
         91659330 q_3^2) - 2 \rmi \sqrt{266} (6445964502 q_2^3\\
& +
         2105853303 \sqrt{15} q_2^2 q_3 + 2294558280 q_2 q_3^2 -
         11826850 \sqrt{15} q_3^3))\\
& +
     90 (-4080172176 \rmi \sqrt{3990} q_1^6 - 1666737 q_1^5
        (251098 \sqrt{15} q_2 + 901885 q_3) \\
&+ 48735 \rmi \sqrt{266} q_1^4
        (1599714 \sqrt{15} q_2^2 + 13818360 q_2 q_3 + 449075 \sqrt{15}
          q_3^2) \\
&+ 97470 q_1^3 (10497416 \sqrt{15} q_2^3 +
         409089485 q_2^2 q_3 + 17776120 \sqrt{15} q_2 q_3^2 +
         7253595 q_3^3)\\
& + 570 \rmi \sqrt{266} q_1^2
        (110678928 \sqrt{15} q_2^4 - 6424828110 q_2^3 q_3 -
         745058085 \sqrt{15} q_2^2 q_3^2\\
& - 22200210 q_2 q_3^3 -
         5745035 \sqrt{15} q_3^4) + 228 q_1 (7552766358 \sqrt{15} q_2^5 -
         161449719700 q_2^4 q_3\\
& - 40151682900 \sqrt{15} q_2^3 q_3^2 -
         15760699275 q_2^2 q_3^3 + 17413200 \sqrt{15} q_2 q_3^4 +
         521797500 q_3^5)\\
& - 4 \rmi \sqrt{266} (9011136556 \sqrt{15} q_2^6 -
         114005567430 q_2^5 q_3 - 56141394100 \sqrt{15} q_2^4 q_3^2\\
& -
         52084382850 q_2^3 q_3^3 - 2311027175 \sqrt{15} q_2^2 q_3^4 +
         3038175000 q_2 q_3^5 + 59134250 \sqrt{15} q_3^6))))
 \end{split}
\]

\bibliographystyle{ajmplain}
%\bibliography{mathreva,ajm,yoshida,morales,books,ziglin,churchill,dgt,mp,grammaticos,oldies,audin,moulin,newton,ploski,gramat,hietarinta,rauch,bertrand,darboux,halphen}

\begin{thebibliography}{10}

\bibitem{Aizenberg:83::}
A{\u\i}zenberg, I.~A. and Yuzhakov, A.~P., \emph{Integral representations and
  residues in multidimensional complex analysis}, volume~58 of
  \emph{Translations of Mathematical Monographs}, American Mathematical
  Society, Providence, RI, 1983.

\bibitem{Almeida:98::}
{Almeida}, M.~A., {Moreira}, I.~C., and {Santos}, F.~C., {On the Ziglin-Yoshida
  Analysis for Some Classes of Homogeneous Hamiltonian Systems}, \emph{Brazil.
  J. Phys.}, 28:470--480, 1998.

\bibitem{Audin:01::c}
Audin, M., \emph{Les syst\`emes hamiltoniens et leur int\'egrabilit\'e}, Cours
  Sp\'ecialis\'es 8, Collection SMF, SMF et EDP Sciences, Paris, 2001.

\bibitem{Baum:70::}
Baum, P.~F. and Bott, R., On the zeros of meromorphic vector-fields, in
  \emph{Essays on Topology and Related Topics (M\'emoires d\'edi\'es \`a
  Georges de Rham)}, pages 29--47, Springer, New York, 1970.

\bibitem{Biernat:89::}
Biernat, G., Reduction of two-dimensional residues to the one-dimensional case,
  \emph{Bull. Soc. Sci. Lett. \L \'od\'z}, 39(15):14, 1989.

\bibitem{Biernat:91::}
Biernat, G., La repr\'esentation param\'etrique d'un r\'esidu
  multidimensionnel, \emph{Rev. Roumaine Math. Pures Appl.}, 36(5-6):207--211,
  1991.

\bibitem{Biernat:92::}
Biernat, G., On the {J}acobi-{K}ronecker formula for a polynomial mapping
  having zeros at infinity, \emph{Bull. Soc. Sci. Lett. \L \'od\'z S\'er. Rech.
  D\'eform.}, 14(131-140):103--111, 1992/93.

\bibitem{Cox:97::}
Cox, D., Little, J. and O'Shea, D., \emph{Ideals, varieties, and algorithms.An introduction to computational algebraic geometry and commutative algebra},  \emph{Undergraduate Texts in Mathematics},
  Springer-Verlag, New York, second edition,1997.

\bibitem{Cox:05::}
Cox, D., Little, J. and O'Shea, D., \emph{Using algebraic geometry}, volume 185 of \emph{Graduate Texts in Mathematics},
Springer-Verlag, New York, second edition, 2005.

\bibitem{Duval:08::}
Duval, G. and Maciejewski, A. J., Jordan obstruction to the integrability of homogeneous potentials, 2008, preprint.

\bibitem{Guillot:01::}
Guillot, A., \emph{Champs quadratiques uniformisables}, Ph.D. thesis, \`Ecole
  Normale Sup\'erieure de Lyon, France, 2001.

\bibitem{Guillot:04::}
Guillot, A., Un th\'eor\`eme de point fixe pour les endomorphismes de l'espace
  projectif avec des applications aux feuilletages alg\'ebriques, \emph{Bull.
  Braz. Math. Soc. (N.S.)}, 35(3):345--362, 2004.


\bibitem{Griffiths:78::}
Griffiths, P. and Harris, J., \emph{Principles of algebraic geometry},
  Wiley-Interscience [John Wiley \& Sons], New York, 1978.

\bibitem{Griffiths:76::}
Griffiths, P.~A., {Variations on a Theorem of Abel.}, \emph{Invent. Math.},
  35:321--390, 1976.

\bibitem{Hietarinta:83::}
Hietarinta, J., A search for integrable two-dimensional {H}amiltonian systems
  with polynomial potential, \emph{Phys. Lett. A}, 96(6):273--278, 1983.

\bibitem{Hietarinta:87::}
Hietarinta, J., Direct methods for the search of the second invariant,
  \emph{Phys. Rep.}, 147(2):87--154, 1987.

\bibitem{Khimshiashvili:06::}
Khimshiashvili, P.~A., G., { Multidimensional Residues and Polynomial
  Equations}, \emph{J. Math. Sci.}, 132(6):757--804, 2006.

\bibitem{Kimura:69::}
Kimura, T., On {R}iemann's equations which are solvable by quadratures,
  \emph{Funkcial. Ekvac.}, 12:269--281, 1969/1970.

\bibitem{Kozlowski:78::}
Koz{\l}owski, A., Remark on systems of algebraic equations, \emph{Bull. Soc.
  Sci. Lett. \L \'od\'z}, 28(11):1--4, 1978.

\bibitem{mp:04::c}
Maciejewski, A.~J. and Przybylska, M., Darboux polynomials and first integrals of natural polynomial {H}amiltonian systems, \emph{Phys.  Lett. A}, 326(3-4):219--226, 2004.

\bibitem{mp:04::d}
Maciejewski, A.~J. and Przybylska, M., All meromorphically integrable 2{D}
  {H}amiltonian systems with homogeneous potentials of degree 3, \emph{Phys.
  Lett. A}, 327(5-6):461--473, 2004.

\bibitem{mp:05::c}
Maciejewski, A.~J. and Przybylska, M., Darboux points and integrability of
  {H}amiltonian systems with homogeneous polynomial potential, \emph{J. Math.
  Phys.}, 46(6):062901--1--33, 2005.

\bibitem{Morales:99::c}
Morales~Ruiz, J.~J., \emph{Differential {G}alois theory and non-integrability
  of {H}amiltonian systems}, volume 179 of \emph{Progress in Mathematics},
  Birkh\"auser Verlag, Basel, 1999.

\bibitem{Morales:00::a}
Morales-Ruiz, J.~J., Kovalevskaya, {L}iapounov, {P}ainlev\'e, {Z}iglin and the
  differential {G}alois theory, \emph{Regul. Chaotic Dyn.}, 5(3):251--272,
  2000.

\bibitem{Morales:01::b1}
Morales-Ruiz, J.~J. and Ramis, J.~P., Galoisian obstructions to integrability
  of {H}amiltonian systems. {I}, \emph{Methods Appl. Anal.}, 8(1):33--95, 2001.

\bibitem{Morales:01::b2}
Morales-Ruiz, J.~J. and Ramis, J.~P., Galoisian obstructions to integrability
  of {H}amiltonian systems. {II}, \emph{Methods Appl. Anal.}, 8(1):97--111,
  2001.

\bibitem{Morales:01::a}
Morales-Ruiz, J.~J. and Ramis, J.~P., A note on the non-integrability of some
  {H}amiltonian systems with a homogeneous potential, \emph{Methods Appl.
  Anal.}, 8(1):113--120, 2001.

\bibitem{Morales:06::}
Morales-Ruiz, J.~J., Ramis, J.~P., and Sim{\'o}, C., Integrability of
  {H}amiltonian systems and differential {G}alois groups of higher variational
  equations,  \emph{Ann. Sci. \'Ecole Norm. Sup.},  40(6):845--884, 2007.

\bibitem{Morales:07::}
Morales-Ruiz, J.~J., Ramis, J.~P., Integrability of Dynamical Systems through Differential Galois Theory: a practical guide, 2007, preprint.

\bibitem{mp:05::d}
Nakagawa, K., Maciejewski, A.~J., and Przybylska, M., New integrable
  {H}amiltonian system with quartic in momenta first integral, \emph{Phys.
  Lett. A}, 343(1-3):171--173, 2005.

\bibitem{Ploski:07::}
P{\l}oski, A., O nier\'owno\'sci {B}ezouta ({O}n {B}ezout inequality),
  \emph{Materia{\l}y na XXVIII Konferencj\c{e} z geometrii analitycznej i
  zespolonej}, pages 41--49, 2007, in Polish.

\bibitem{mp:07::h}
Przybylska, M., Darboux points and integrability of {H}amiltonian systems with
  homogeneous polynomial potential {II}, in preparation.

\bibitem{mp:07::a}
Przybylska, M., Finiteness of integrable $n$-dimensional homogeneous polynomial
  potentials, \emph{Phys Lett A.}, 369(3):180--187, 2007.

\bibitem{Shafarevich:77::}
Shafarevich, I. R., \emph{Basic algebraic geometry},
  Springer-Verlag, Berlin, 1977.

\bibitem{Schwarz:1872::}
Schwarz, H.~A., Ueber diejenigen f\"alle, in welchen die gaussische
  hypergeometrische reihe eine algebraische function ihres vierten elementes
  darstellt, \emph{J. f\"ur die reine und angew. Math.}, 75:292--335, 1872.

\bibitem{Tsikh:92::}
Tsikh, A.~K., \emph{Multidimensional residues and their applications}, volume
  103 of \emph{Translations of Mathematical Monographs}, American Mathematical
  Society, Providence, RI, 1992.

\bibitem{Yoshida:87::a}
Yoshida, H., A criterion for the nonexistence of an additional integral in
  {H}amiltonian systems with a homogeneous potential, \emph{Phys. D},
  29(1-2):128--142, 1987.

\bibitem{Yoshida:88::b}
Yoshida, H., Nonintegrability of the truncated {T}oda lattice {H}amiltonian at
  any order, \emph{Comm. Math. Phys.}, 116(4):529--538, 1988.

\bibitem{Ziglin:82::b}
Ziglin, S.~L., Branching of solutions and non-existence of first integrals in
  {H}amiltonian mechanics. {I}, \emph{Functional Anal. Appl.}, 16:181--189,
  1982.

\bibitem{Ziglin:83::b}
Ziglin, S.~L., Branching of solutions and non-existence of first integrals in
  {H}amiltonian mechanics. {II}, \emph{Functional Anal. Appl.}, 17:6--17, 1983.

\end{thebibliography}
\newcommand{\noopsort}[1]{}\def\polhk#1{\setbox0=\hbox{#1}{\ooalign{\hidewidth
  \lower1.5ex\hbox{`}\hidewidth\crcr\unhbox0}}} \def\cprime{$'$}
  \def\cydot{\leavevmode\raise.4ex\hbox{.}} \def\cprime{$'$} \def\cprime{$'$}
  \def\polhk#1{\setbox0=\hbox{#1}{\ooalign{\hidewidth
  \lower1.5ex\hbox{`}\hidewidth\crcr\unhbox0}}} \def\cprime{$'$}
  \def\cprime{$'$} \def\cprime{$'$} \def\cprime{$'$} \def\cprime{$'$}
  \def\cprime{$'$}

\end{document}